\documentclass{aa} %

\usepackage{graphicx}
\usepackage{txfonts}
\begin{document} 

\title{Inference of magnetic field strength and density \\from damped transverse coronal waves}

\titlerunning{Inference of magnetic field strength and density from damped transverse coronal waves}
\authorrunning{Arregui et al.}

   \author{I. Arregui
          \inst{1,2}
          \and
          M. Montes-Sol\'{\i}s\inst{1,2}
          \and
          A. Asensio Ramos\inst{1,2}
          }

   \institute{Instituto de Astrof\'{\i}sica de Canarias, E-38205 La Laguna, Tenerife, Spain\\
              \email{iarregui@iac.es}
         \and
            Departamento de Astrof\'{i}sica, Universidad de La Laguna, E-38206 La Laguna, Tenerife, Spain\\
             }

   \date{Received ; accepted}

 
 \abstract{A classic application of coronal seismology uses transverse oscillations of waveguides to obtain estimates of the magnetic field strength. The procedure requires information on the density of the structures. Often, it ignores the damping of the oscillations. We computed marginal posteriors for parameters such as the waveguide density; the density contrast; the transverse inhomogeneity length-scale; and the magnetic field strength, under the assumption that the oscillations can be modelled as standing magnetohydrodynamic (MHD) kink modes damped by resonant absorption. Our results show that the magnetic field strength can be properly inferred, even if the densities inside and outside the structure are largely unknown. Incorporating observational estimates of plasma density further constrains the obtained posteriors. The amount of information one is willing to include (a priori) for the density and the density contrast influences their corresponding posteriors, but very little the inferred magnetic field strength. The decision to include or leave out the information on the damping and the damping time-scales have a minimal impact on the obtained magnetic field strength.  In contrast to the classic method which provides with numerical estimates with error bars or possible ranges of variation for the magnetic field strength, Bayesian methods offer the full distribution of plausibility over the considered range of possible values. The methods are applied to available datasets of observed transverse loop oscillations, can be extended to prominence fine structures or chromospheric spicules and implemented to propagating waves in addition to standing oscillations.} 

   \maketitle
   %
%

\section{Introduction}             

Coronal seismology uses observed and theoretically predicted properties of magnetohydrodynamic (MHD) waves and oscillations to infer plasma and field properties. The method was suggested decades ago by \cite{uchida70}, \cite{rosenberg70}, and \cite{roberts84}. 
The method was first applied to the inference of the magnetic field strength in coronal loops by \cite{nakariakov01}. In observations made by the Transition Region and Coronal Explorer (TRACE) and reported by \cite{aschwanden99} and \cite{nakariakov99}, the observed lateral displacements of coronal loops were interpreted in terms of the fundamental MHD kink mode of a magnetic flux tube. By estimating the phase speed of the waves and associating this observable to the theoretical kink speed in the thin tube approximation, the magnetic field strength could be determined, upon making a number of assumptions on the values of the plasma density inside and outside the coronal loops. The main shortcoming of the method is that values for physical parameters that cannot be directly measured are assumed. Since then, the same method has been widely used to obtain information on the local magnetic field strength in structures such as prominence threads \citep{lin09}, chromospheric spicules \citep{zaqarashvili09}, or coronal streamers \citep{chen11}; in  applications of global seismology using EIT waves  \citep{ballai07,west11,long13,long17}; or in the first application of radio seismology in the outer corona by \cite{zaqarashvili13}. Extended reviews on coronal seismology can be found in \cite{demoortel05,banerjee07,nakariakov08,arregui12a,demoortel12}.

Besides the use of the kinematics of the oscillations, the availability of stereoscopic or spectroscopic information has led to improvements when constraining the magnetic field strength. \cite{verwichte09} presented the first seismological analysis of a transverse loop oscillation observed by both Solar TErrestrial RElations Observatories (STEREO) spacecraft. \cite{vandoorsselaere08} used Hinode/EIS to obtain information on the mass density to infer the local magnetic field strength in a coronal loop with unprecedented accuracy. 
The calculations from simplified theoretical models have been found to be consistent with those deduced from magnetic extrapolation and spectral methods by \cite{verwichte13b}. They have also been compared to results obtained from improved models and realistic numerical simulations in order to assess their reliability.  The non-planarity of coronal loops and the density variation along the loop only weakly affect the estimates of the magnetic field magnitude \citep{scott12}. The presence of field aligned flows can cause the underestimation of the magnetic field strength in coronal loops when using the traditional seismological methods \citep{terradas10}. The studies by \cite{demoortel09} and \cite{pascoe14} have shown that the combined effect of the loop curvature, the density ratio, and aspect ratio can lead to ambiguous estimates of the magnetic field strength, because of the appearance of a secondary mode.  \cite{chen15} have used results from a three-dimensional coronal simulation in which loop oscillations are present to test the inversions based on coronal seismology. The field derived by coronal seismology is found to be about 15\% to 20\% smaller that the average field strength in their simulation.

Coronal loop oscillations display time damping and this information has been widely used to infer the characteristic spatial scales for the variation of the mass density across the magnetic field in these structures \citep[see e.g.,][for a number of examples]{arregui07a,goossens08a,pascoe13,arregui14,arregui15c,pascoe16}.  The observed damping has been considered in previous inversions, see e.g., \cite{pascoe16}, but whether or not the consideration of damping time and spatial scales influence the seismological inversion of the magnetic field strength remains unknown.

The purpose of the present study is twofold. First, we applied Bayesian methods to the solution of inverse problems to infer the magnetic field strength \citep[see][for a recent review on Bayesian coronal seismology]{arregui18}.  The reason to use Bayesian analysis is the lack of direct access to the physical conditions of interest which forces us to use indirect observational information which is always incomplete and uncertain. 
Extracting information on physical parameters by comparison of theoretical predictions with observed data has therefore to be carried out in a probabilistic framework. This means that our conclusions will at best be probabilities, in the form of posterior probability density functions. These posteriors arise from a principled way of combining prior information, model predictions, and observations, providing inferences that are conditional on the data. Second, we compared results on the inference of the magnetic field strength with and without considering the damping of transverse oscillations to see to what extent the inclusion of damping time scales has an impact on the inference results. The analysis is performed for standing transverse waves under coronal conditions, but can easily be generalised to propagating waves and to plasmas with chromospheric or prominence properties.

The layout of the paper is as follows. In Section~\ref{inference} our inference method is described. Our results are shown in Section~\ref{results} where different inference problems are solved under different knowledge circumstances. We present our conclusions in Section~\ref{conclusions}.

\section{Inference method}\label{inference} 

We adopt the methods of probabilistic inference in the Bayesian framework which consider any inversion problem as the task of 
estimating the degree of belief on statements about parameter values, conditional on observed data. The methods rely on the use of Bayes theorem, 

\begin{equation}\label{bayes}
p(\mbox{\boldmath{$\theta$}} | D, M) =\frac{p(D| \mbox{\boldmath$\theta$}, M) p(\mbox{\boldmath$\theta$})}{p(D)},
\end{equation}
which says that our state of knowledge on a given parameter set, $\mbox{\boldmath$\theta$}$, conditional on the observed data, $D$, and the assumed theoretical model, $M$, is a combination of what we know independently of the data, the so-called prior $p(\mbox{\boldmath$\theta$} | M)$, and the likelihood of obtaining the observed data as a function of the parameter vector, $p(D| \mbox{\boldmath$\theta$}, M)$. Their combination leads to the posterior, $p(\mbox{\boldmath$\theta$} | D, M)$, which contains all the available information about the unknown parameters of interest. The denominator is the so-called evidence, a factor that accounts for the full probability of the data. As this quantity is independent of the parameter vector, it just serves as a normalising constant and does not affect the shape of the posteriors. In this study, unless otherwise stated, all probability densities will be normalised so that the full integral is unity.

Once the full posterior is computed, information on a particular parameter can be obtained by performing an integral of the posterior with respect to the remaining parameters to obtain the so-called marginal posterior

\begin{figure*}
\center
\includegraphics[width = 0.49\textwidth]{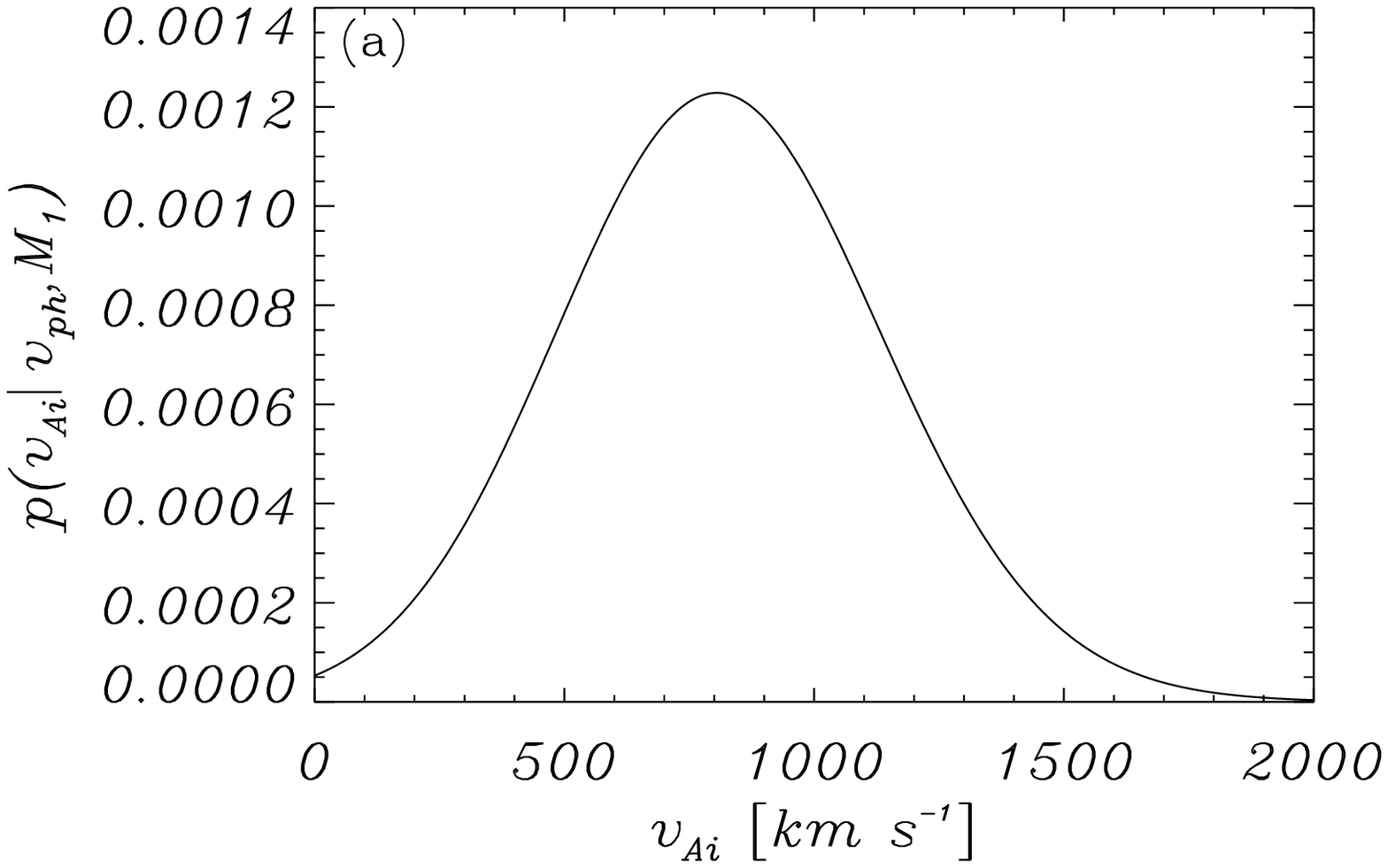} 
\includegraphics[width = 0.49\textwidth]{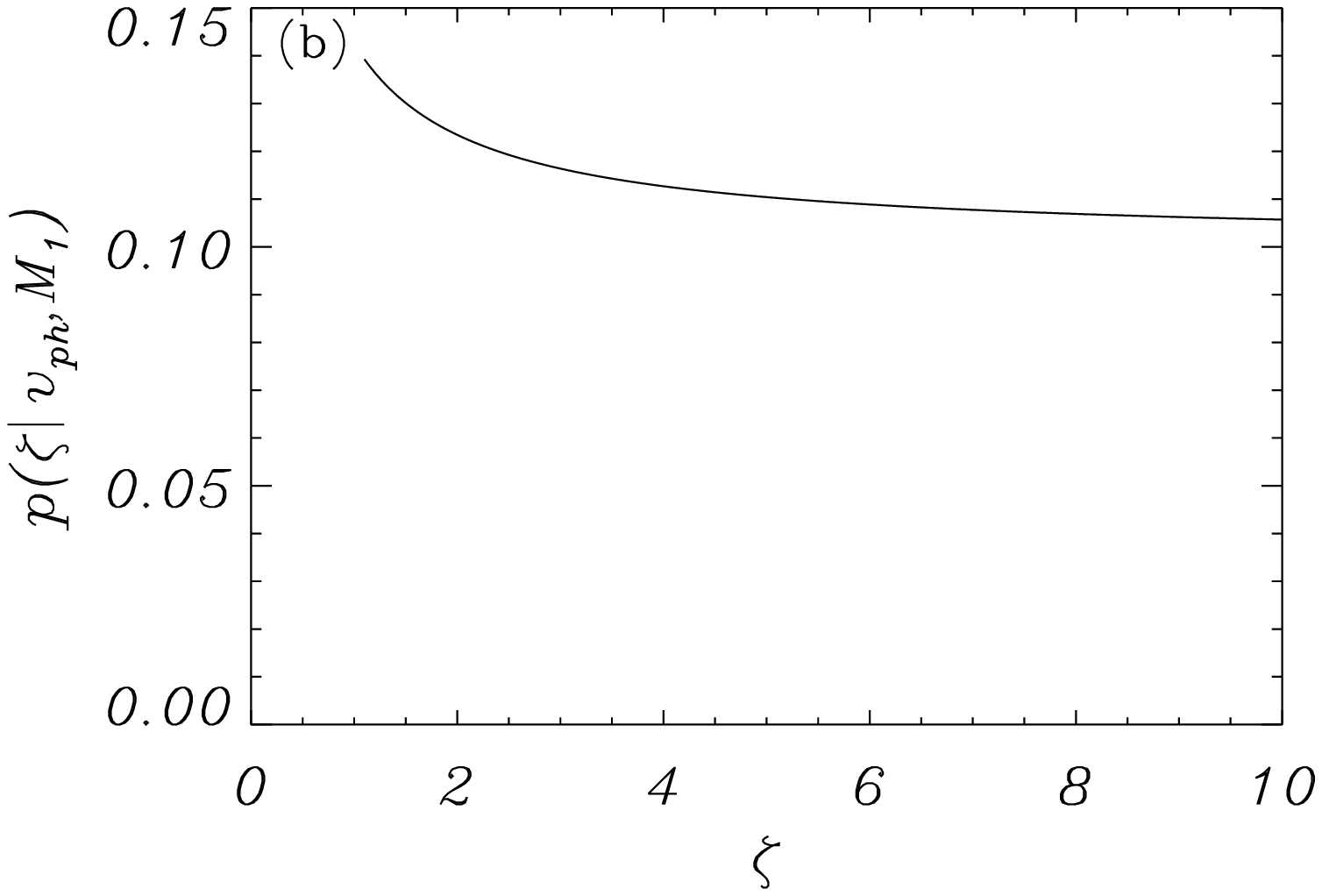} \\
\includegraphics[width = 0.49\textwidth]{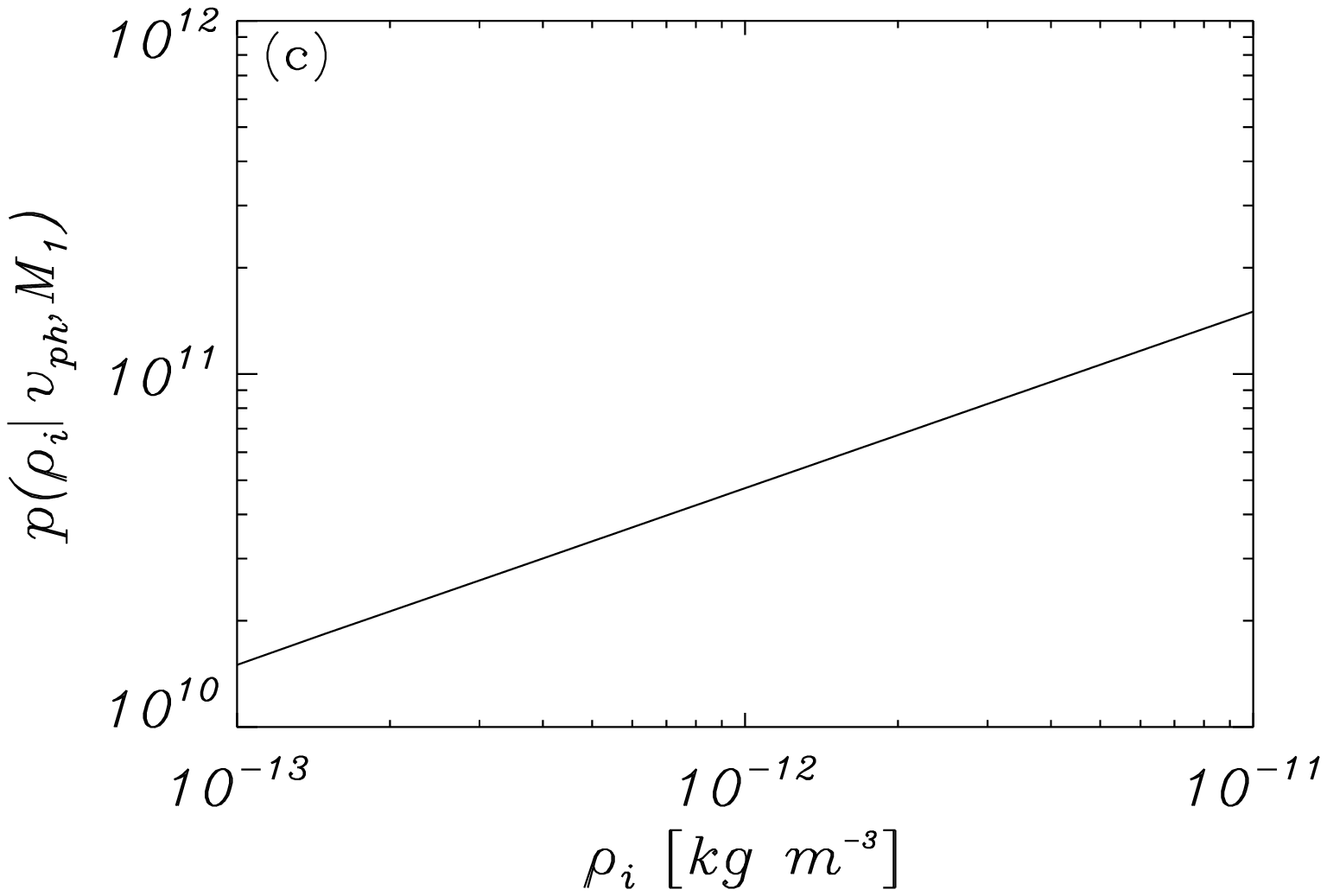} 
\includegraphics[width = 0.49\textwidth]{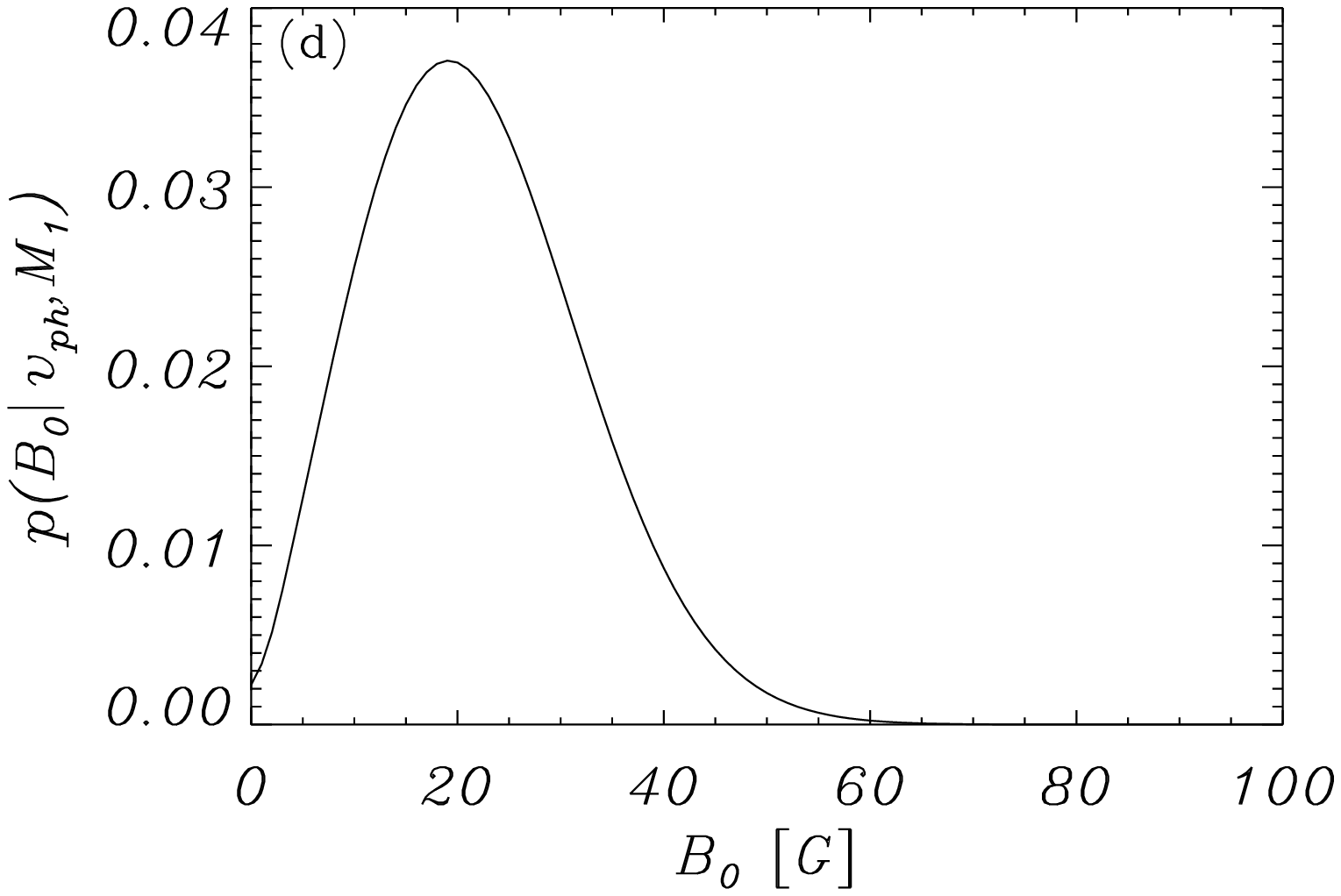} 
\caption{\label{fig1} Posterior probability distributions for (a) the internal Alfv\'en speed, (b) the density contrast, (c) the internal density, and (d) the magnetic field strength for a loop oscillation event with observed phase speed $v_{\rm ph}= 1030\pm 410$ km s$^{-1}$ under model $M_1$, given by Eqs.~(\ref{vphase}) and (\ref{vphaseb0}). The inferred median values for the Alfv\'en speed and the magnetic field strength are $v_{\rm Ai} = 813^{+ 330}_{-317}$ km s$^{-1}$ and $B_0 = 21^{+12}_{-9}$ G respectively, with uncertainties given at the 68\% credible interval. Uniform priors were considered in the intervals $v_{\rm Ai}\in[1,2000]$ km s$^{-1}$, $\zeta\in[1.1,10]$, $\rho_i\in[10^{-13},10^{-11}]$ kg m$^{-3}$, and $B_0\in[0.1,100]$ G. We note that despite the visual impression, the marginal probabilities for $v_{\rm Ai}$ and $B_0$ at their zero values are zero, because of the considered ranges.}
\end{figure*}

\begin{equation}
p(\theta_i | D, M) = \int p(\mbox{\boldmath{$\theta$}} | D, M)\, d\theta_1\ldots d\theta_{i-1}d\theta_{i+1}\ldots d\theta_N.
\end{equation}
To perform the inference and compute either the full posterior or the marginal posterior distribution of a given parameter, different alternatives are available. In low-dimensional parameter spaces, one can still compute the full posterior for different combinations of parameters and then perform a direct numerical integration to obtain the marginal posterior for a given parameter. This approach ceases to be feasible as we increase the complexity and dimensionality of our problem. Then, alternative numerical methods need to be employed to evaluate the relevant distributions by e.g., performing a Markov Chain Monte Carlo (MCMC) sampling of the posterior \citep[see e.g.,][for a recent review]{sharma17}. The simplicity of the models considered in this work still make possible the use of direct integration over a grid of numerical points, although MCMC methods are also used to further confirm our results and in the prior dependency analysis shown in Appendix~\ref{appa}  . 

The posterior probability density in Eq.~(\ref{bayes}) is a derived quantity, while both the likelihood and the prior have to be assigned when constructing the statistical model. The prior probability distribution contains our state of belief on the values the unknown parameters can take before considering the observed data. This information usually comes from past knowledge and experience from which a guess is usually made. The physical model can also impose limitations to the particular value or range of values a parameter can take on. In this study, different prior distributions are employed. When our information on a given parameter is limited to a plausible range of variation, we use a uniform prior probability distribution over the considered range of the form

\begin{equation}
p(\theta_i)=H(\theta_i, \theta^{\mathrm{min}}_i, \theta^{\mathrm{max}}_i),
\end{equation}

\noindent
where $H(x, a, b)$ is the top-hat function

\begin{equation}
H(x, a, b)=\left\{\begin{array}{ll}
\frac{1}{b-a}\hspace{0.8cm} \textrm{$a\le x \le b$},\\\\
0  \hspace{1.1cm} \textrm{otherwise}.\\
\end{array}\right.
\end{equation}


When more specific information on a given parameter value is available from observations, this knowledge will be used to construct a more informative prior using the measured value of the parameter and the error reported from observations to define the mean, $\mu_{\theta_i}$, and standard deviation, $\sigma_{\theta_i}$, of a Gaussian prior of the form

\begin{equation}\label{gaussp}
p(\theta_i)=(2	\pi\sigma^2_{\theta_i})^{-1/2} \exp\left[\frac{-(\theta_i-\mu_{\theta_i})^2}{2\sigma^2_{\theta_i}}\right].
\end{equation}
Finally, it is important to assess the dependence of the inversion result on the prior information. In Appendix~\ref{appa}, another prior distributions are employed in an analysis of the effect of prior information on the obtained results. They include e.g., a Jeffreys type prior to assign a decreasing probability distribution for increasing values of the parameter or Cauchy functions based priors.

Regarding the likelihood function, we will consider that observations are corrupted with Gaussian noise and that they are statistically independent. Then, a given observed variable $D$ and its theoretical prediction $D^{\rm model}$ can be compared by adopting a Gaussian likelihood of the form

\begin{equation}\label{likegeneral}
p(D|\mbox{\boldmath$\theta$}) = \frac{1}{\sqrt{2\pi}\sigma} \exp \left\{-\frac{\left[D-D^{\mathrm{model}}(\mbox{\boldmath$\theta$})\right]^2}{2 \sigma^2} \right\},
\end{equation}
with $\sigma$ the uncertainty associated to the measured $D$. 

\section{Results}\label{results} 

The methods described above are next applied to a number of problems in which the forward and inverse problems, as well as the base-knowledge information is different. In this way, a step-by-step knowledge is acquired on the amount of information that we can gather on the unknown parameters, conditional on the assumptions of each physical model and the available data with their uncertainty.

\begin{figure*}
\center
\includegraphics[width = 0.49\textwidth]{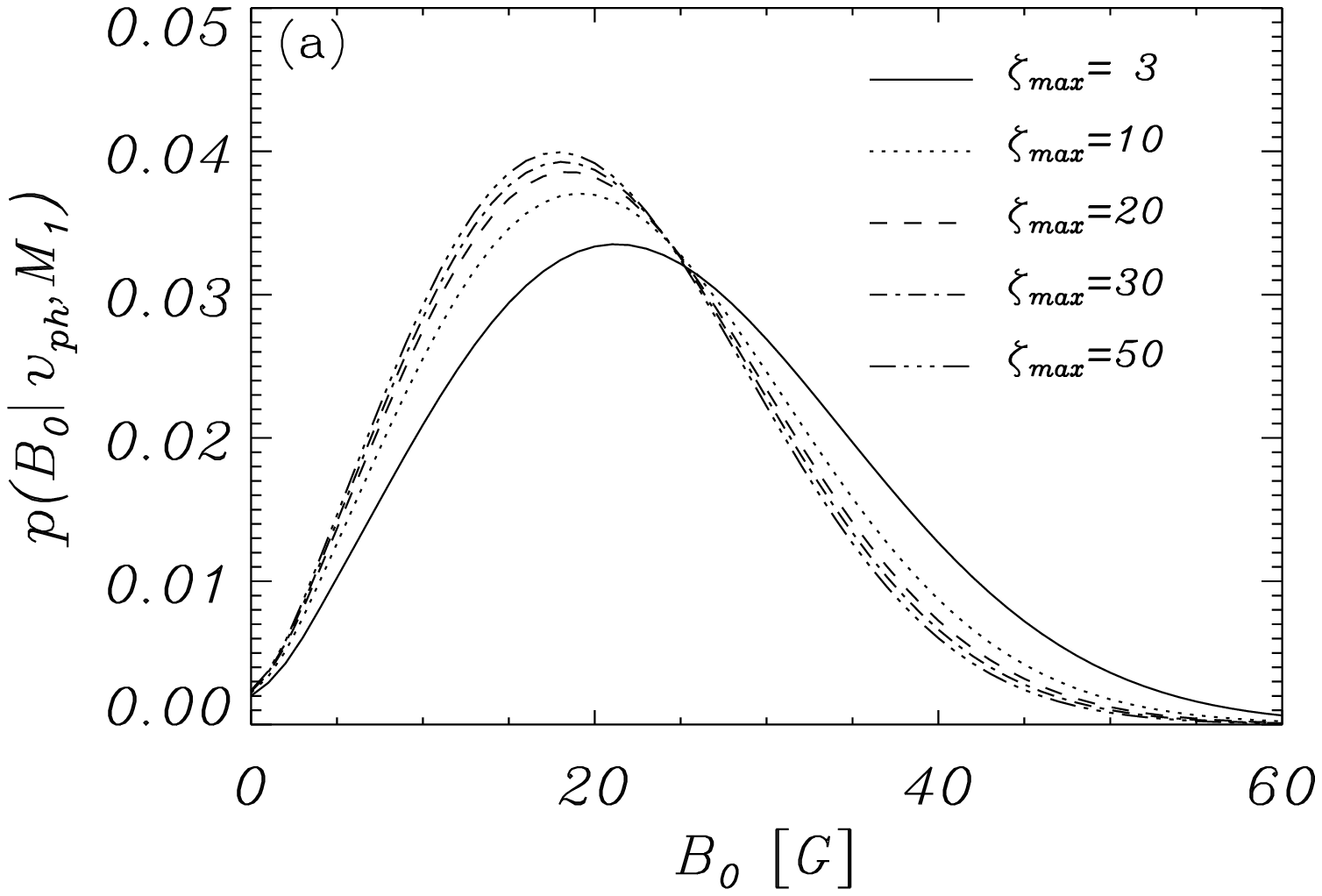} 
\includegraphics[width = 0.49\textwidth]{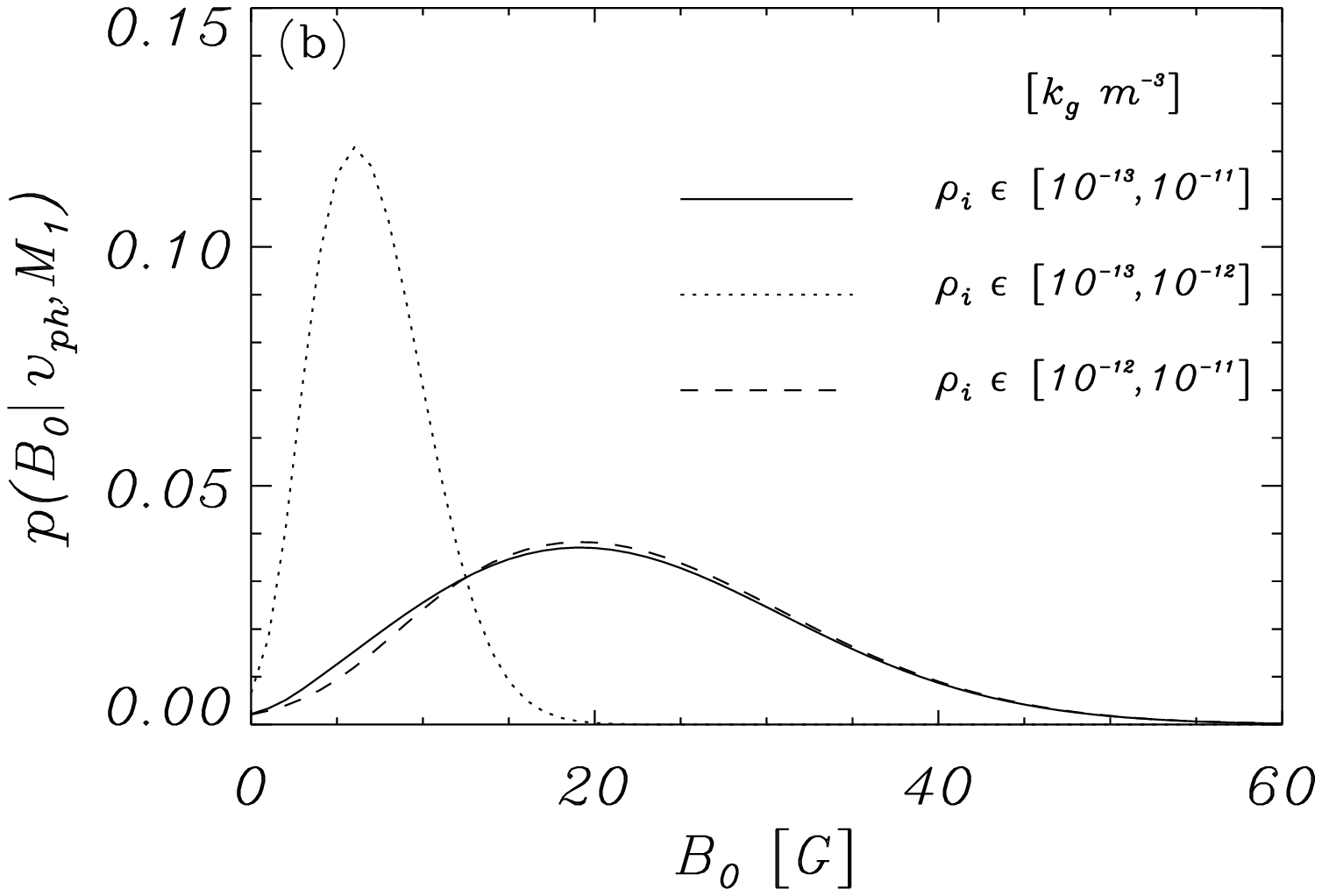} 
\caption{\label{fig2}Marginal posterior distributions for the magnetic field strength for the same loop oscillation event and model as in Fig.~\ref{fig1}, using different ranges for the prior distribution on (a) density contrast and (b) internal density.}
\end{figure*}

\begin{figure*}
\center
\includegraphics[width = 0.49\textwidth]{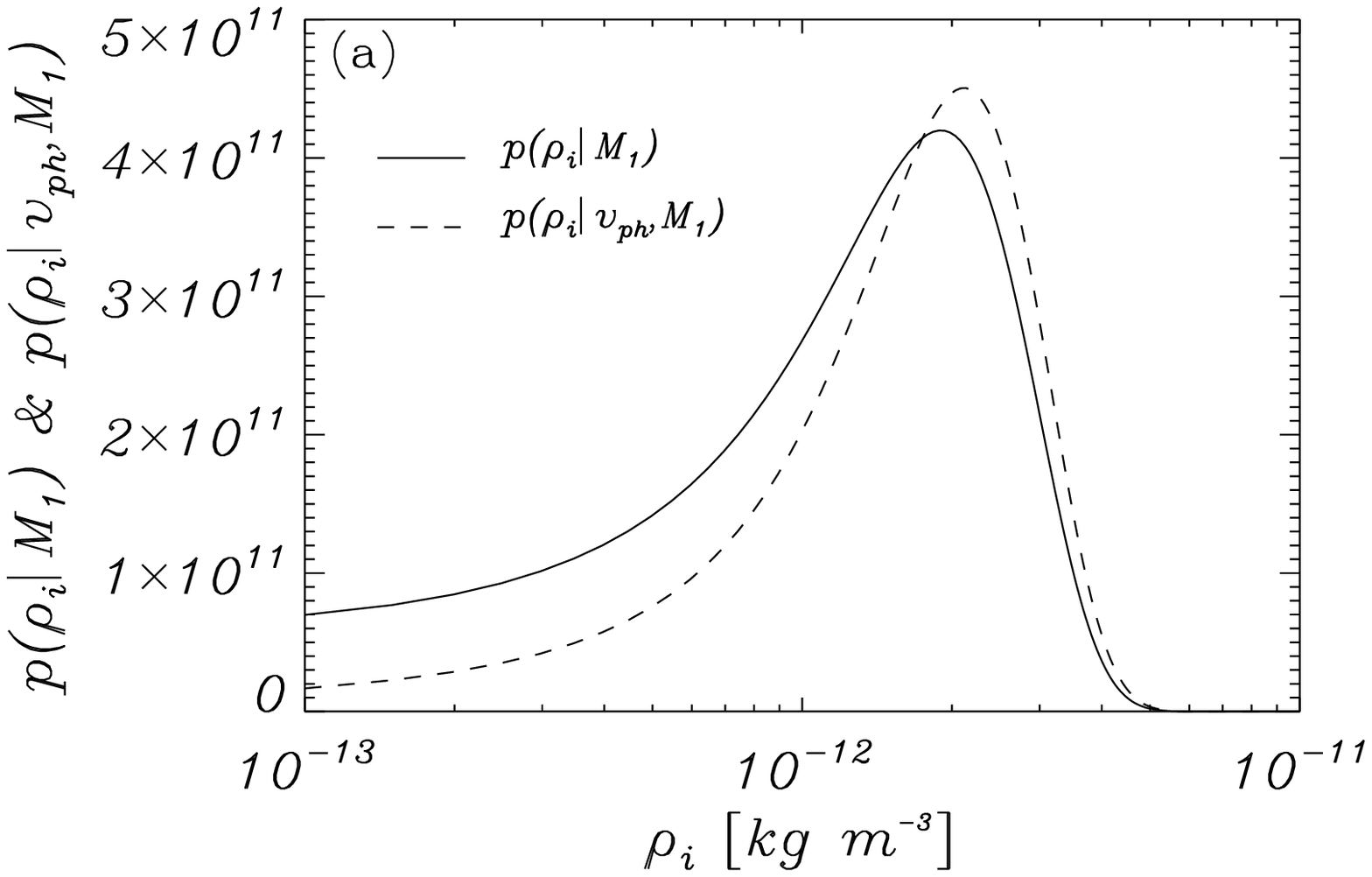} 
\includegraphics[width = 0.49\textwidth]{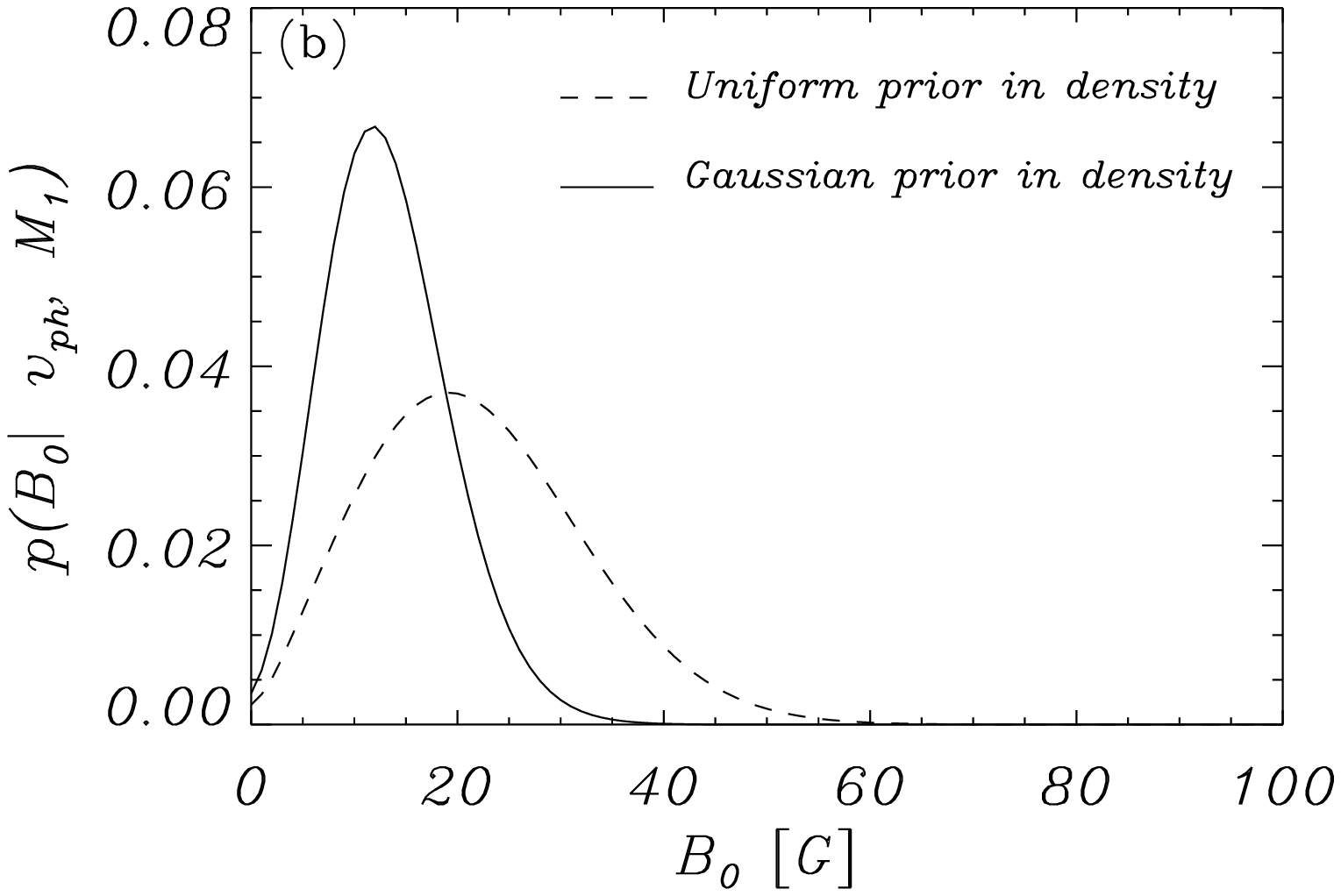} \\
\includegraphics[width = 0.49\textwidth]{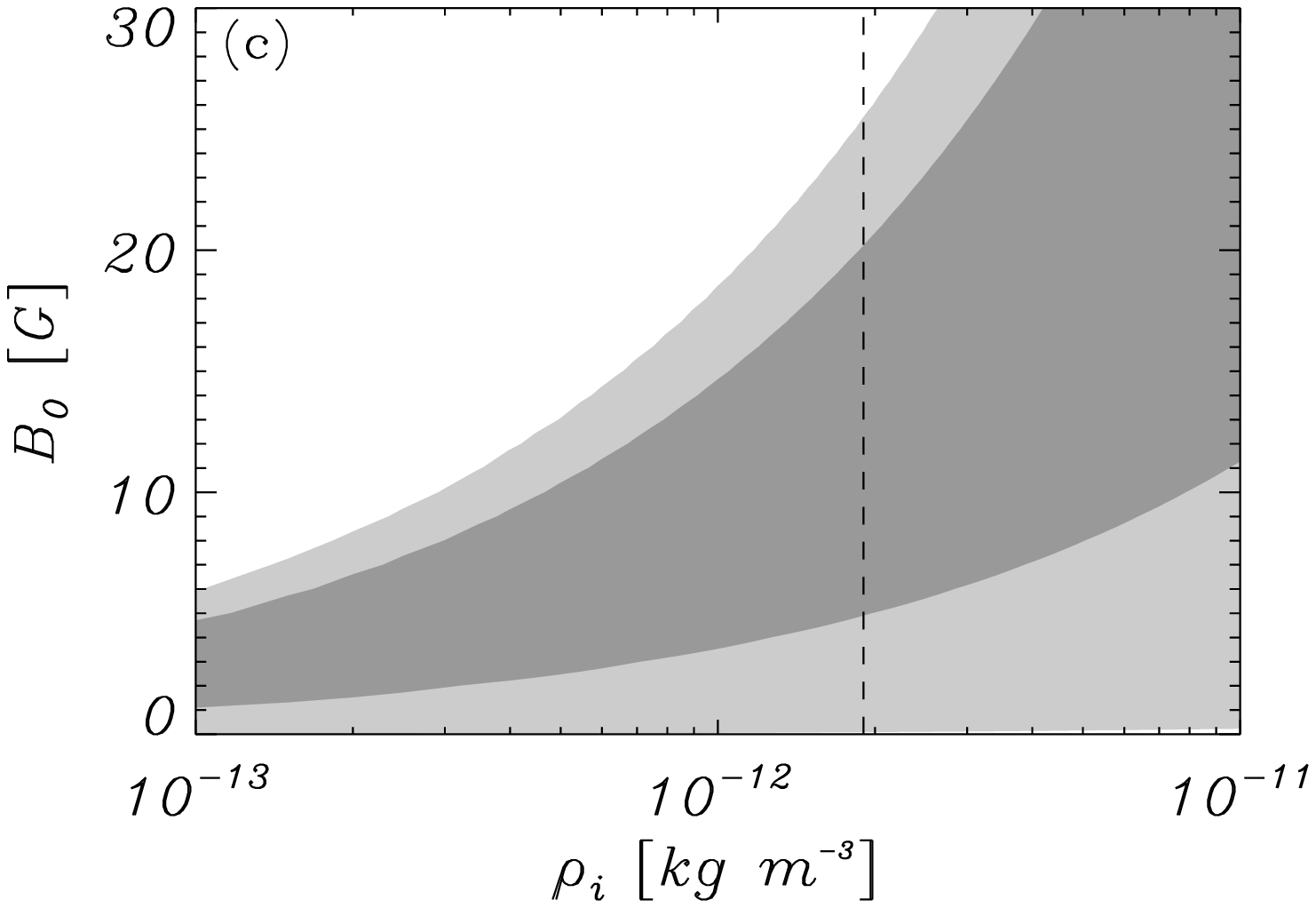} 
\includegraphics[width = 0.49\textwidth]{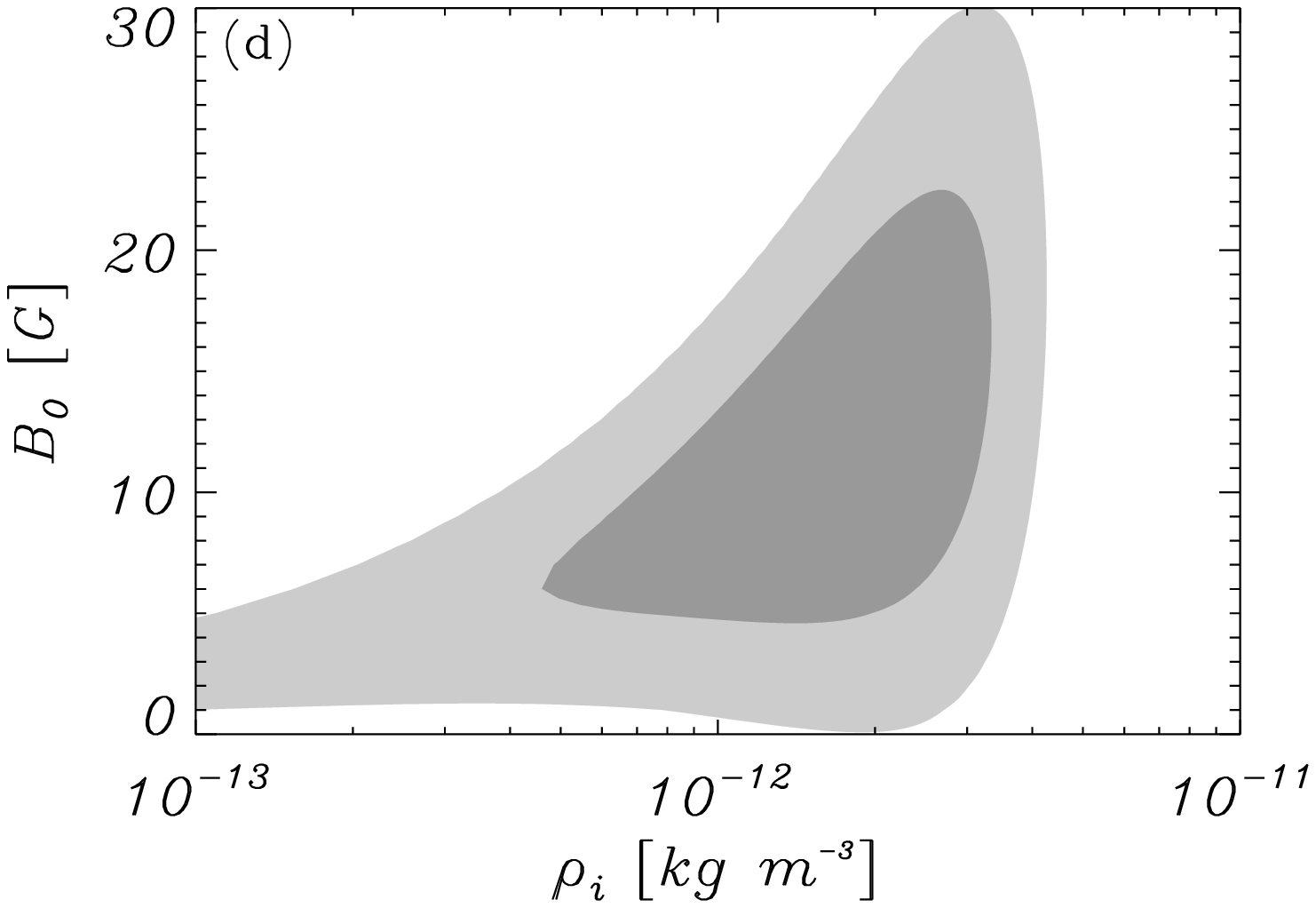} 
\caption{\label{fig3} (a) Prior and posterior distributions for the waveguide density in the inversion of Eq.~(\ref{vphaseb0}) with $v_{\rm ph}= 1030\pm 410$ km s$^{-1}$, under model $M_1$. (b) Comparison between the marginal posteriors for magnetic field strength in the same inversion for the cases of uniform and Gaussian prior on the waveguide density. (c) and (d) Comparison between the joint two-dimensional posterior distributions for the internal density of the waveguide and the magnetic field strength obtained for the inference with $v_{\rm ph}= 1030\pm 410$ km s$^{-1}$, under model $M_1$, for the cases of uniform priors (left) and a Gaussian prior for the internal density with $\mu_{\rho_{\rm i}}=1.9\times 10^{-12}$ kg m$^{-3}$ and $\sigma_{\rho_{\rm i}}=0.5\mu_{\rho_{\rm i}}$ (right). The inference with the more informative prior on density leads to $B_0=13^{+7}_{-6}$ G and $\rho_{\rm i}= (2.2^{+0.9}_{-0.9})\times 10^{-12}$ kg m$^{-3}$. In the bottom panels, the outer boundaries of the light grey and dark grey shaded regions indicate the 95\% and 68\% credible regions.}
\end{figure*}

\begin{figure}
\center
\includegraphics[width = 0.49\textwidth]{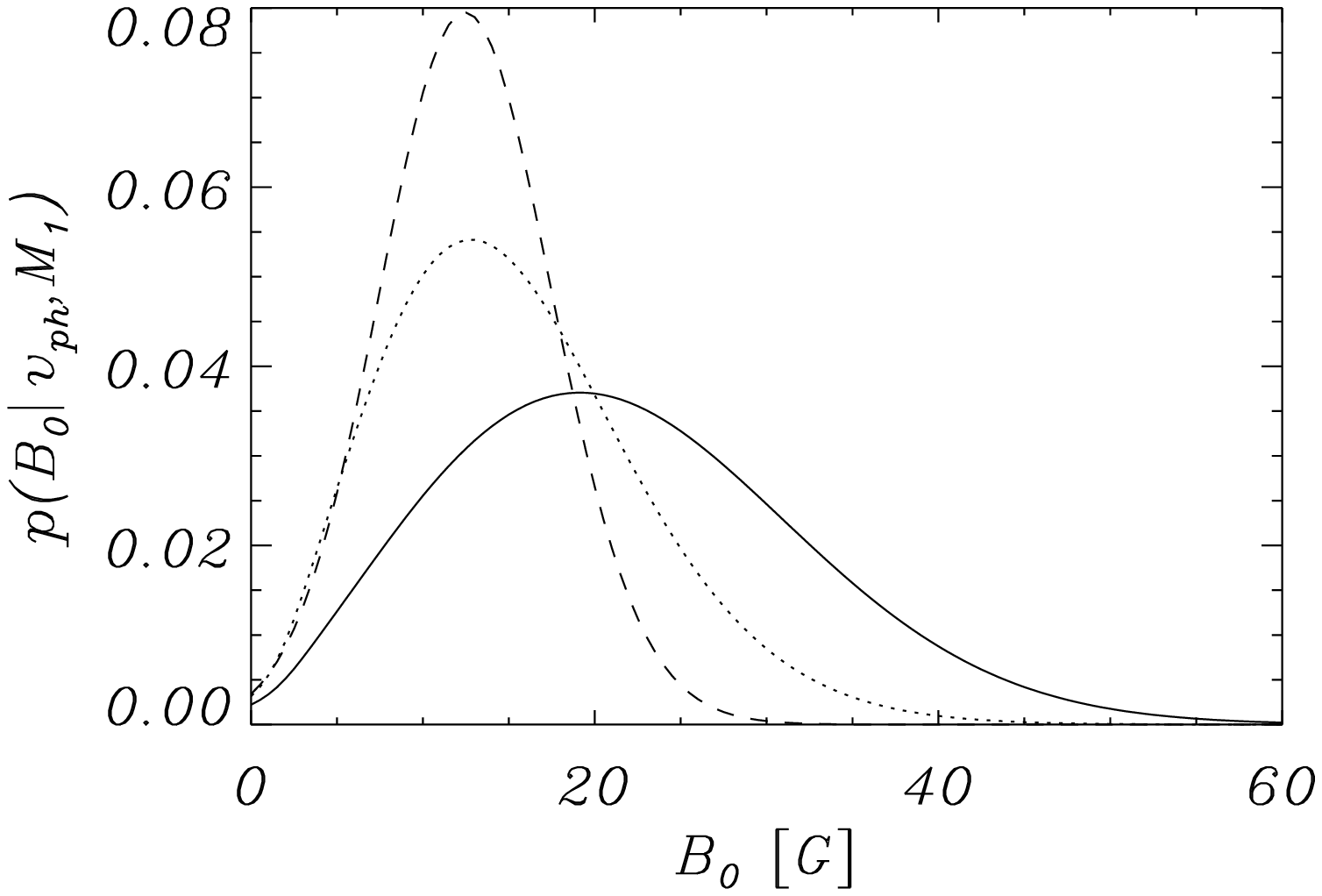} 
\caption{\label{postsb0} Comparison between marginal posterior distributions for the magnetic field strength computed by solving the same inversion problem as in Fig.~\ref{fig1}, under model $M_1$ and using a uniform prior over the full range of plasma density values, $\rho_i\in[10^{-13},10^{-11}]$ kg m$^{-1}$ and then marginalising (solid-line); taking a cut of the two-dimensional joint posterior at the value $\mu_{\rho_i}= 1.9\times 10^{-12}$ kg m$^{-1}$ (dashed-line) and solving the problem using a Gaussian prior with  $\mu_{\rho_i}= 1.9\times 10^{-12}$ kg m$^{-1}$ and $\sigma_{\rho_i}= 0.5 \mu_{\rho_i}$ (dotted-line).}
\end{figure}

\subsection{Internal Alfv\'en speed and magnetic field strength}\label{b0inf}

Let us first consider the simplest inversion problem consisting of inferring the internal Alfv\'en speed in a coronal loop undergoing undamped transverse oscillations interpreted as the fundamental MHD kink mode. Assuming that coronal loops can be modelled as one-dimensional density enhancements in cylindrical geometry and under the thin tube approximation, theory relates the observable phase speed, $v_{\rm ph}$, to the internal Alfv\'en speed, $v_{\rm Ai}$, and the contrast between the internal $\rho_i$ and external density $\rho_e$, $\zeta=\rho_{\rm i}/\rho_{\rm e}$, in the following manner

\begin{equation}\label{vphase}
v_{\rm ph} \sim v_{\rm Ai}\left(\frac{2\zeta}{1 + \zeta}\right)^{1/2}.
\end{equation}
\noindent
We will refer to this as model $M_1$.
\begin{figure*}
\center
\includegraphics[width = 0.33\textwidth]{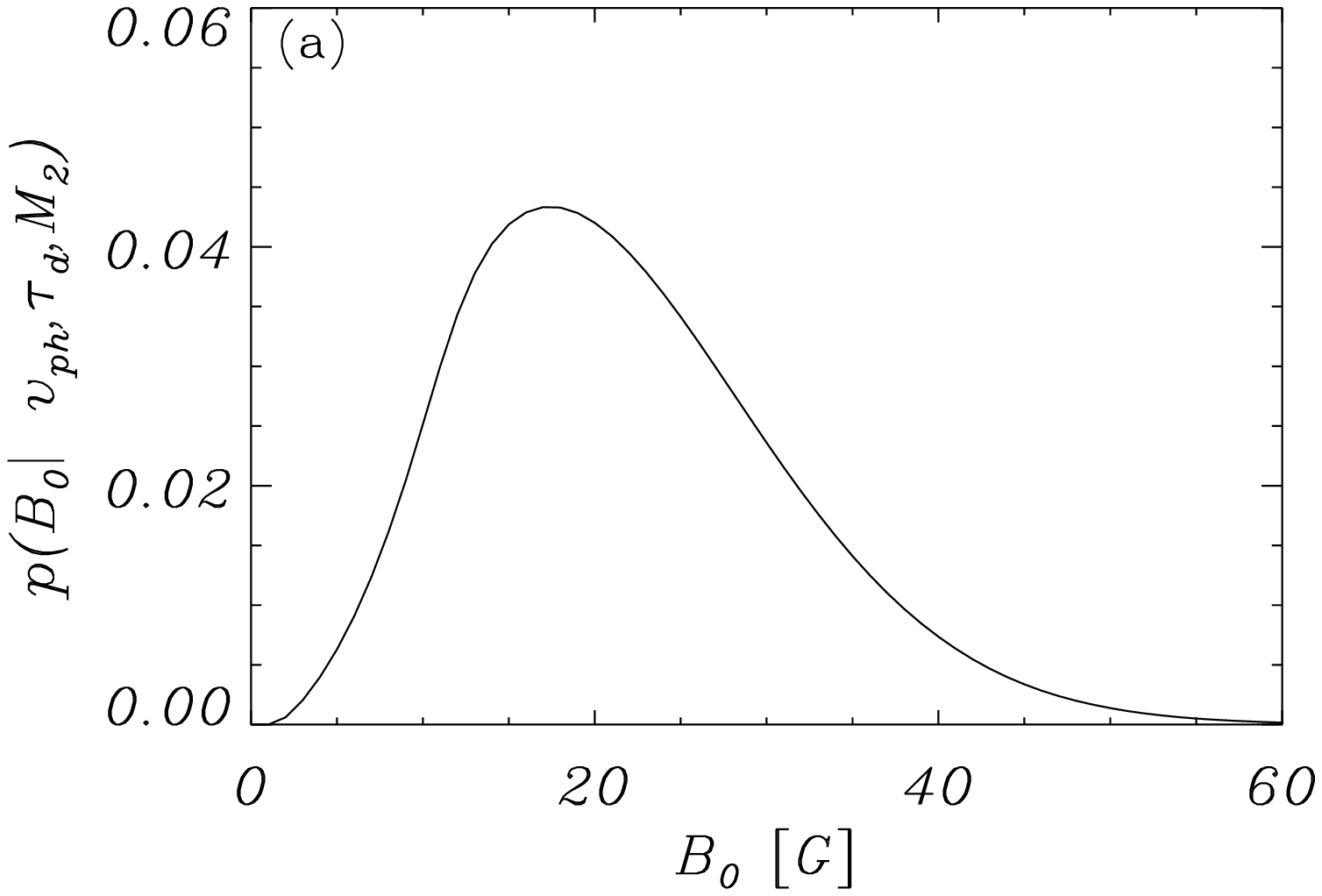} 
\includegraphics[width = 0.33\textwidth]{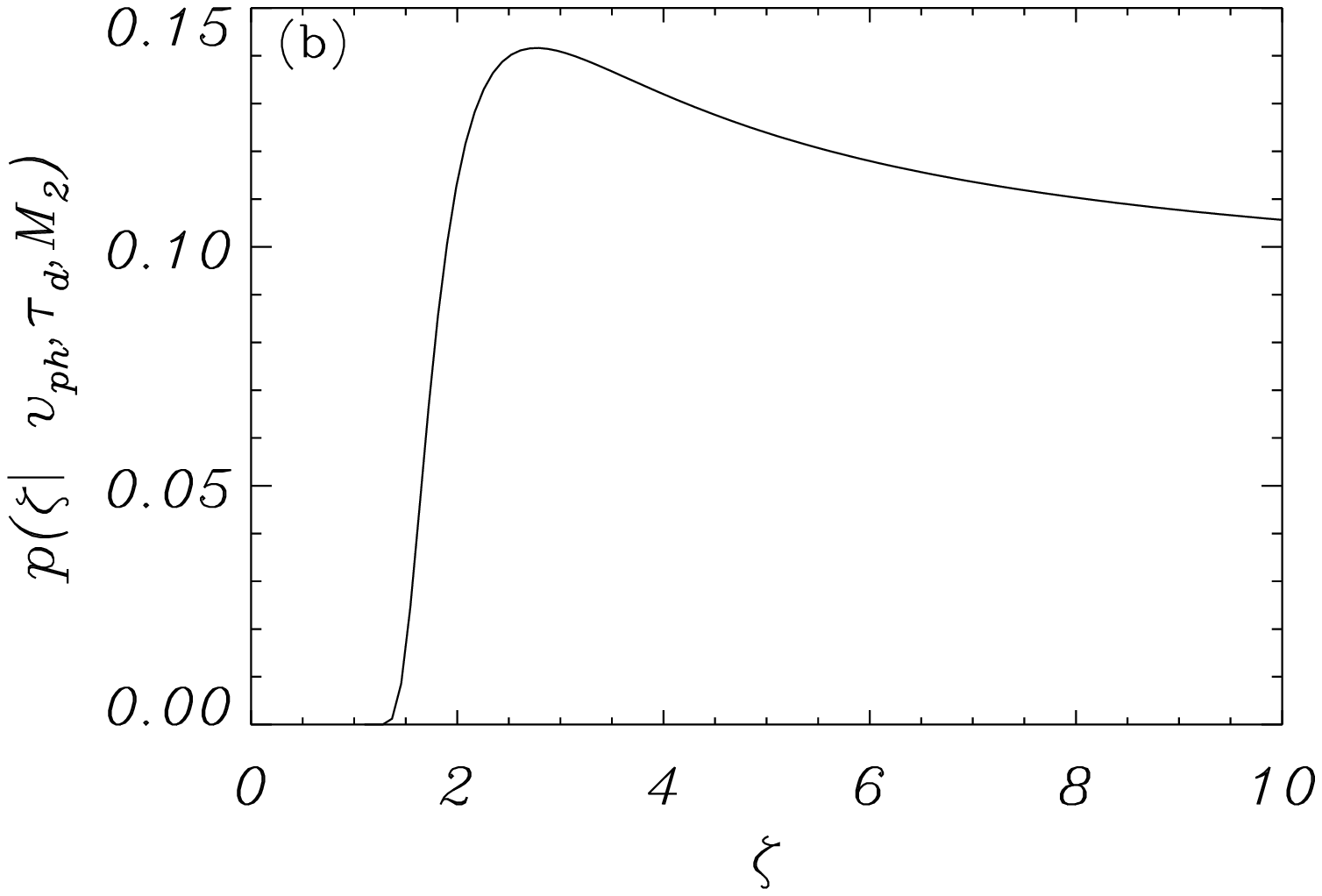} 
\includegraphics[width = 0.33\textwidth]{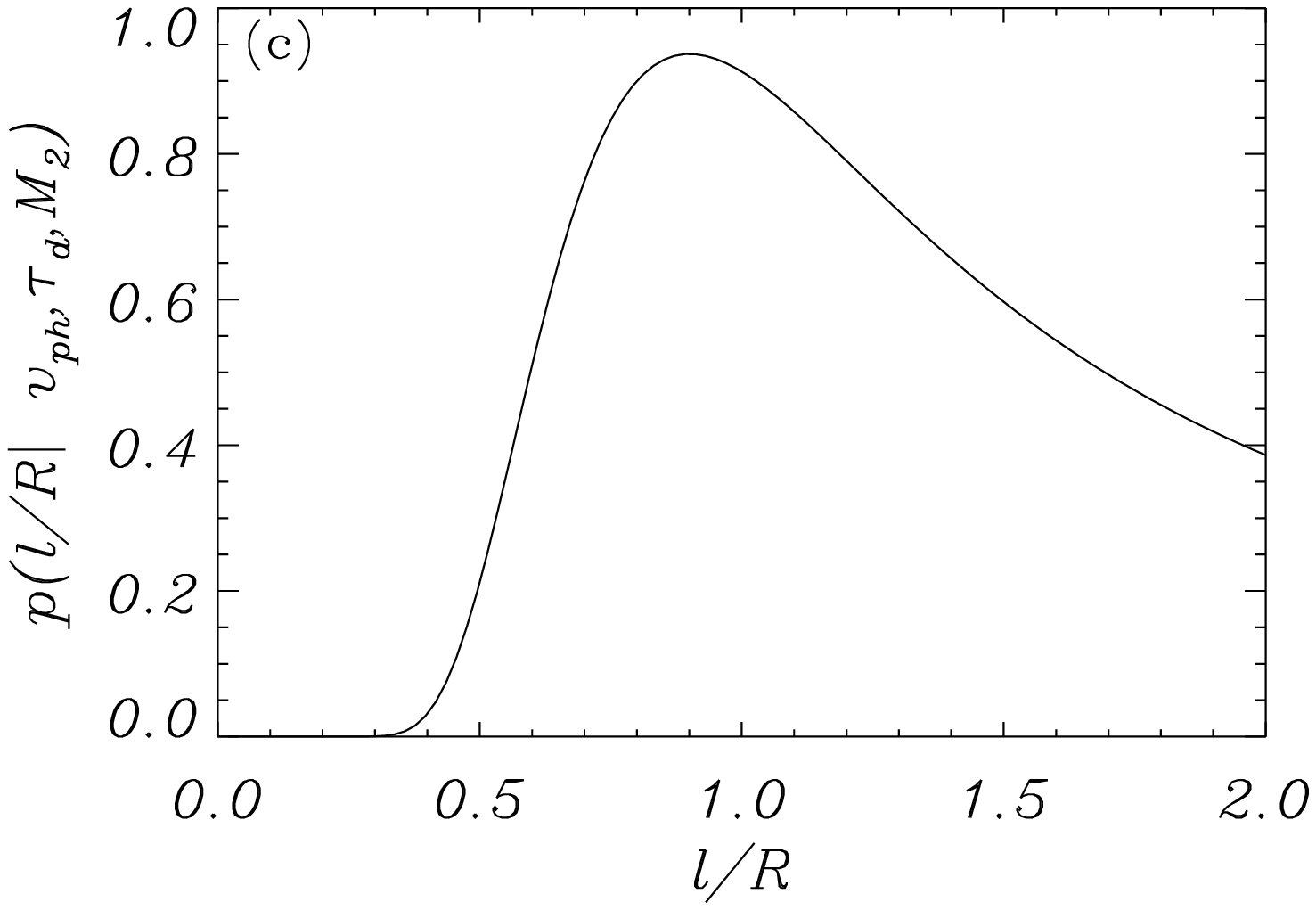} \\
\includegraphics[width = 0.33\textwidth]{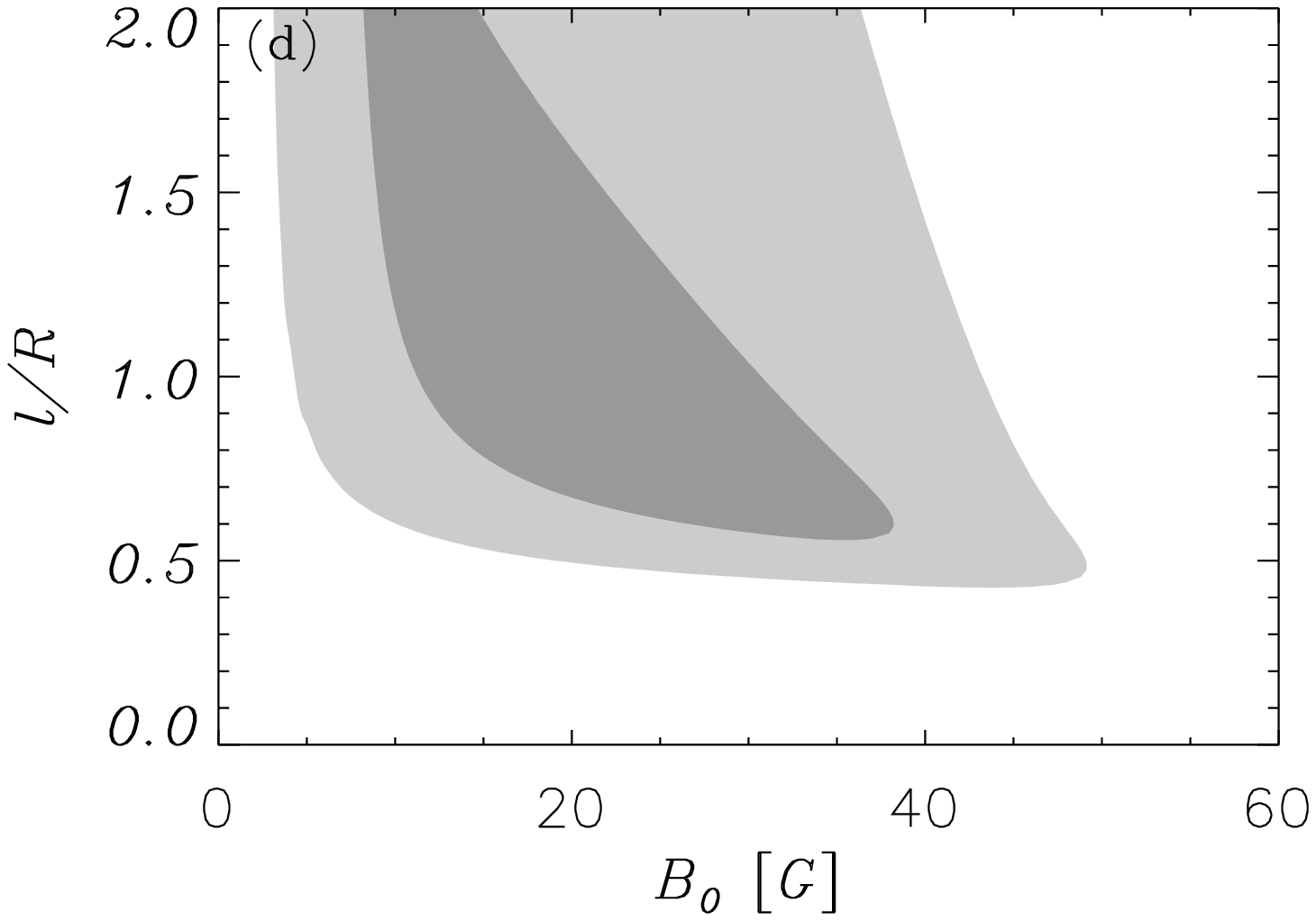} 
\includegraphics[width = 0.33\textwidth]{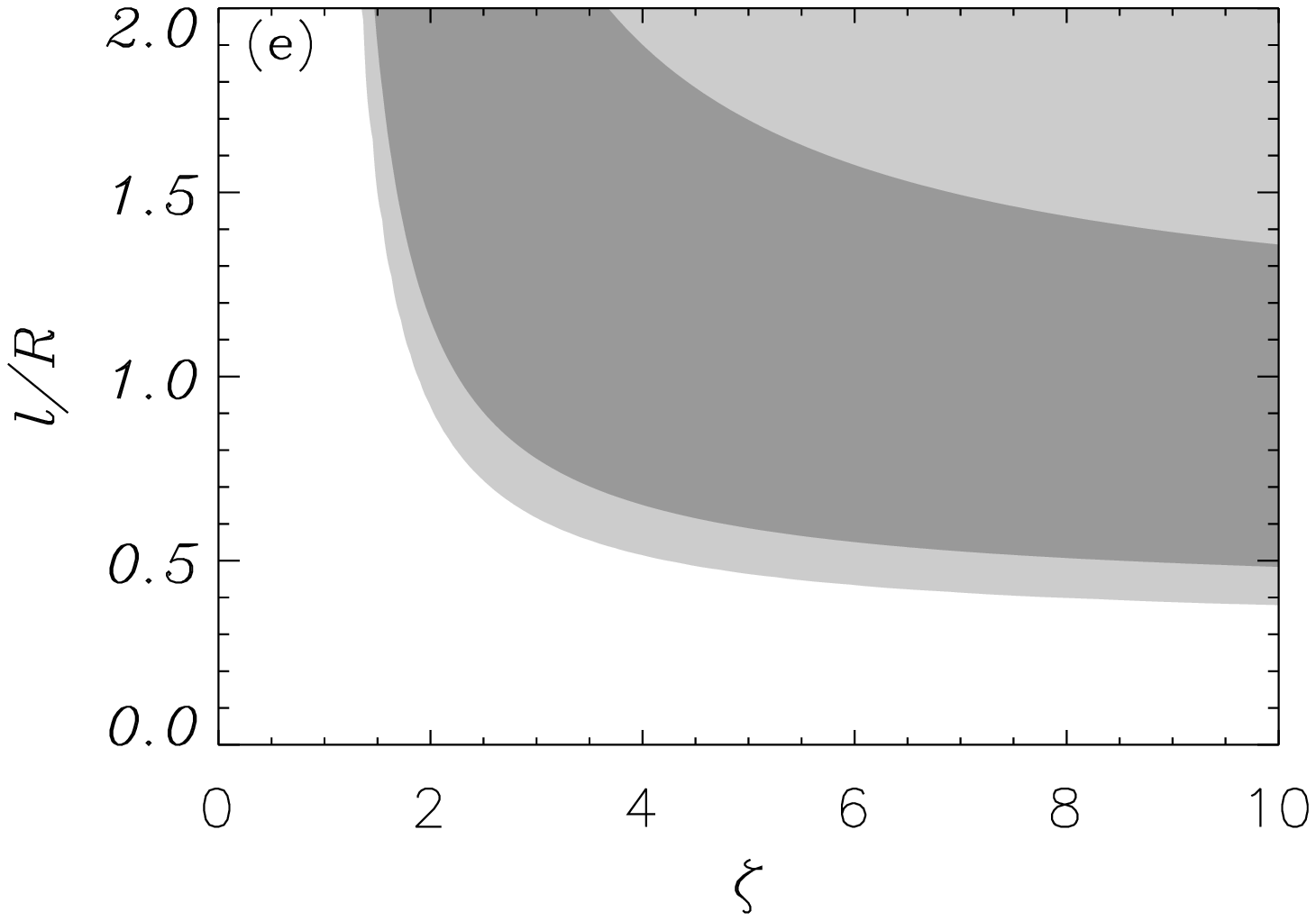} 
\includegraphics[width = 0.33\textwidth]{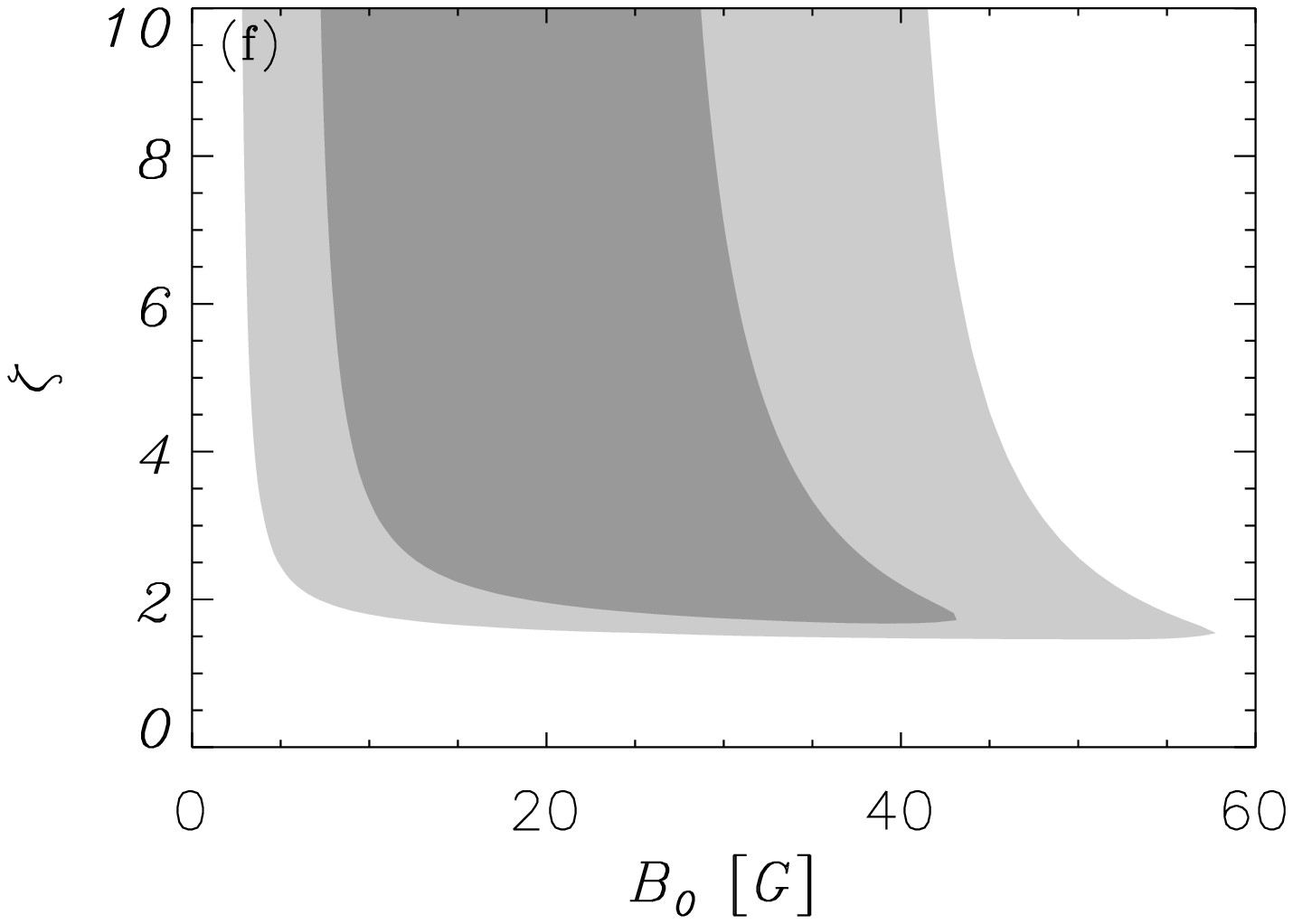} 
\caption{\label{fig5} Top row: marginal posterior distributions for (a) magnetic field strength; (b) density contrast; and (c) transverse inhomogeneity length scale for the inversion of problem with forward model $M_2$, given by Eqs.~(\ref{vphasedamp}) and (\ref{damp2}), and a transverse oscillation with $v_{\rm ph}=1030 \pm 410$ km s$^{-1}$ and a damping time $\tau_{\rm d}=500 \pm 50$ s. Bottom row: joint two-dimensional posterior distributions for  (d) magnetic field strength and transverse density inhomogeneity length-scale; (e) density contrast and transverse  density inhomogeneity length-scale; and (f) magnetic field strength and density contrast. The outer boundaries of the light grey and dark grey shaded regions indicate the 95\% and 68\% credible regions. The inferred median values are $B_{0} = 20^{+ 11}_{-8}$ G and $l/R = 1.2^{+0.5}_{-0.4}$, with uncertainties given at the 68\% credible interval. A fixed value for the loop length, $L=1.9\times 10^{10}$ cm, was considered in this computation.}
\end{figure*}

We first applied our Bayesian scheme to the inference of the internal Alfv\'en speed of coronal loops, considering a particular event, first described by \cite{nakariakov99}, and subsequently analysed by \cite{nakariakov01}.  In the event of the 4th of July, 1999, a period of $P=360$ s and a loop length of $L=1.9\times10^{10}$ cm were measured. The authors report  an estimated phase speed of $v_{\rm ph}=1030\pm 410$ km s$^{-1}$ after considering the uncertainty on the measured variables. Bayes theorem applied to this particular problem tells us that the posterior for the two 
unknowns, $\mbox{\boldmath$\theta$} = \{v_{\rm Ai}, \zeta\}$, conditional on the measured phase speed, $D = v_{\rm ph}$, and the assumed model, $M_1$, is a combination of the likelihood of the data as a function of the unknowns and the prior distributions. Explicitly, 

\begin{equation}\label{bayes1}
p(\{v_{\rm Ai}, \zeta\} | v_{\rm ph},M_1)=\frac{p(v_{\rm ph} | \{v_{\rm Ai}, \zeta\},M_1) p(\{v_{\rm Ai}, \zeta\} |M_1)}{Z_1},
\end{equation}
with $Z_1=\int p(v_{\rm ph} | \{v_{\rm Ai}, \zeta\},M_1) p(\{v_{\rm Ai}, \zeta\} |M_1)dv_{\rm Ai} d\zeta$ the evidence. Considering a  Gaussian likelihood function and uniform prior distributions for the unknowns over plausible ranges leads to the marginal posterior distributions shown in Figs.~\ref{fig1}a and \ref{fig1}b, which indicate that the internal Alfv\'en speed can be properly inferred (Fig.~\ref{fig1}a). Notice that the factor with the square root in Eq.~(\ref{vphase}) is allowed to vary in between 1 and $\sqrt{2}$, when $\zeta$ is allowed to vary in between just a little more than 1 and $\infty$. The classic result is therefore that one could expect $v_{\rm Ai}$ to be constrained to a narrow range, as pointed out by \cite{arregui07a}. However all those values are not equally probable. What our Bayesian result offers is the probability distribution of those possible values within the range. On the other hand, the density contrast cannot be inferred with the information on the phase speed alone (Fig.~\ref{fig1}b).

Equation~(\ref{vphase}) for the wave phase speed can be expanded to incorporate the magnetic field strength, $B_0$, to the inversion. The forward problem for model $M_1$ can now be formulated as

\begin{equation}\label{vphaseb0}
v_{\rm ph} (\zeta, \rho_{\rm i}, B_{\rm 0}) = \frac{B_0}{\sqrt{\mu_0\rho_{	\rm i}}}\left(\frac{2\zeta}{1+\zeta}\right)^{1/2},
\end{equation}
with $\mu_0$ the magnetic permeability. In this case, Bayes theorem provides us with the posterior for three unknowns, $\mbox{\boldmath$\theta$} =\{\rho_{\rm i}, \zeta, B_0\}$ - internal density, density contrast, and magnetic field strength. The explicit expression for the Bayes theorem now reads

\begin{equation}\label{bayes2}
p(\{\rho_{\rm i}, \zeta, B_0\} | v_{\rm ph},M_1)\sim p(v_{\rm ph} | \{\rho_{\rm i}, \zeta, B_0\},M_1) p(\{\rho_{\rm i}, \zeta, B_0\} |M_1).
\end{equation}
where we have omitted the explicit expression for the evidence for brevity.

Figures~\ref{fig1}c and \ref{fig1}d  show the marginal posterior distributions for the loop density and for the magnetic field strength, respectively. The marginal posterior for the density contrast is the same as the one shown in Fig.~\ref{fig1}b. The results now show  a well-constrained distribution for the magnetic field strength, that can therefore be properly inferred.  The same is not true for the loop density, for which no information can be gathered.  Let us compare our result with that by \cite{nakariakov01}. \cite{nakariakov01} obtained a numerical estimate of $B_0=13\pm9$ G, upon employing a density of 10$^{9.3\pm0.3}$ cm$^{-3}$. When allowing the density to vary from $1\times10^{9}$ to $6\times10^{9}$ cm$^{-3}$, a range of variation for $B_0$ was obtained. The main advantage of our Bayesian result is that, when considering a range of possible values for the density, the marginal posterior in Fig.~\ref{fig1}d tells us how the plausibility of the corresponding possible values of the magnetic field strength is distributed. If we wish to use the estimated density, the Bayesian method enables to mimic this by employing a Gaussian prior, as will be discussed in Sect.~\ref{infodensity}. 

In summary, by adopting the simplest possible model for transverse loop oscillations in the long wavelength approximation, measuring the phase speed of the waves, and considering uniform priors for the unknown parameters, the internal Alfv\'en speed and the magnetic field strength can be inferred, even if the densities inside and outside the waveguide are largely unknown.

\subsection{Information on plasma density}\label{infodensity}

Our knowledge on the plasma density inside the waveguide turns out to be an important matter when inferring the magnetic field strength. In the results above, a large range of possible values for the waveguide density was considered, with typical coronal loop densities in the range $\rho_i\in[10^{-13}-10^{-11}]$ kg m$^{-3}$,  corresponding to particle densities in the range $n\sim[10^{8}-10^{10}$] cm$^{-3}$ \citep{priest82}. Also, a density contrast in the range $\zeta\in[1.1-10]$ was fixed. Uniform prior distributions for $\rho_i$ and $\zeta$ were considered over those ranges. 

\begin{figure}[!t]
\center
\includegraphics[width = 0.49\textwidth]{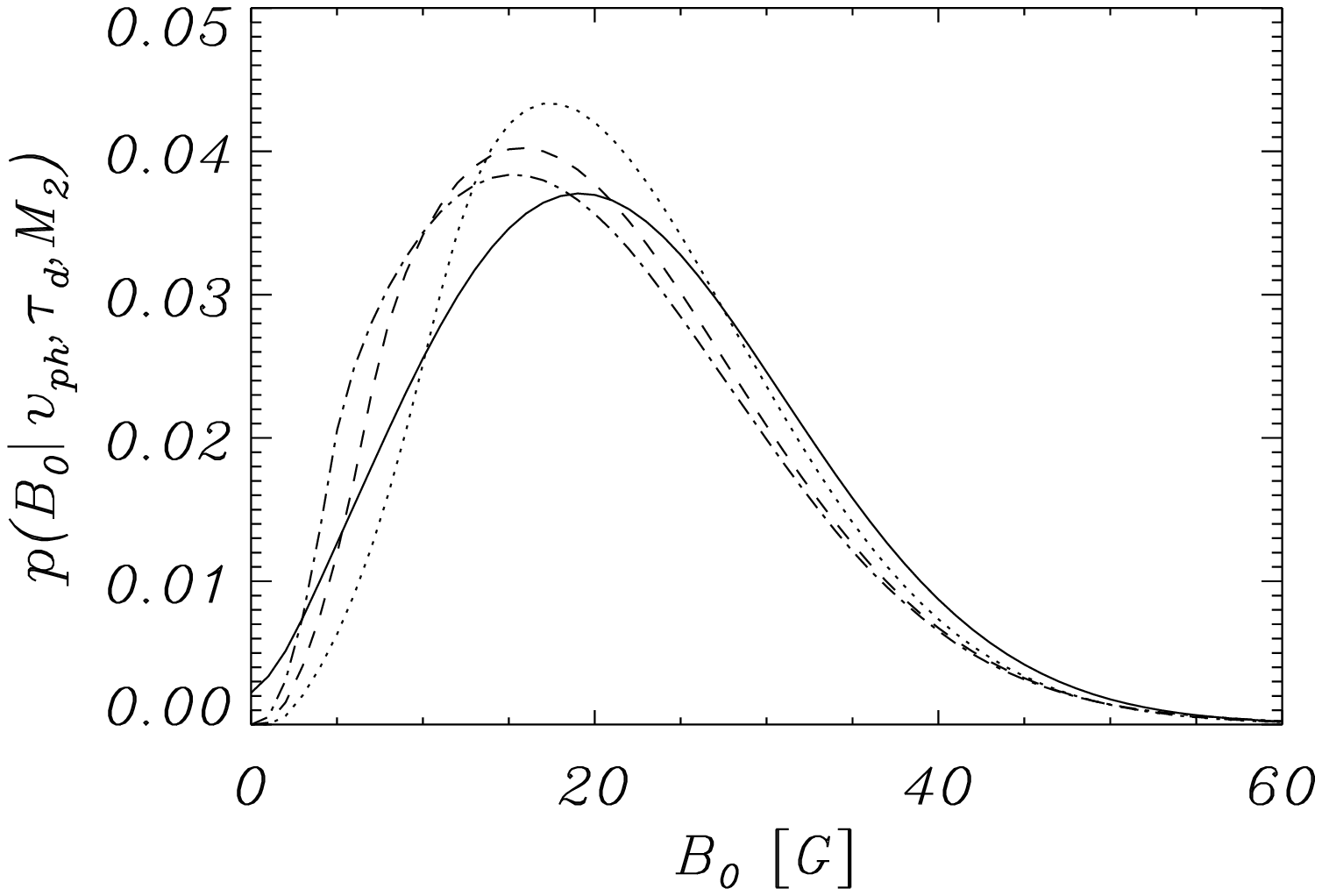} 
\includegraphics[width = 0.49\textwidth]{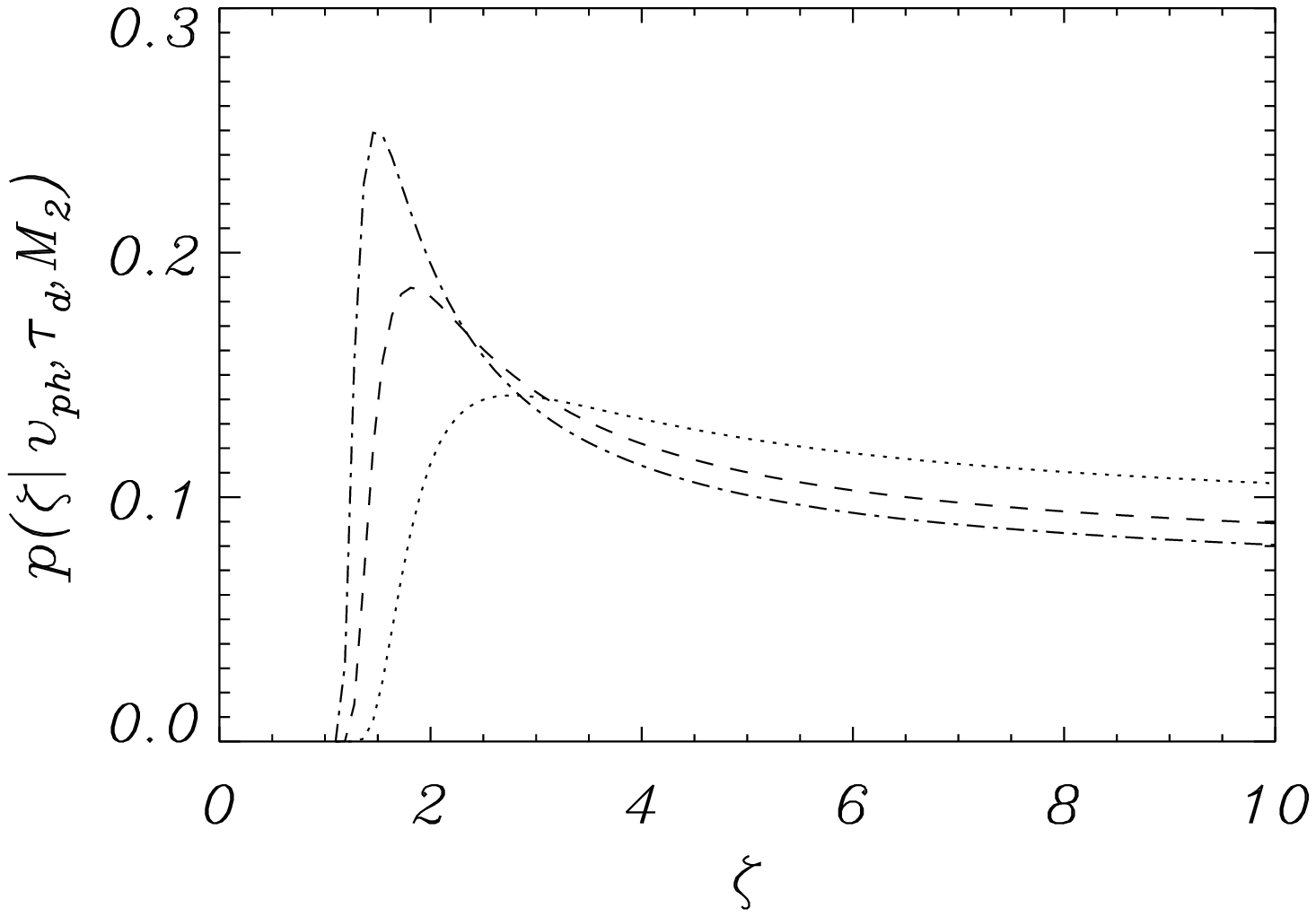} 
\includegraphics[width = 0.49\textwidth]{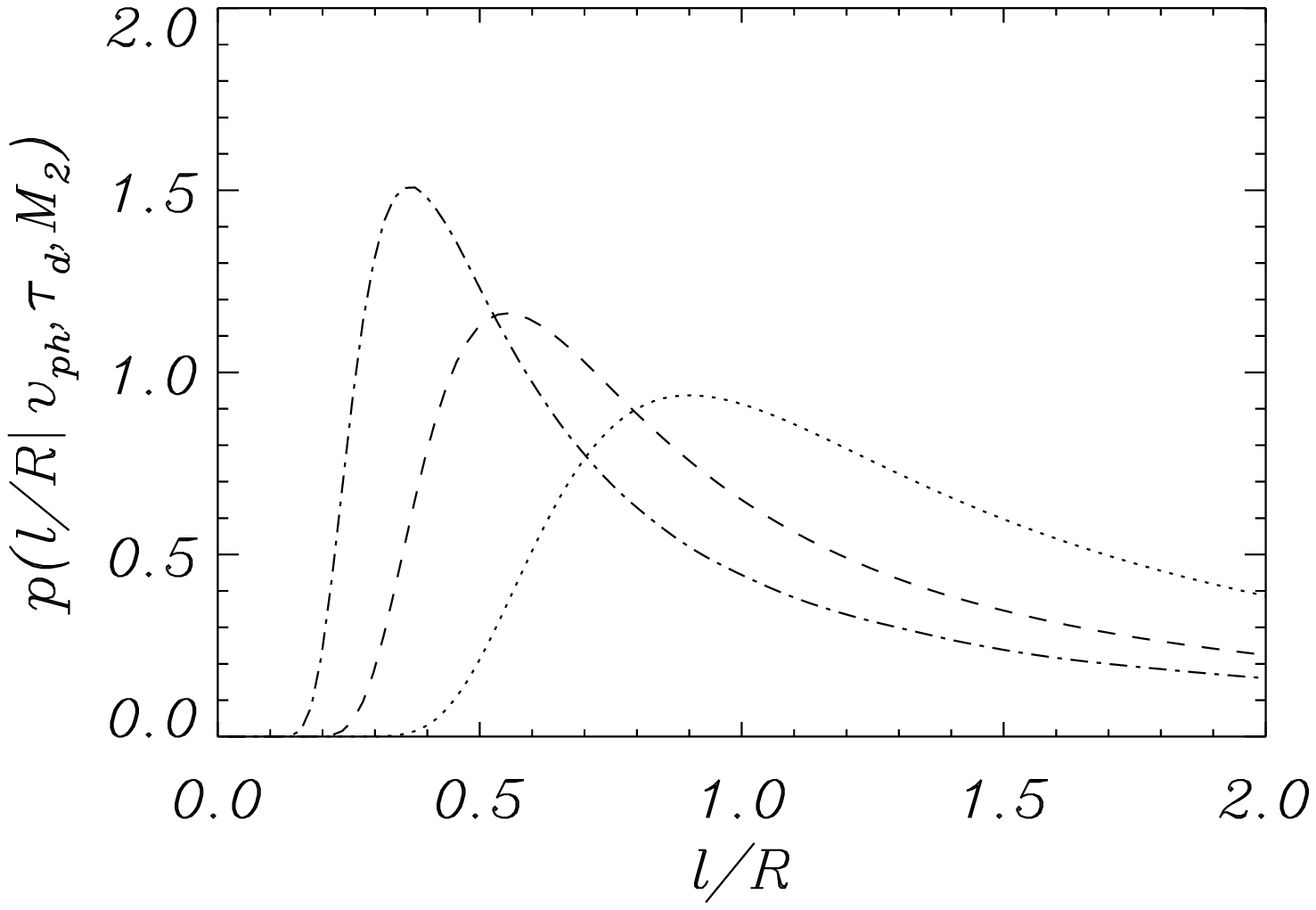} 
\caption{\label{fig6} Marginal posterior distributions for (a) magnetic field strength; (b) density contrast; and (c) transverse inhomogeneity length scale for the inversion of problem with forward model given by Eqs.~(\ref{vphasedamp}) and (\ref{damp2}) and a transverse oscillation with $v_{\rm ph}=1030 \pm 410$ km s$^{-1}$ and three values for the damping time: no damping (solid-line); $\tau_{\rm d}=500$ s (dotted-line);  $\tau_{\rm d}=800$ s (dashed-line); and $\tau_{\rm d}=1200$ s (dash-dotted-line) with an associated uncertainty of 50 s in all cases. The inferred medians with errors at the 68\% credible interval are:  $B_{0} = 21^{+ 12}_{-9}$ G for the undamped case;  $B_{0} = 20^{+ 11}_{-8}$ G for $\tau_{\rm d}= 500 $s;  $B_{0} = 19^{+ 11}_{-8}$ G for $\tau_{\rm d}= 800 $s; and $B_{0} = 18^{+ 11}_{-9}$ G for $\tau_{\rm d}= 500 $s. A fixed value for the loop length, $L=1.9\times 10^{10}$ cm, was considered in all computations.}
\end{figure}

We repeated the inversion of magnetic field strength by  first considering uniform priors over different ranges for both parameters, $\rho_i$ and $\zeta$. First, the maximum value of the density contrast was modified by considering several values from $\zeta=3$ up to $\zeta=50$. Figure~\ref{fig2}a shows the obtained results. We can see that the extent on density contrast over which a uniform prior is considered does not influence much the magnetic field strength inference. A different result is obtained when the inversion is performed considering uniform prior distributions over different ranges for the internal density. Figure~\ref{fig2}b shows marginal posterior distributions for the magnetic field strength computed by considering a uniform prior for the fixed range of $\zeta\in[1.1-10]$ and uniform priors over three different ranges for the waveguide density.
We can see that considering the lower density half-range with $\rho_i\in[10^{-13}-10^{-12}]$ kg m$^{-3}$ or the higher density half-range with $\rho_i\in[10^{-12}-10^{-11}]$ kg m$^{-3}$ produce rather different results. When the full range $\rho_i\in[10^{-13}-10^{-11}]$ kg m$^{-3}$ is taken, results similar to those obtained with the higher density half are found. This range is in close correspondence to the one considered by \cite{nakariakov01} and was chosen for this reason in the results above. A note of warning is in order here. When a parameter is left to vary over several decades, using a uniform prior is not the best choice. As the result above shows, the contribution to the marginal posterior from the integral over the range [10$^{-13}$ -- 10$^{-12}$] is insignificant in comparison to the contribution from the integral over the range [10$^{-12}$ -- 10$^{-11}$]. In this case, considering the density a scale parameter and using a Jeffreys prior, which gives equal probabilities over decades in a logarithmic scale might be more appropriate, as discussed in Appendix~\ref{appa}.

Besides changing the range of variation over which uniform prior distributions are defined, the Bayesian formalism enables us to incorporate more informative priors. Spectroscopy or the analysis of the Differential Emission Measure (DEM) enables us to obtain some properties of the emitting coronal plasma, such as the density \citep[see e.g.,][]{vandoorsselaere08,su18}. When additional knowledge like this becomes available,  the classic approach has been to use it to obtain a numerical estimate for the magnetic field strength. For instance, \cite{nakariakov01} used a density of $10^{9.3\pm0.3}$ cm$^{-3}$ in their calculation. The Bayesian approach enables to self-consistently incorporate this information to update the posteriors, i.e., our state of belief. The proper way to proceed is to recalculate the posterior using this more informative prior.

As an example, let us use the same estimate as \cite{nakariakov01}, which considering the proton mass and a mean molecular mass of around 1.16 in the corona translates into $\sim 1.9\times10^{12}$ kg m$^{-3}$, with its corresponding uncertainty. We can use this information to construct a Gaussian prior for the density, by considering Eq.~(\ref{gaussp}) for the unknown parameter $\theta_i=\rho_i$, centred on the numerical estimate $\mu_{\theta_i}=\mu_{\rho_i}$, and with uncertainty $\sigma_{\theta_i}=\sigma_{\rho_i}$.

Figure~{\ref{fig3}} shows the result of such an inversion for the magnetic field strength and the plasma density inside the waveguide.
Figure~\ref{fig3}a shows the Gaussian prior for the density and its corresponding posterior. The resulting posterior closely resembles the assumed prior, although the information on the data has slightly altered the posterior. Figure~\ref{fig3}b shows a comparison between the marginal posterior distributions for the magnetic field strength using the uniform prior and the Gaussian prior in density. Using the information  obtained from measuring the plasma density produces a shift in the marginal posterior for the magnetic field towards smaller values and a more constrained distribution.  The summary of this posterior using the median and errors at the 68\% credible interval leads to $B_0=13^{+7}_{-6}$ G, in good agreement with the numerical estimate by \cite{nakariakov01}. Table~\ref{tab_B1} in Appendix~\ref{appb} shows a comparison between previous estimates of magnetic field strength in a number of
reported events and our Bayesian posterior summaries. In all cases, a good agreement is found. In addition, our results provide us with the full probability distributions. The last two panels show a comparison between the joint posteriors for internal density and magnetic field strength obtained by employing the uniform prior (Fig.~\ref{fig3}c) and the Gaussian prior (Fig.~\ref{fig3}d) for the internal density.  As can be seen, the addition of information enables us to further constrain our estimates for both unknowns, waveguide density and magnetic field strength. We note the lack of symmetry of the joint posterior in Fig~\ref{fig3}d, which explains why assumptions in density might have an impact on the inferred magnetic field strength and why considering a range in density from 10$^{-12}$ to 10$^{-11}$  kg m$^{-3}$ leads to a very similar result to considering the full range (see Fig~\ref{fig2}b), since that range already covers most of the region where the joint posterior is large. 

By looking at Fig.~\ref{fig3}c one may wonder if, instead of using a Gaussian prior on density, it is not possible to solve the inversion problem by considering a uniform prior over the full range of internal density values and then performing a cut of the joint two-dimensional posterior for density and magnetic field at the plasma density values that has been measured (dashed line in Fig.~\ref{fig3}c). Another option could be to simply insert the value of  $\mu_{\rho_{\rm i}}$ in $\rho_i$ in Eq.~(\ref{vphaseb0}) and solve that oversimplified inversion problem for two unknowns, $\mbox{\boldmath$\theta$} =\{\zeta, B_0\}$. Figure~\ref{postsb0} shows the result of a comparison of marginal posteriors for the options being discussed. The solid line is the result of the integration of the full posterior over the full range of values for $\rho_i$ (and $\zeta$) and gives the most uncertain result with the possible values for the magnetic field strength extending up to $\sim 60$ G. The dashed-line shows the result of taking a cut of the joint posterior for $B_0$ and $\rho_i$ at the measured value for the plasma density. This is the most constrained distribution, but it does not take into account the uncertainty on the measured plasma density. It just considers that this uncertainty is zero. In our view, simply inserting a measurement of $\rho_i$ without its associated uncertainty, even if it provides the more constrained result, is not the best option. The dotted curve is the posterior corresponding to the use of the Gaussian prior for the density.  This exercise shows that what one is willing to assume about the plasma density inside the waveguide may influence the inference of the magnetic field strength.  The advantage of the Bayesian approach is that one is forced to explicitly specify this ``what one is willing to assume'' in the definition of the priors when constructing the statistical model. In Appendix~\ref{appa} a prior dependence analysis of the obtained results is presented. The results indicate that changes on what one is willing to accept a priori for the density and density contrast have an effect on their corresponding posteriors, but not on the inferred magnetic field strength.

\subsection{Information on wave damping}

Time damping is a commonly observed property in transverse loop oscillations, with characteristic damping times of a few oscillatory periods.   Although inferences of magnetic field strength using the damping of the oscillations have been presented \citep[see e.g.,][]{pascoe16}, the influence of this observable on the estimates of the magnetic field strength inferred by seismology is unknown. For this reason,  we performed the inference including the simplest available model for damping by resonant absorption, a plausible mechanism to explain the observed damping time scales \citep{goossens02a,ruderman02}. We therefore consider model $M_2$ in which the previous model $M_1$ is modified by including a non-uniform density layer of length $l$ at the boundary of the waveguide and centred around the radius $R$ of the tube. Under the long wavelength and thin boundary ($l\ll R$) approximations analytical relationships can be obtained for the observable phase speed and damping time as a function of four unknown parameters - internal density, density contrast, magnetic field strength, and transverse inhomogeneity length scale  - such that 

\begin{eqnarray}\label{vphasedamp}
v_{\rm ph}(\rho_{\rm i}, \zeta, B_0) &=& \frac{B_0}{\sqrt{\mu_0\rho_{\rm i}}}\left(\frac{2\zeta}{1+\zeta}\right)^{1/2},\label{damp1}\\
\tau_{\rm d}(\rho_{\rm i}, \zeta, B_0, l/R) &=& \frac{2}{\pi} \left(\frac{\zeta + 1}{\zeta - 1}\right) \left(\frac{1}{l/R}\right) \left(\frac{2L}{v_{\rm ph}}\right)\label{damp2}.
\end{eqnarray}
Here,  $L$ is the length of the loop, a magnitude that is assumed to be measurable. The expression for the damping time contains a factor $2/\pi$ as a result of the particular assumption of a sinusoidal density profile for the density at the non-uniform layer. Considering different alternatives for the density profile leads to differences in the forward predictions and also in the inverse solutions. The latter may be important in strong damping regimes, as shown by \cite{arregui15c}. 

Equations~(\ref{damp1}) and (\ref{damp2}) show that even using the simplest model for resonantly damped oscillations the expressions for the phase speed and the damping time are coupled and, hence, some degree of influence of the information on the damping time can be expected when inferring the magnetic field strength. We note that the phase speed is independent of the transverse inhomogeneity length scale, a consequence of the adoption of the thin boundary approximation. \cite{vandoorsselaere04a} and \cite{arregui05} have shown that,  outside this approximation, the period of the fundamental kink mode depends on the transverse inhomogeneity length-scale, producing significant variations especially for values above $l/R=1$.

As the event of 4th of July used in our previous calculations does not come with the corresponding damping time, we first repeated the inference presented in Sect.~\ref{b0inf}, by including some reasonable value for the damping time and using the forward model given by (\ref{vphasedamp}) and (\ref{damp2}).  Bayes theorem now includes additional parameters, $\mbox{\boldmath$\theta$} =\{ \rho_i, \zeta, B_0\}$, and observables, $D = \{v_{\rm ph}, \tau_{\rm d}, L\}$, and can be written as

\begin{eqnarray}\label{bayes3}
&&p(\{\rho_{\rm i}, \zeta, B_0, l/R\} | \{v_{\rm ph}, \tau_{\rm d},L\},M_2)\sim\nonumber\\
&&p(\{v_{\rm ph}, \tau_{\rm d},L\} | \{\rho_{\rm i}, \zeta, B_0, l/R\},M_2) p(\{\rho_i, \zeta, B_0\} |M_2).
\end{eqnarray}

An example inversion result is shown in Fig.~{\ref{fig5}}, which displays marginal posterior distributions and  joint probability distributions for the magnetic field strength, the density contrast, and the transverse density inhomogeneity length scale.  As in Section~\ref{b0inf}, we used uniform priors over given ranges for all parameters. Again, the magnetic field strength can be properly constrained. The inclusion of a damping time leads to a constraint on the lower limit of density contrast and transverse density inhomogeneity length scale, but no restriction can be found on the upper limit beyond the assumed prior.

The posterior for the magnetic field strength shows a very similar probability distribution to the inversion result without the use of the damping time (Sect.~{\ref{b0inf}}).  Both the median of the probability density and the dispersion are similar to the case without damping. There is a slight decrease in the median for $B_0$ for larger damping times. This is more clearly seen in Fig.~\ref{fig6}a, were posteriors for the magnetic field strength for different damping regimes are displayed. The posteriors between the case with no damping and the three cases with damping show  minimal differences. Also, once a value for the damping time is considered, it does not seem to matter that much the precise value of the damping time. These results point to  a negligible importance of considering the information on the damping time scale of coronal loop oscillations when inferring the magnetic field strength, when the thin tube and thin boundary approximations are used. Table~\ref{tab_B2} in Appendix~\ref{appb} shows that this conclusion holds when analysing a large sample of events with damping presented by \cite{goddard16}.  Outside these approximations, the period becomes dependent on the transverse inhomogeneity length-scale \citep{vandoorsselaere04a,arregui05,soler14a} and a stronger coupling between period and damping time is expected, which could lead to a more effective influence of the damping on the inferred  magnetic field strength. The remaining panels in Fig.~\ref{fig6} show that the two parameters that define the cross-field variation of the density, $\zeta$ and $l/R$ can be constrained better or worse, depending on the actual value of the damping time. The constraint on the lower limit of $l/R$ is much more sensitive to the choice of damping time than the lower limit on $\zeta$. Constrained posterior distributions for $\zeta$ and $l/R$ can also be obtained using two resonant damping regimes, as shown by \cite{arregui13b} for synthetic data and by \cite{pascoe17a} for observational data.

\section{Summary and conclusions}\label{conclusions}

We applied Bayesian inversion methods to the inference of the magnetic field strength in transversely oscillating coronal waveguides. 
The classic approach to the problem has been to use measured values for the period or phase speed together with an analytical approximation to the kink speed in the long wavelength limit to extract information on the magnetic field strength, upon inserting numerical estimates, either assumed or as a result of an indirect measurement, for parameters such as the density contrast of the waveguide and the internal density \citep{nakariakov01} .  The result is a numerical estimate with its corresponding error bar or a range of variation for the possible values the magnetic field strength can take on. In some cases, information on the density gathered from spectroscopic measurements has enabled to further constrain the estimates \citep[see e.g.,][]{vandoorsselaere08}.  Our approach consist of adopting Bayesian methods to obtain the global probability density distribution, or posterior, for the unknowns. This posterior can then be marginalised to obtain information on a particular parameter of interest, e.g., the magnetic field strength, which assigns a level of plausibility to each considered parameter value. The process self-consistently propagates uncertainty from data and between the unknowns of the problem.

\begin{figure*}
\center
\includegraphics[width = 0.47\textwidth]{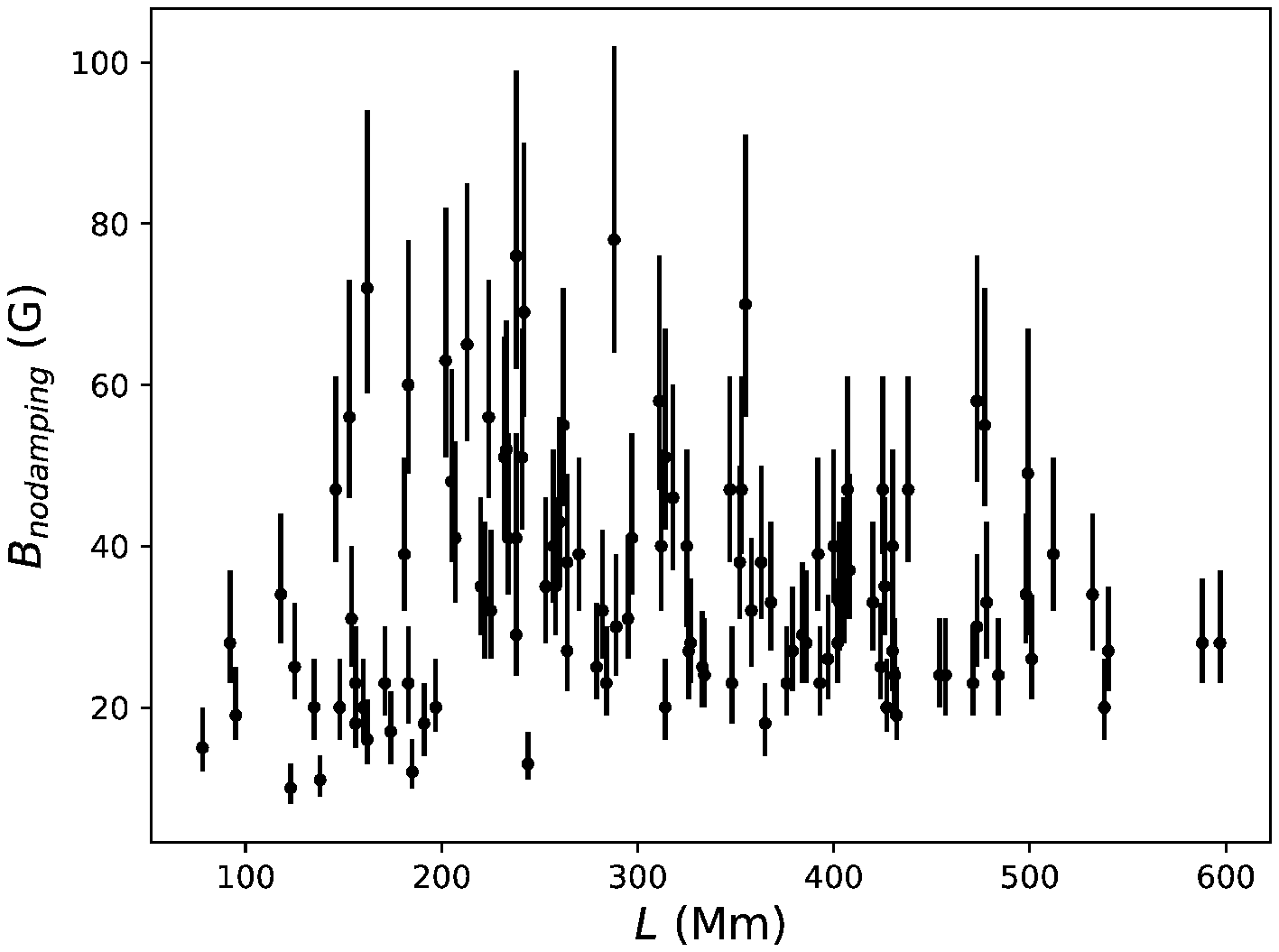} 
\includegraphics[width = 0.47\textwidth]{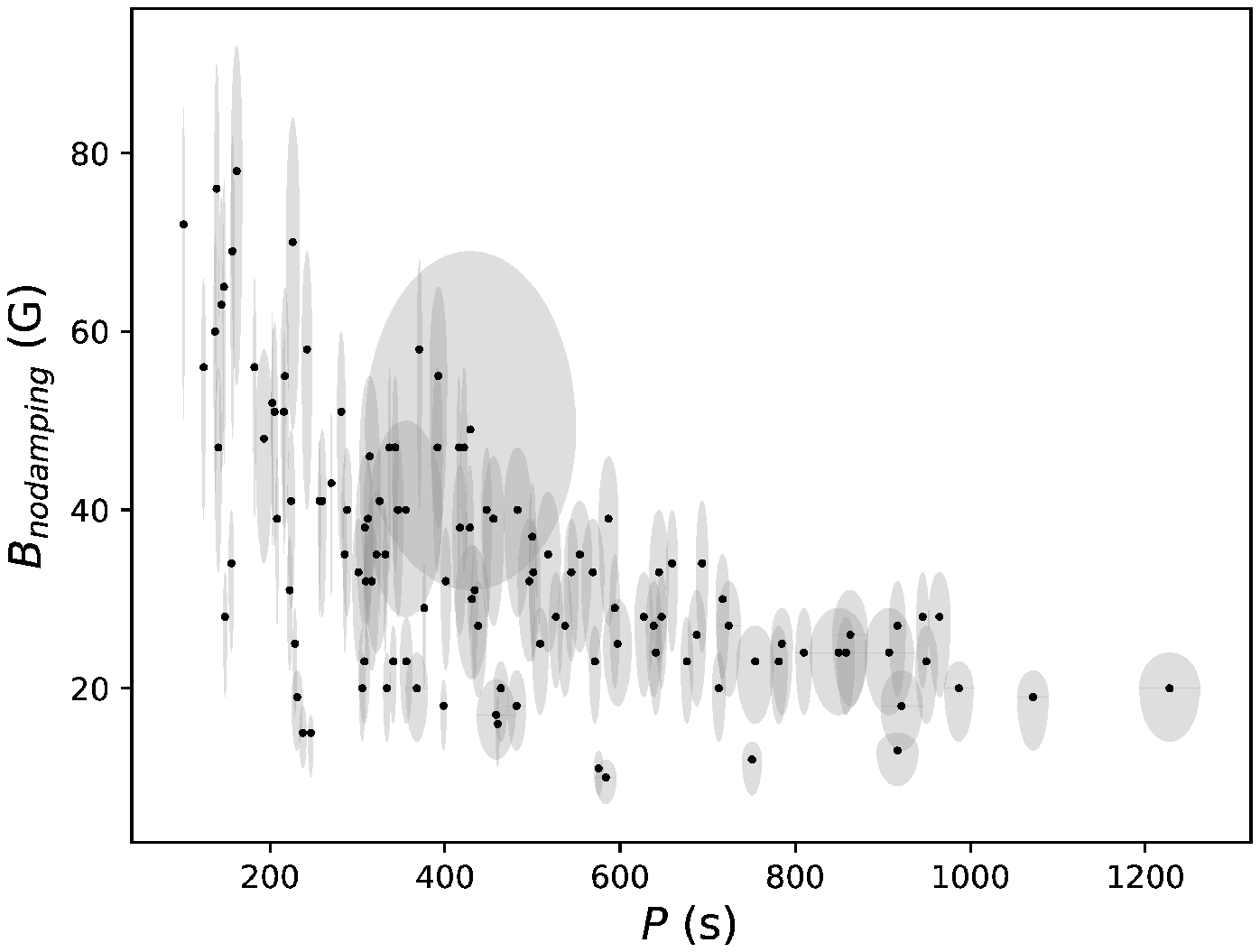} 
\caption{Summary of values for the magnetic field strength in Table~\ref{tab_B2} and their corresponding uncertainties as a function of the loop length (left) 
and the oscillation period (right). We note that the loop lengths are given without errors. The errors in the magnetic field strength are given at the 68\% credible interval. \label{statistics}}
\end{figure*}

 These methods have been applied to compute probability density distributions for the magnetic field strength to study in detail the possible impact on the variability of density and density contrast and of the observed damping time. This was done by solving different inversion problems in which our base-knowledge is different to analyse how the uncertainty on the density and the density contrast affects the magnetic field strength inversion and whether or not the consideration of the damping modifies the obtained posteriors significantly.
 
We found that the magnetic field strength can be inferred, even if the densities inside and outside and their ratio are largely unknown. The obtained marginal posteriors show well constrained distributions.  In comparison to e.g., \cite{nakariakov01}, who obtained numerical estimates or a range of possible values as a function of the considered density, our results offer how the relative plausibility between those possible values is distributed. When spectroscopic information on plasma density is available, the method enables to incorporate this knowledge in a self-consistent manner, further constraining the inference. In the inference process, both the density in the waveguide and the density contrast with respect to the coronal plasma density need to be considered. By considering uniform prior distributions over different ranges in these two parameters the results indicate that the  posterior for the magnetic field strength inference is very little dependent on the density contrast range that is assumed, but the range of variation for the density of the waveguide has an impact on the obtained magnetic field strength distribution. A sensitivity analysis considering other alternative priors for density contrast and loop density shows that these priors influence the corresponding posteriors in these two parameters, but very little the posterior for the magnetic field strength.

The observed oscillation damping is practically irrelevant to the inversion of the magnetic field strength, at least when the thin tube and thin boundary approximations are considered in the forward problem.  However, its inclusion enables to obtain information on the transverse inhomogeneity length scale of the density at the boundary of the waveguide, a parameter directly related to  wave heating processes. 

We applied the methods here presented to a set of observed transverse loop oscillations. The results are displayed in Tables~\ref{tab_B1} and \ref{tab_B2} in Appendix~\ref{appb}. In Table~\ref{tab_B1} we compare magnetic field strength estimates provided by \cite{nakariakov01}, \cite{aschwanden02}, \cite{goossens02a}, \cite{vandoorsselaere08}, \cite{white12}, and \cite{pascoe16} with the corresponding results obtained using the Bayesian method. The field strengths are first computed using a uniform prior on density over the extended range $\rho_{\rm i}\in[10^{-13}-10^{-11}]$ kg m$^{-3}$, which lead to the values $B_{\rm u}$ in Table~\ref{tab_B1}. Then, Gaussian priors on density are employed around the density values estimated or assumed by those authors, which lead to the values $B_{\rm G}$ in Table~\ref{tab_B1}. The summaries of our distributions are in agreement with the previous numerical estimates. In order to confirm our result regarding the minimal influence of the damping on the inferred magnetic field strength, when the thin tube and thin boundary expressions for the forward problem are used, we compared the inference results, with and without damping, for 52 events compiled by \cite{goddard16} from the catalogue by \cite{zimovets15}. The results presented in Table~\ref{tab_B2} in Appendix~\ref{appb} show that practically the same posterior summaries are obtained with or without damping, with differences of only $\pm 1$ G.

Figure~\ref{statistics} shows a summary of the magnetic field strengths in Table~\ref{tab_B2} for the inferences of $B_{\rm no \mbox{\hspace{0.05cm}}damping}$ as a function of the loop length and of the oscillation period. No clear signature of a trend is found in the case of the magnetic field strength as a function of the loop length. On the contrary, the inverse relationship between magnetic field strength and period, present in the model, can be clearly seen from the inferred field strengths. 

The methods here presented can in principle be applied to another magnetic and plasma structures in the solar atmosphere, such as prominence fine structures  \citep[see e.g.][]{montesolis19} or chromospheric spicules.

%
\section*{Acknowledgments}   
We acknowledge financial support by the Spanish Ministry of Economy and Competitiveness (MINECO) through projects 
AYA2014-55456-P (Bayesian Analysis of the Solar Corona) and AYA2014-60476-P (Solar Magnetometry in the Era of Large Solar Telescopes) and FEDER funds. M.M-S. acknowledges financial support through a Severo Ochoa FPI Fellowship under the project SEV-2011-0187-03. 


\begin{appendix}
\section{Influence of prior information}\label{appa}

Our analysis has shown that the inference of the magnetic field strength can be performed regardless of the fact that the density and the density contrast cannot be properly inferred and are therefore largely unknown.  Using different uniform priors for the density contrast does not affect the marginal posterior for the magnetic field strength, as shown in Fig.~\ref{fig2}a. The same is not true for the density of the waveguide, see Fig.~\ref{fig2}b. Issues with priors are evident already from the results shown in Fig.~\ref{fig1}. Figure ~\ref{fig1}b shows that the marginal posterior for density contrast has a long tail and is not going to zero at the extremes. In Fig.~\ref{fig1}c, the range of variation for the internal density spans over two orders of magnitude. Using a uniform prior and integrating over the range [10$^{-13}$ -- 10$^{-12}$] or over the range [10$^{-13}$ -- 10$^{-11}$] will lead to probabilities that are not of the same order. This makes the density a scale parameter for which a Jeffreys prior might be more appropriate.

In this section, a more detailed prior analysis is presented. In addition to the use of a uniform prior for density contrast, the inversion of Eq.~(\ref{vphaseb0}) was performed using different priors for density contrast and waveguide density. 

For density contrast,  the following alternative priors were considered:

\begin{itemize}
\item A Normal distribution,  $p(\zeta)= \mathcal{N}(\mu_\zeta,\sigma^2)$, with $\mu_\zeta = 5.5$ and $\sigma =7.75$.

\item A Cauchy distribution, 

\begin{equation}
p(\zeta)=f(\zeta; \mu_\zeta, \gamma) =\frac{1}{\pi\gamma\left[1 + \left( \frac{\zeta-\mu_\zeta}{\gamma}\right)^2\right]},
\end{equation}
with $\mu_\zeta =5.5$ and $\gamma=1,2,3,4,5$.

\item An exponential, $p(\zeta)= C\exp(-\zeta/4)$, with $\zeta\in[1.1-10]$ and $C$ such that the integral is unity.
\end{itemize}

 In the same manner, the following priors for the density of the waveguide were used, as alternatives to the uniform prior:
 
 \begin{itemize}
 \item A Jeffreys prior, $p(\rho_i)=\left[\rho_i \log\left(\frac{\rho_i^{\rm max}}{\rho_i^{\rm min}}\right)\right]^{-1}$, with $\rho_i^{\rm min}= 10^{-13}$ kg m$^{-3}$ and $\rho_i^{\rm max}= 10^{-11}$ kg m$^{-3}$, corresponding to the full range considered in Fig.~\ref{fig1}.

\item A Normal distribution,  $p(\rho_i)= \mathcal{N}(\mu_{\rho_{i}},\sigma^2)$, with $\mu_{\rho_i} = 10^{-12}$ kg m$^{-3}$  and $\sigma^2 =5\times 10^{-12}$ kg m$^{-3}$.
 \end{itemize}

\begin{figure*}
\center
\includegraphics[width = 0.24\textwidth]{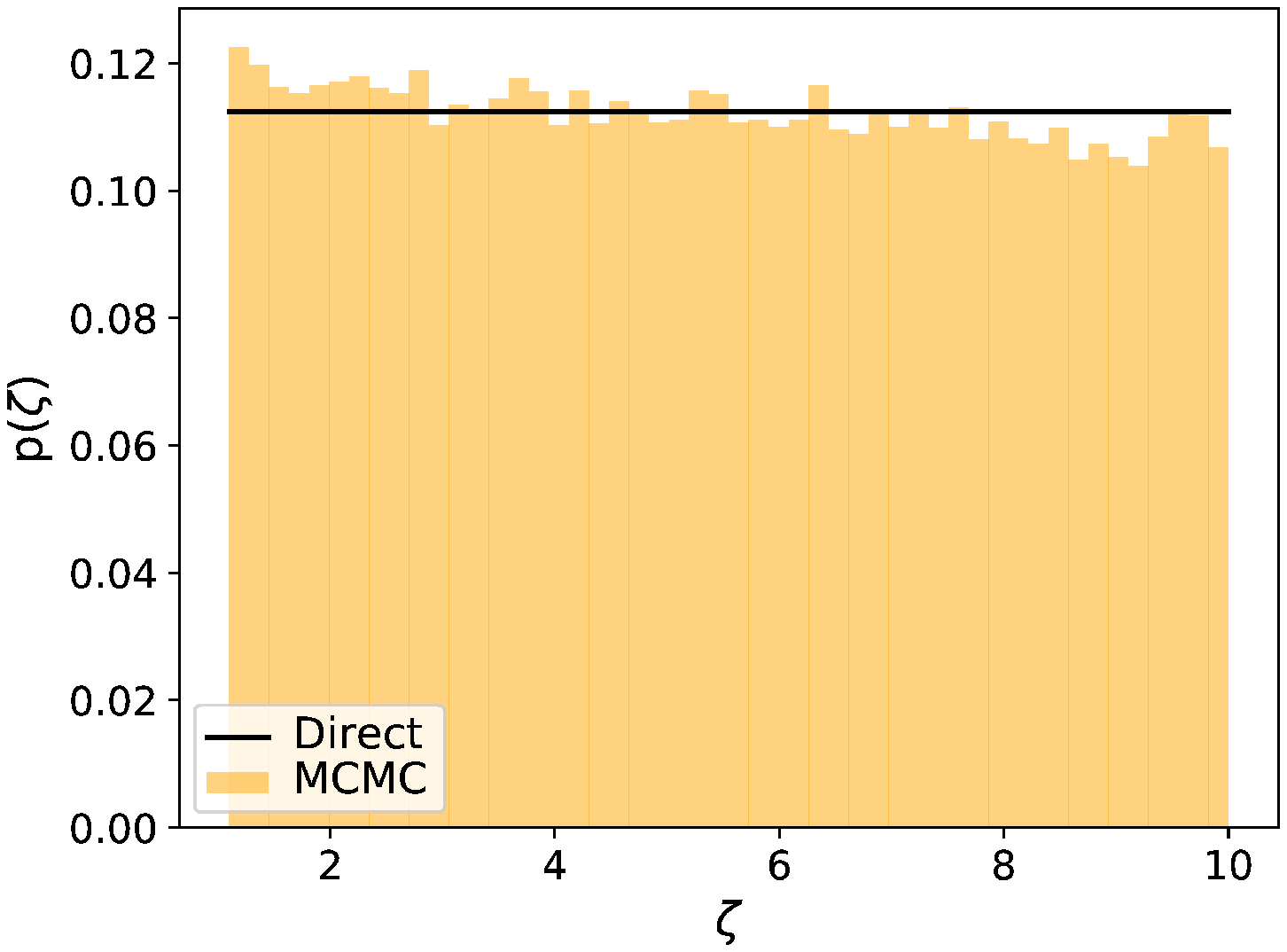} 
\includegraphics[width = 0.24\textwidth]{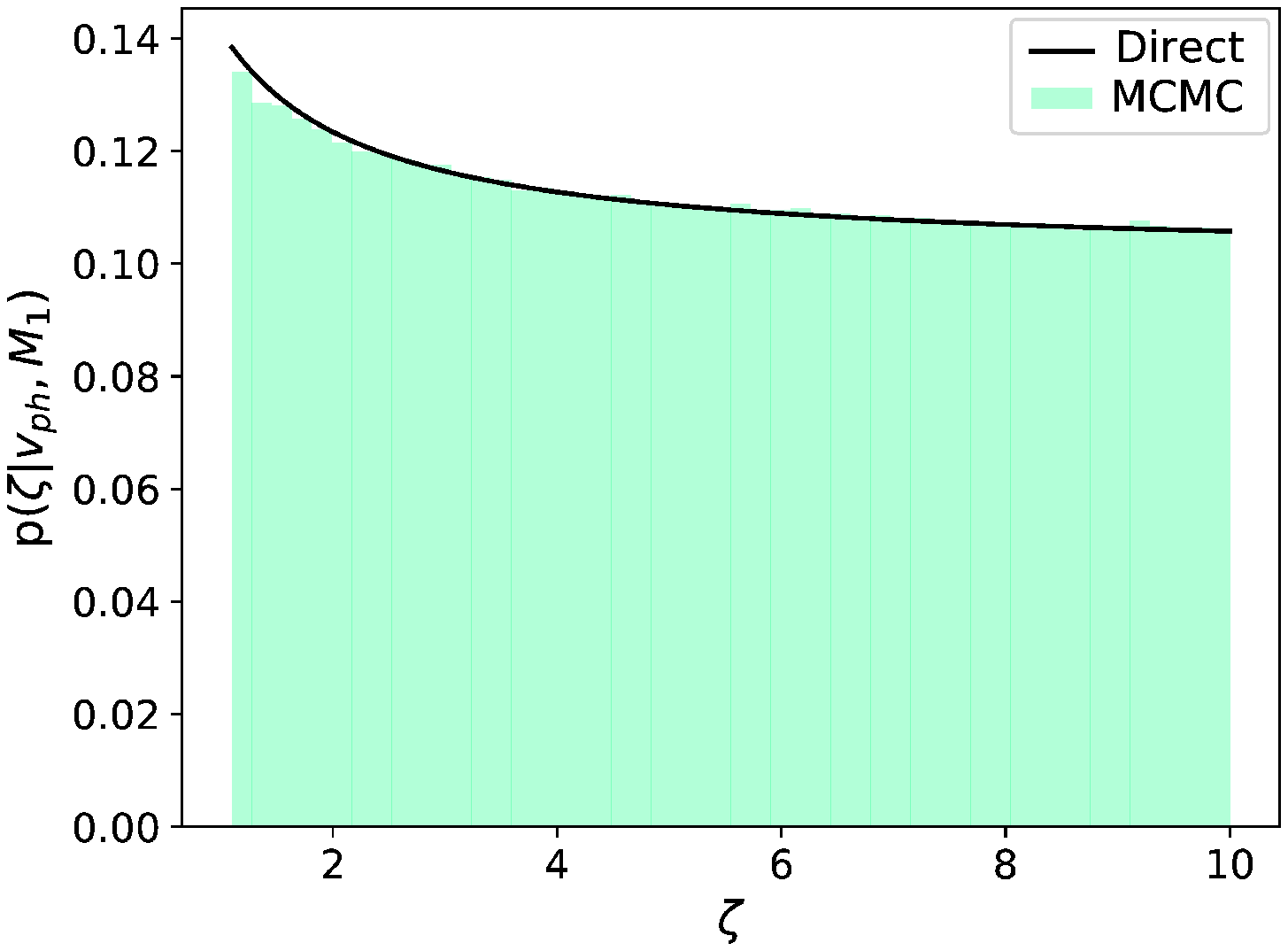} 
\includegraphics[width = 0.24\textwidth]{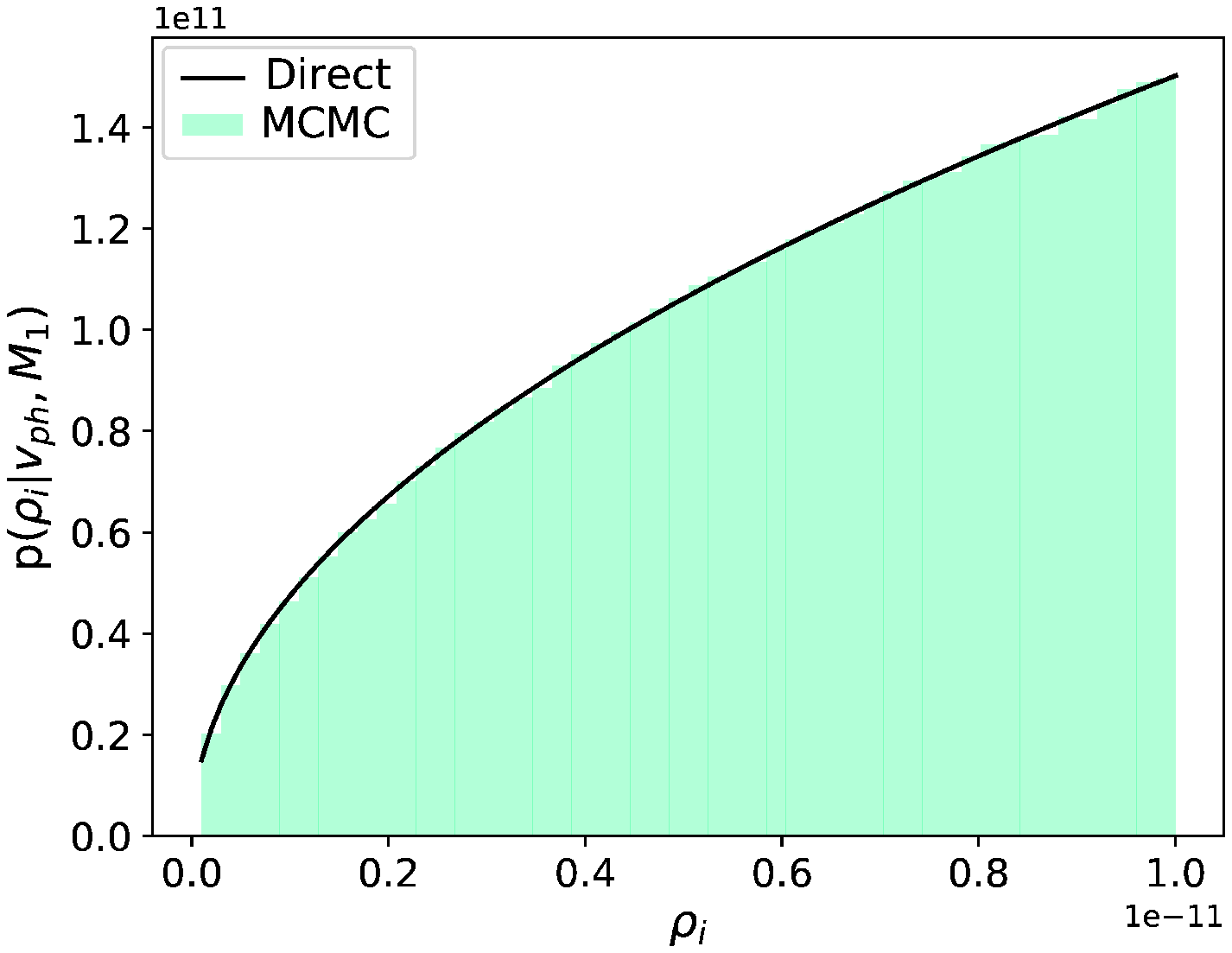} 
\includegraphics[width = 0.24\textwidth]{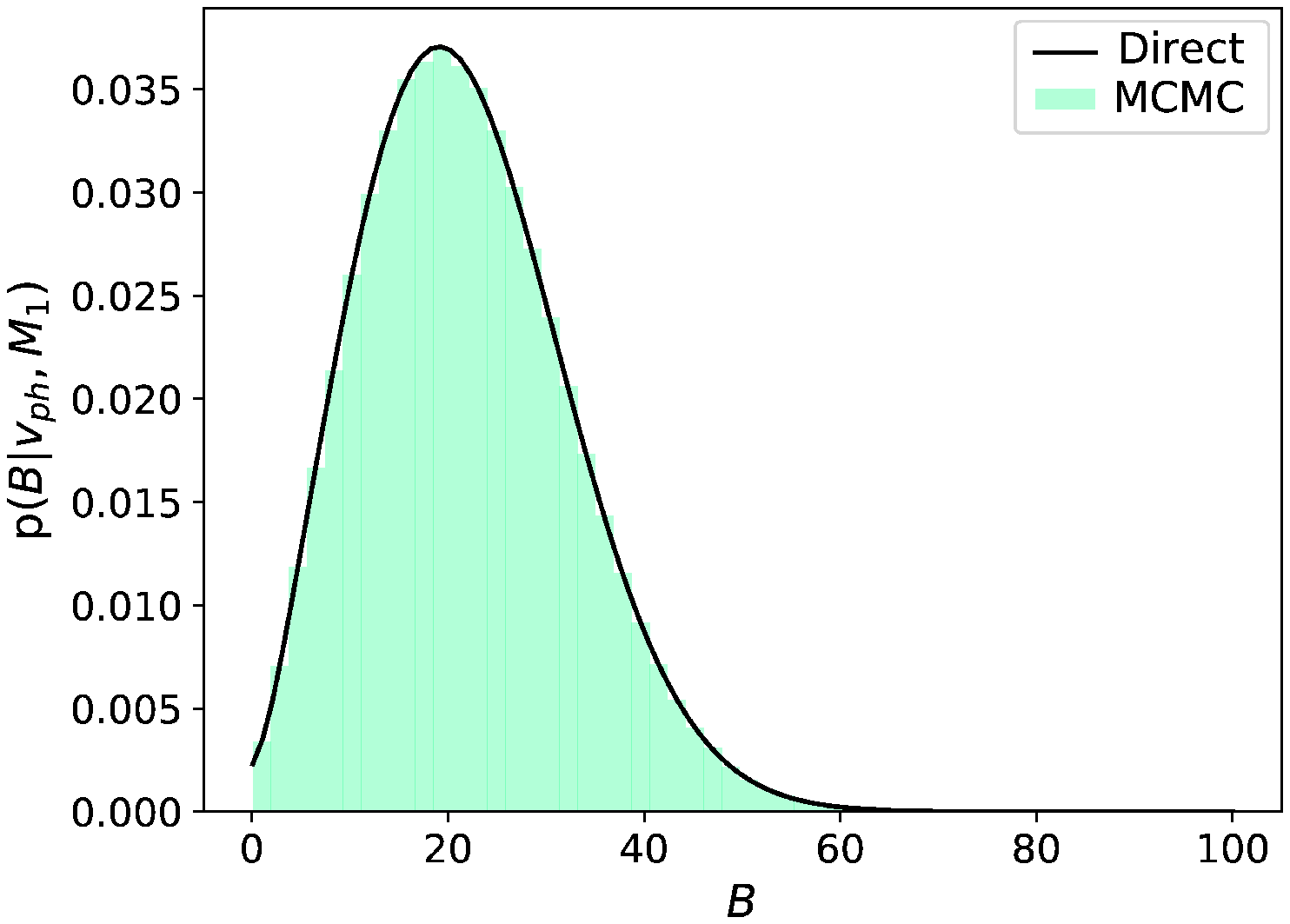} \\
\includegraphics[width = 0.24\textwidth]{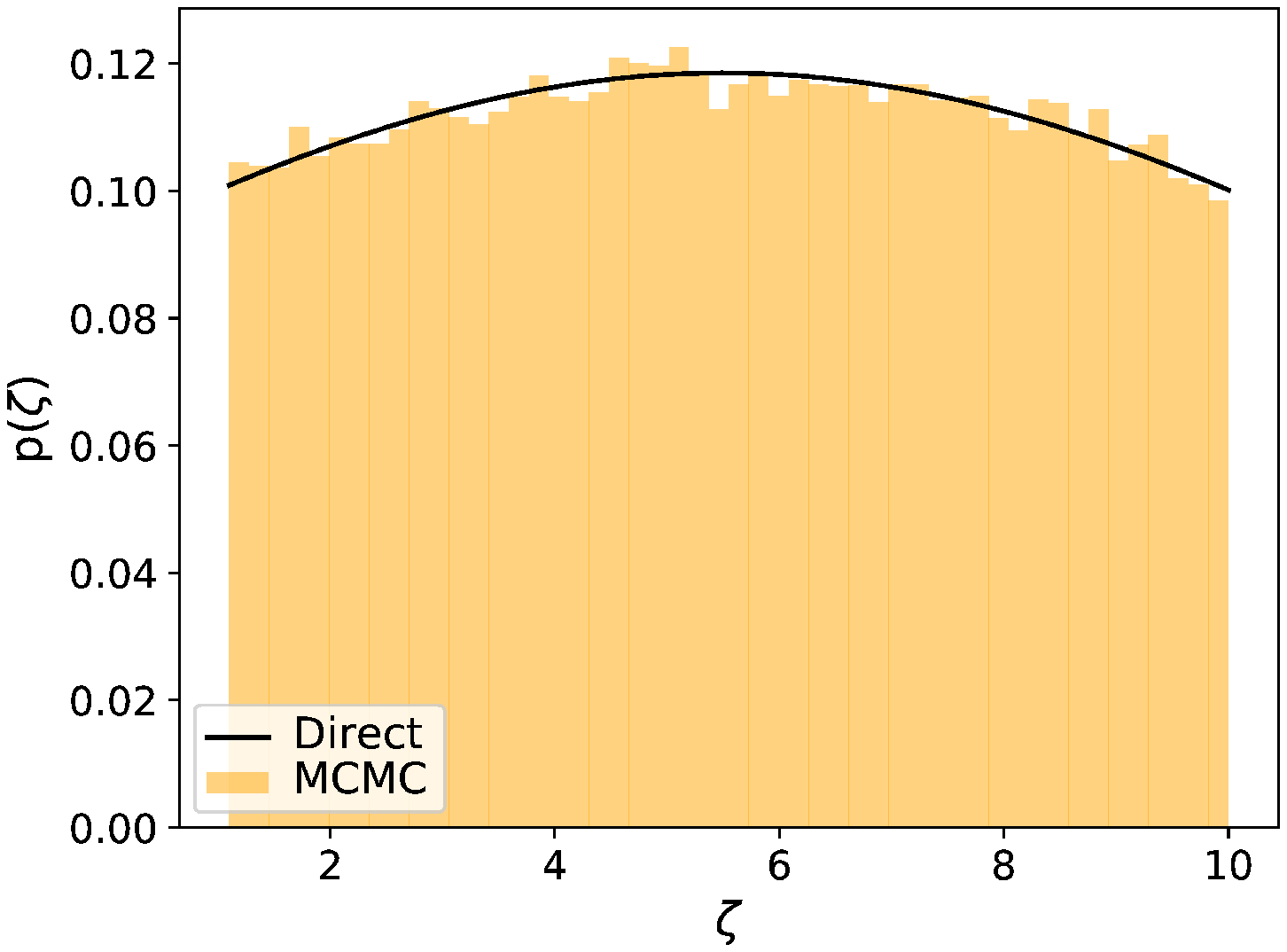} 
\includegraphics[width = 0.24\textwidth]{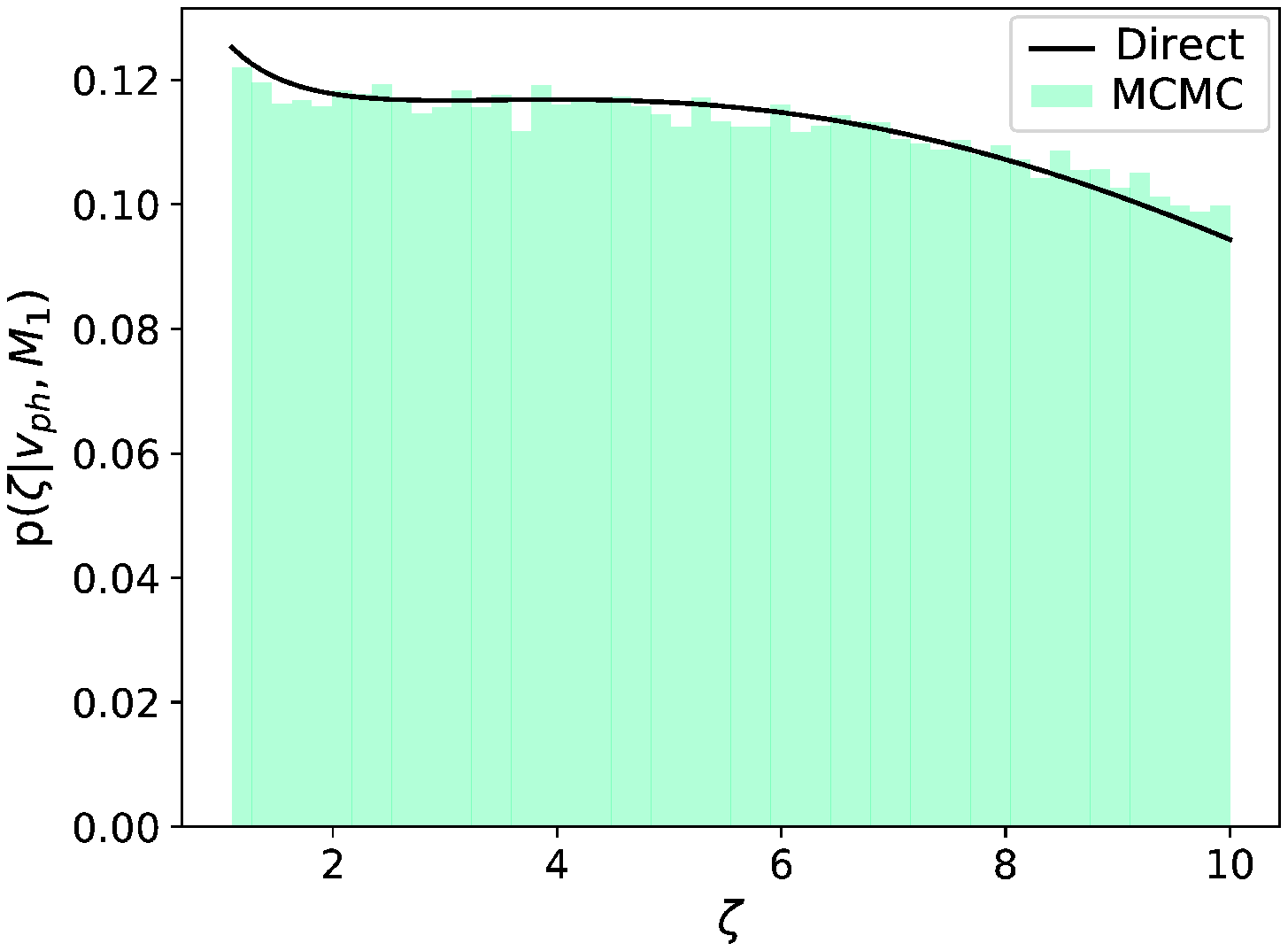} 
\includegraphics[width = 0.24\textwidth]{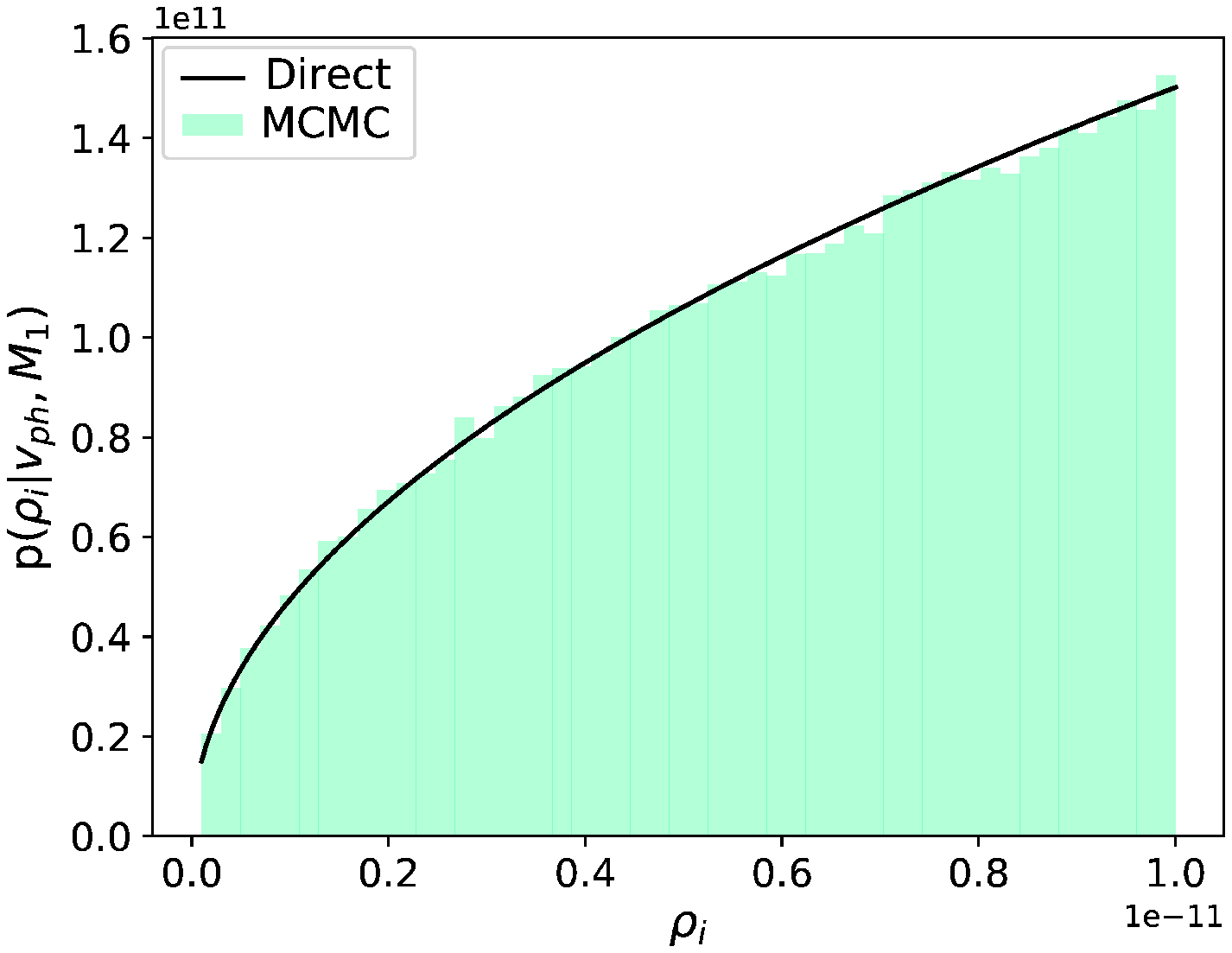} 
\includegraphics[width = 0.24\textwidth]{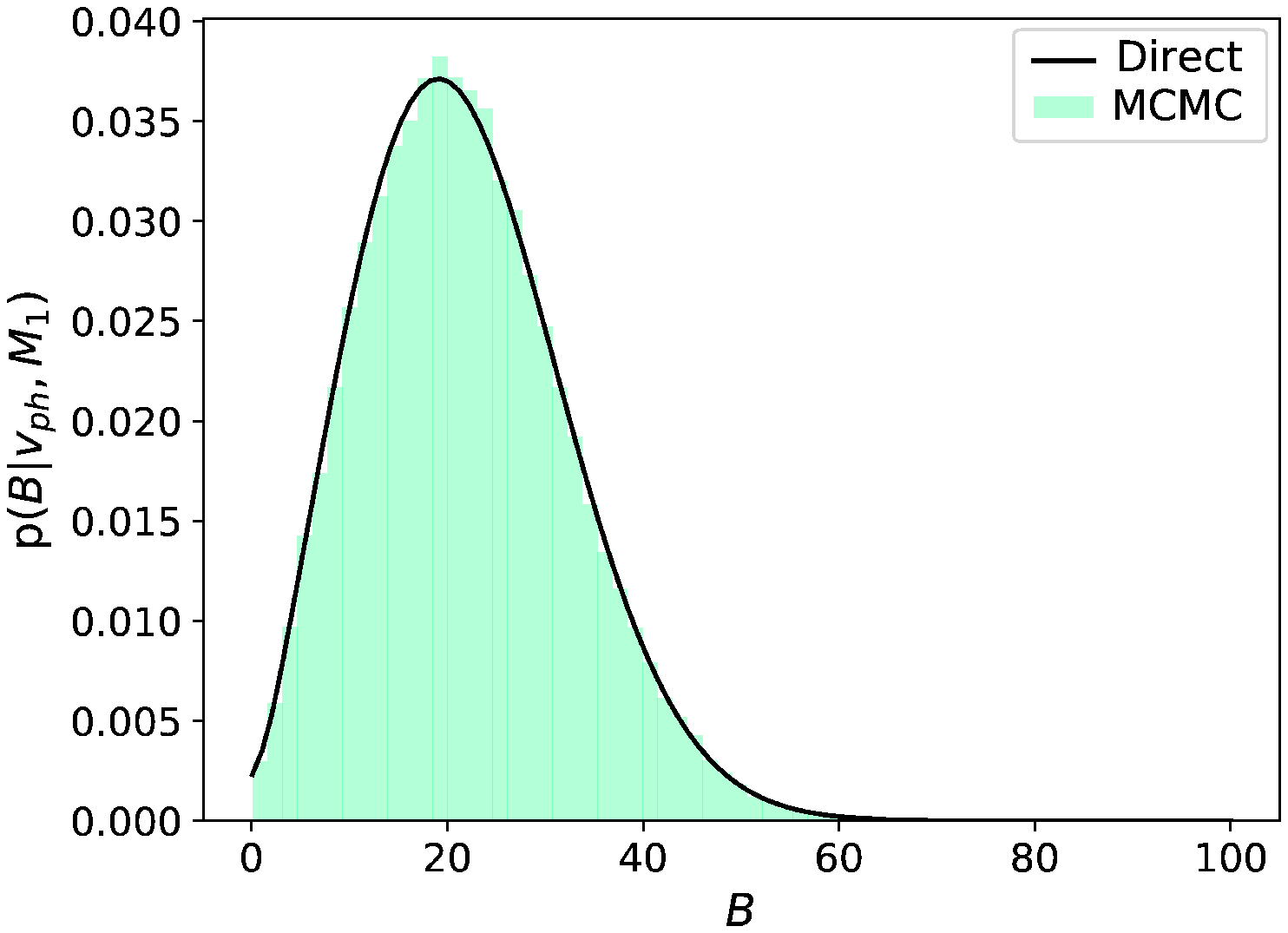} \\
\includegraphics[width = 0.24\textwidth]{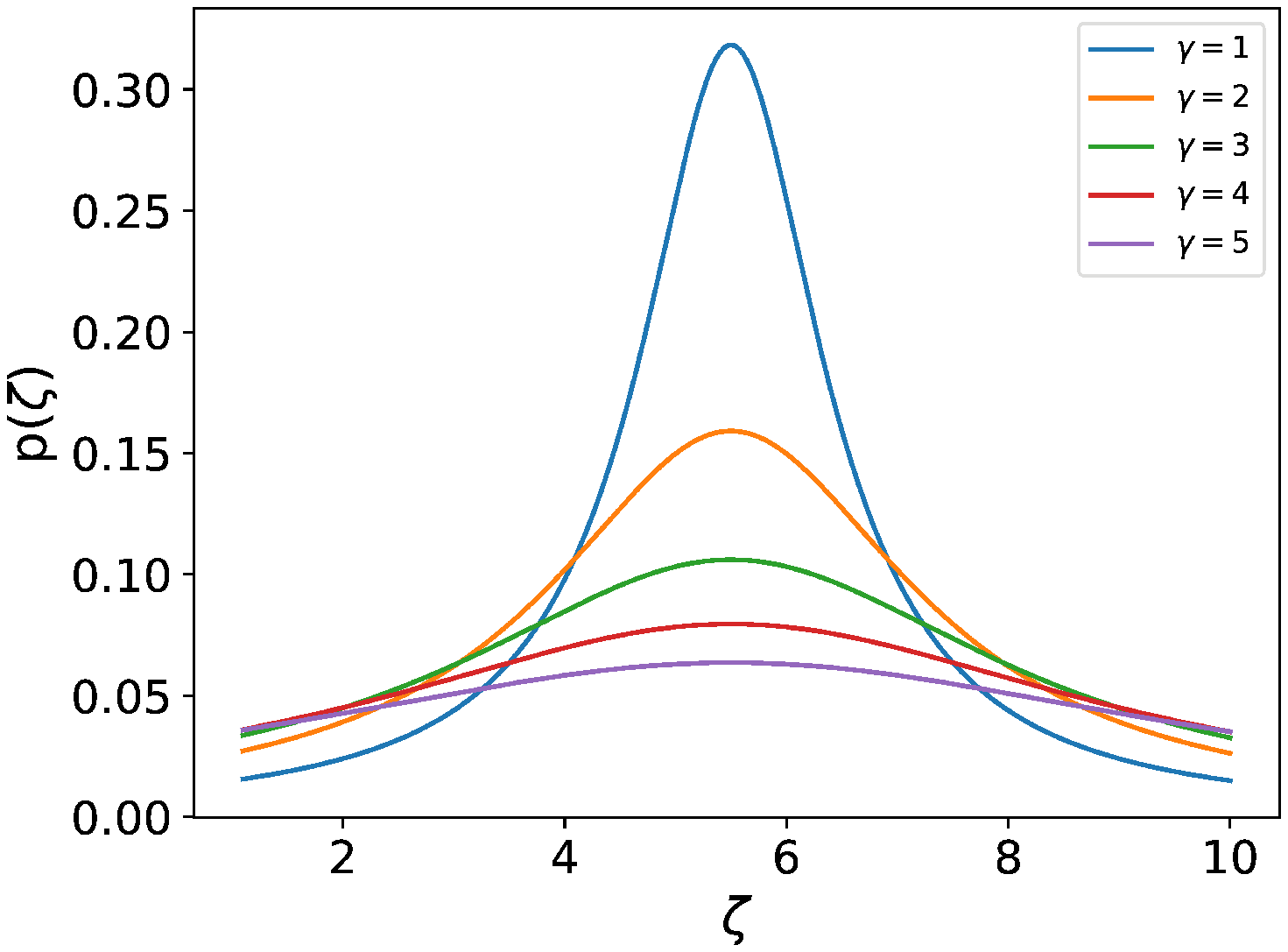} 
\includegraphics[width = 0.24\textwidth]{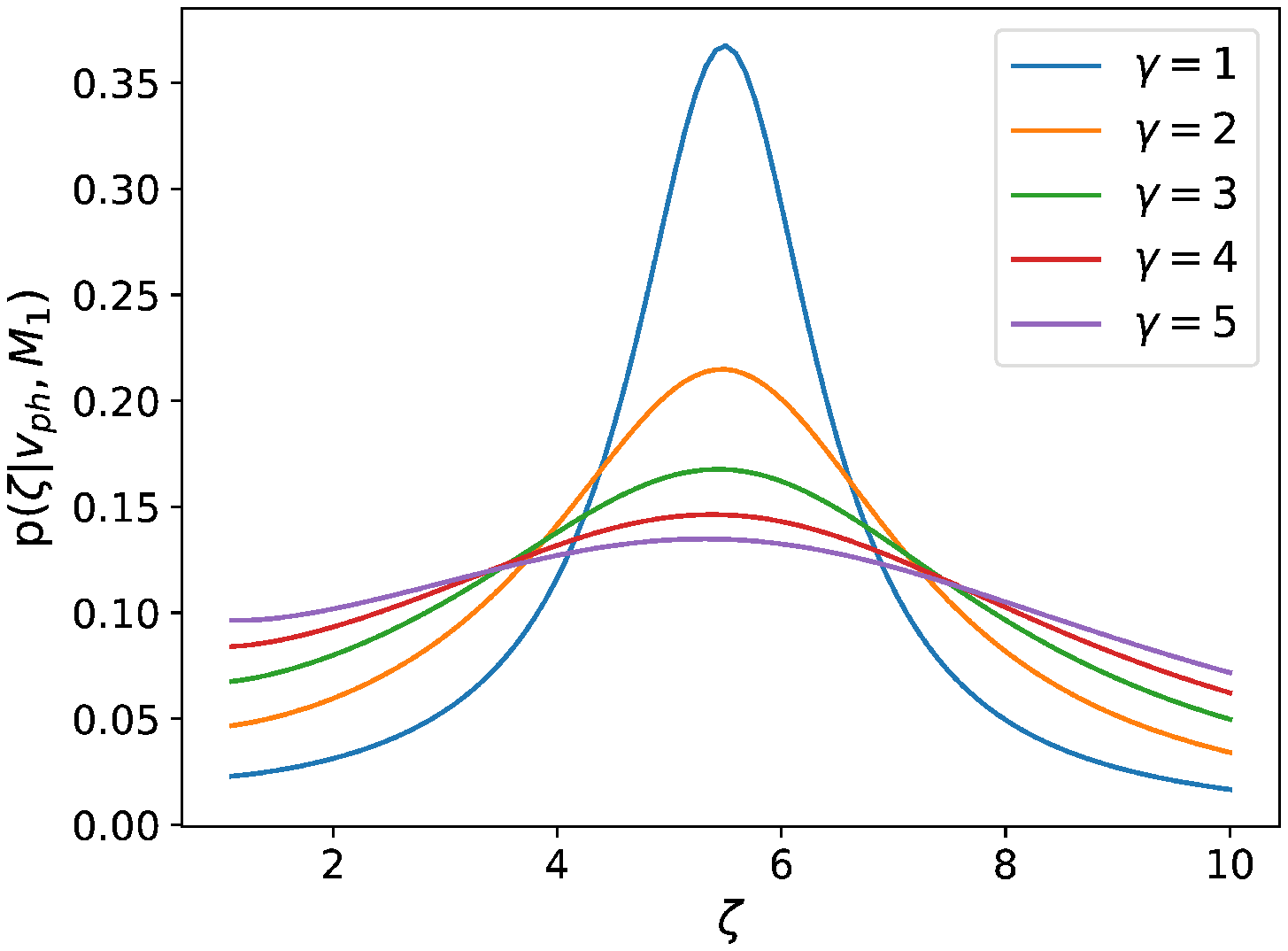} 
\includegraphics[width = 0.24\textwidth]{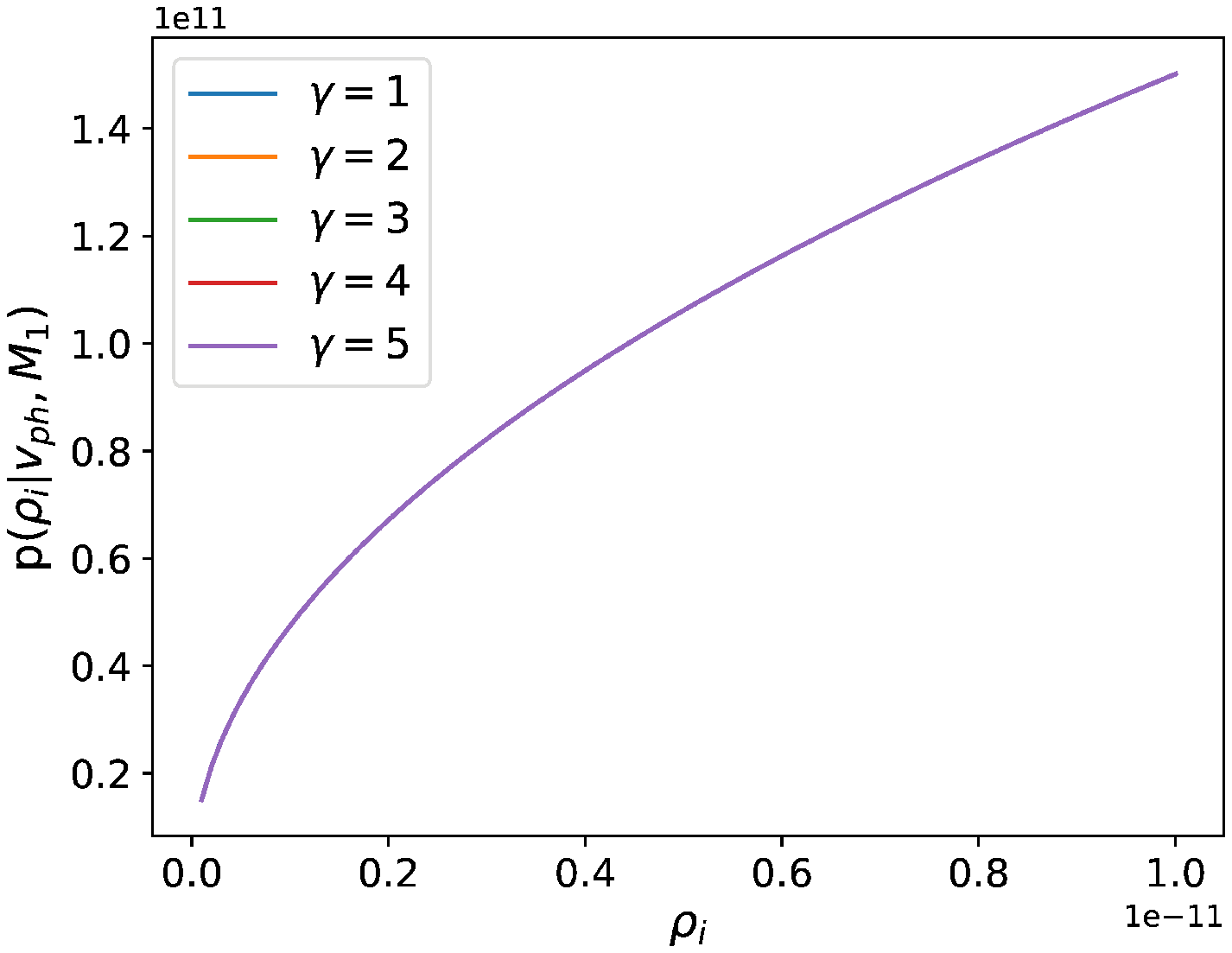} 
\includegraphics[width = 0.24\textwidth]{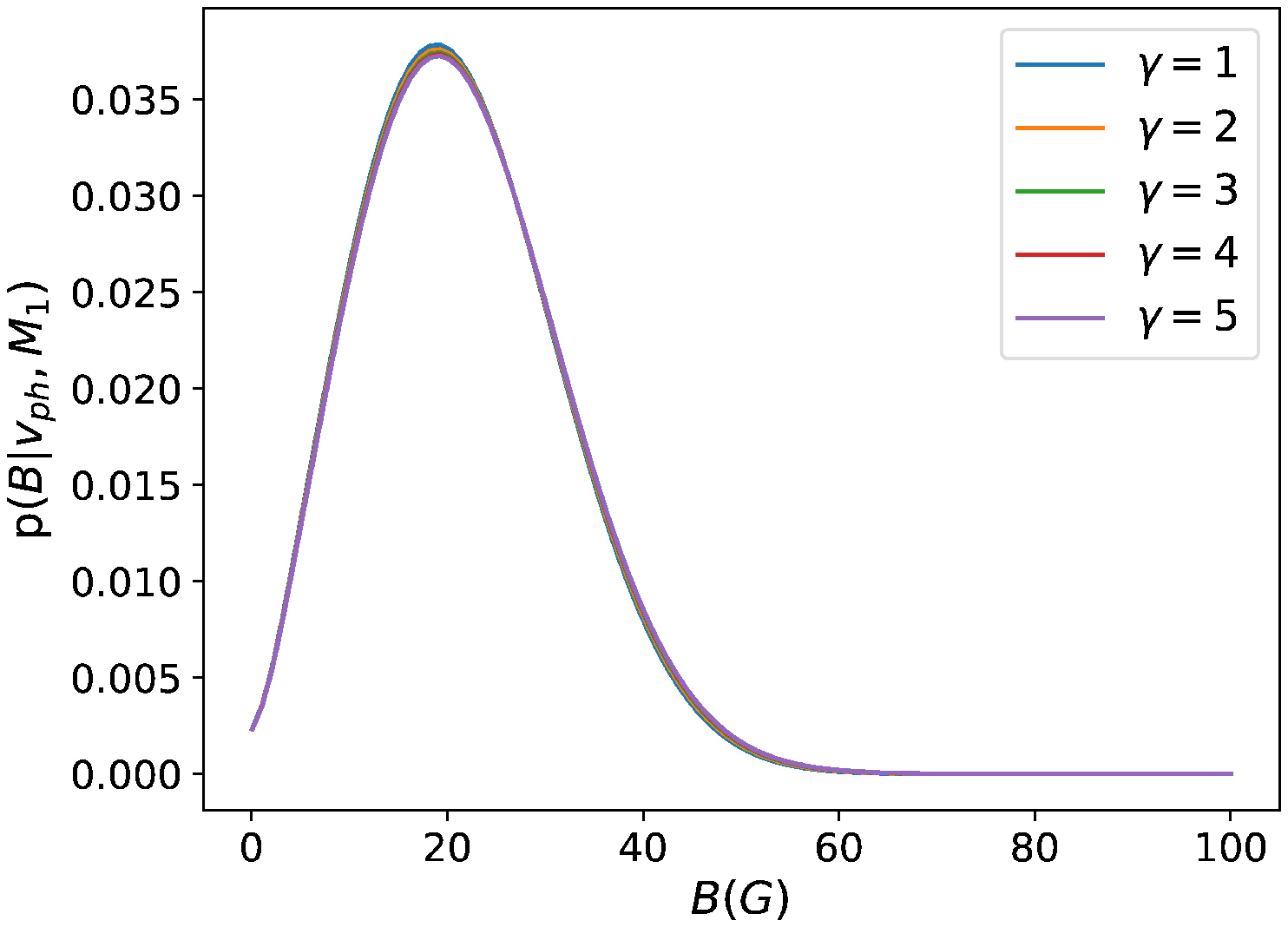} \\
\includegraphics[width = 0.24\textwidth]{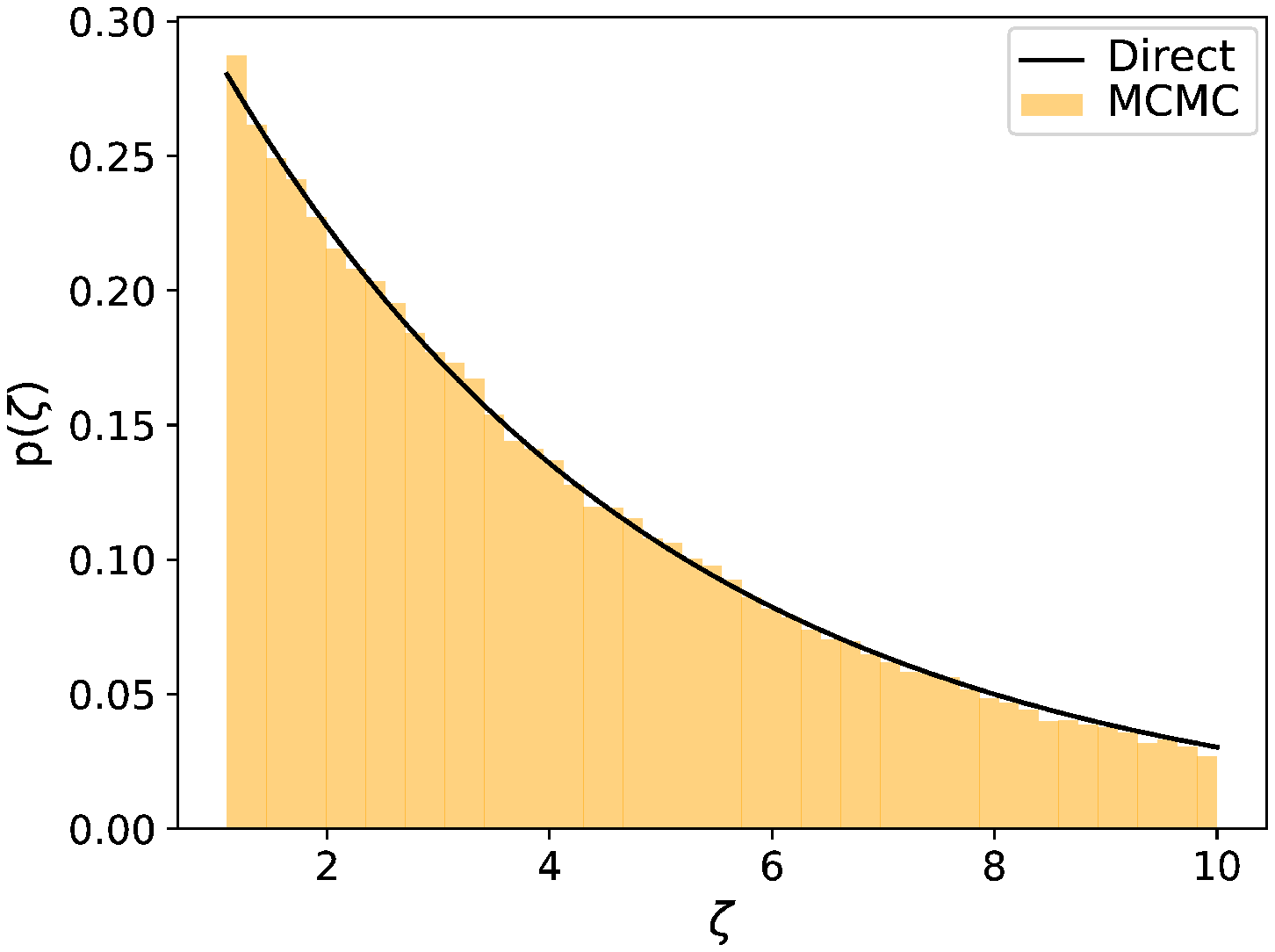} 
\includegraphics[width = 0.24\textwidth]{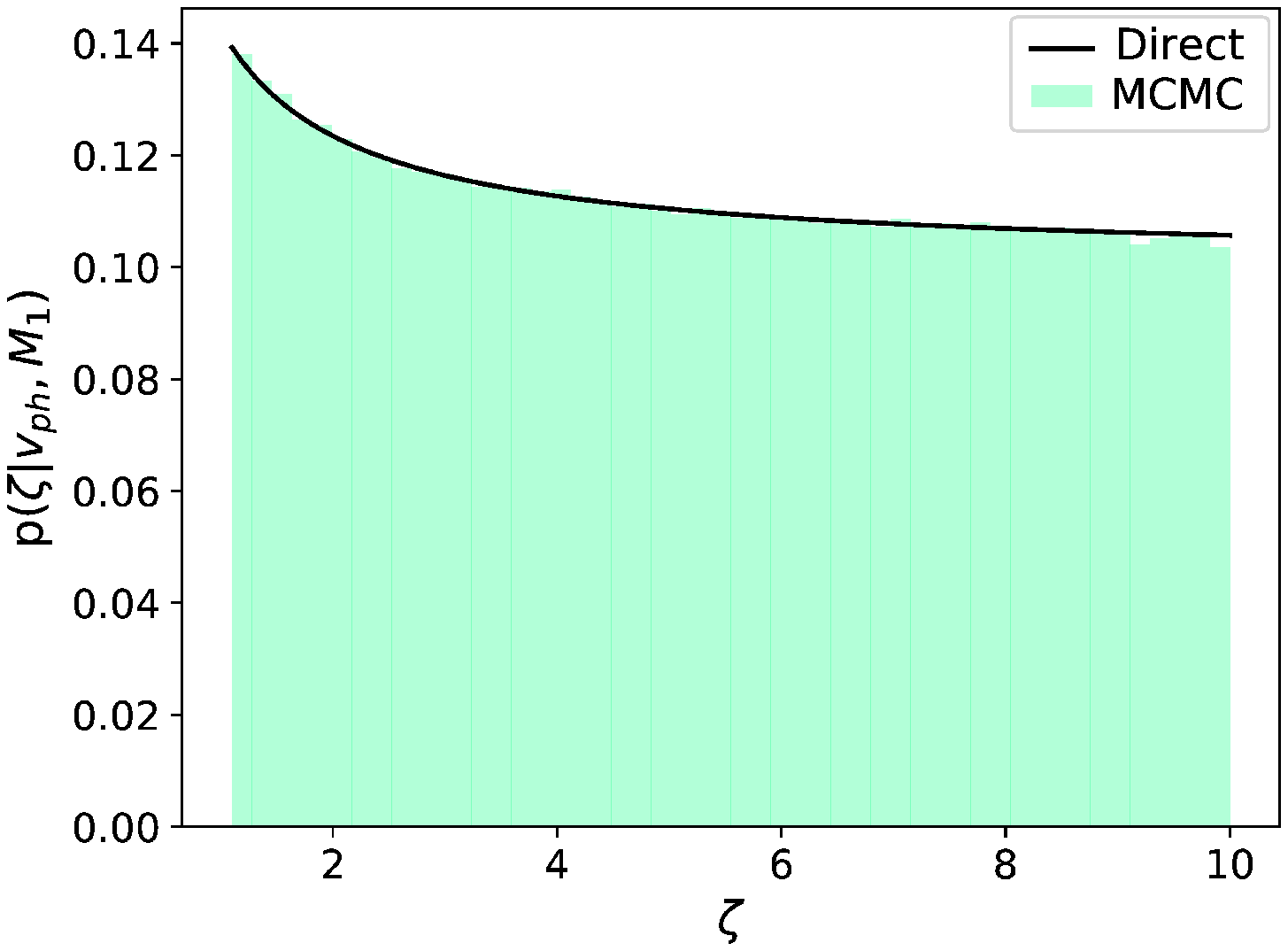} 
\includegraphics[width = 0.24\textwidth]{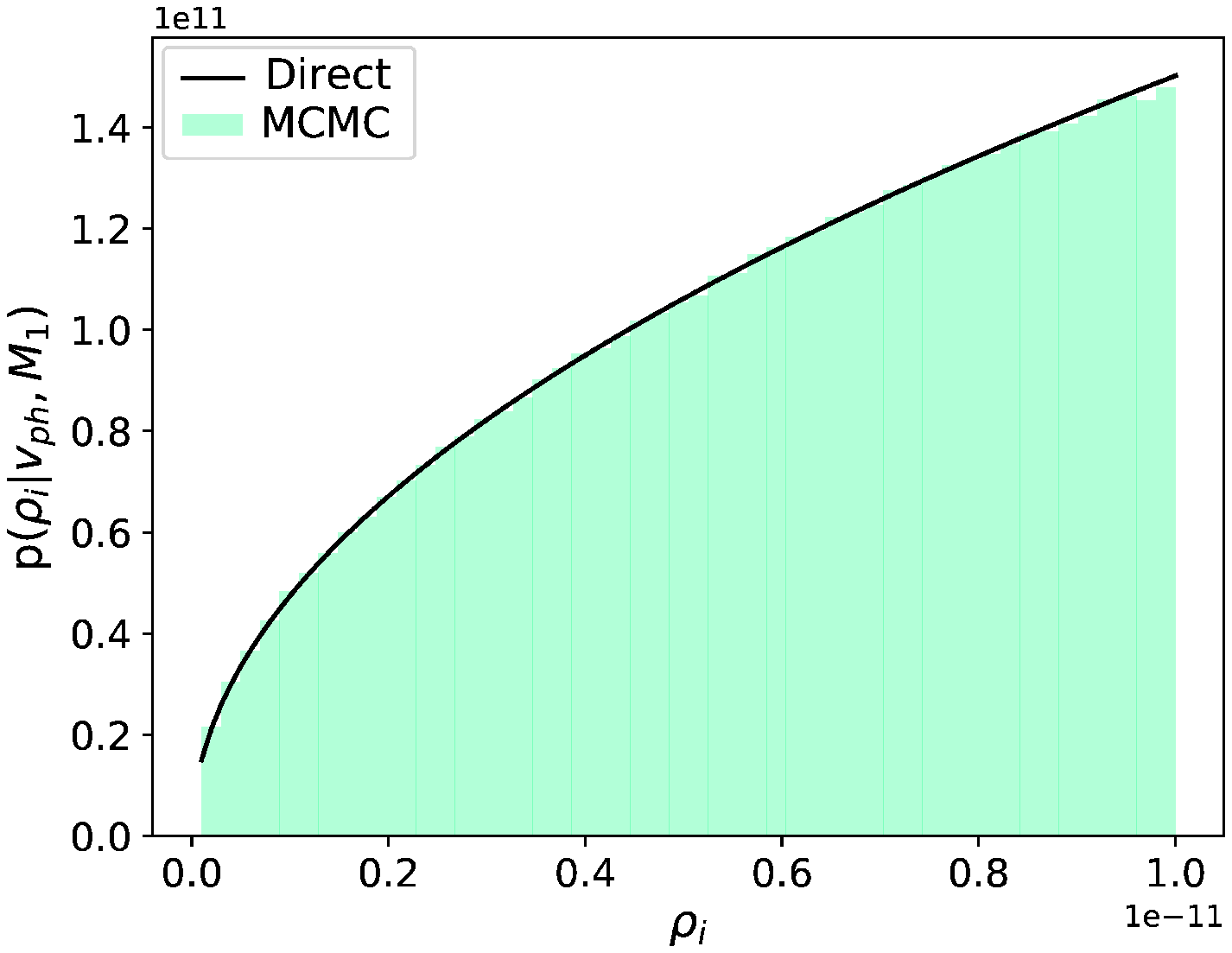} 
\includegraphics[width = 0.24\textwidth]{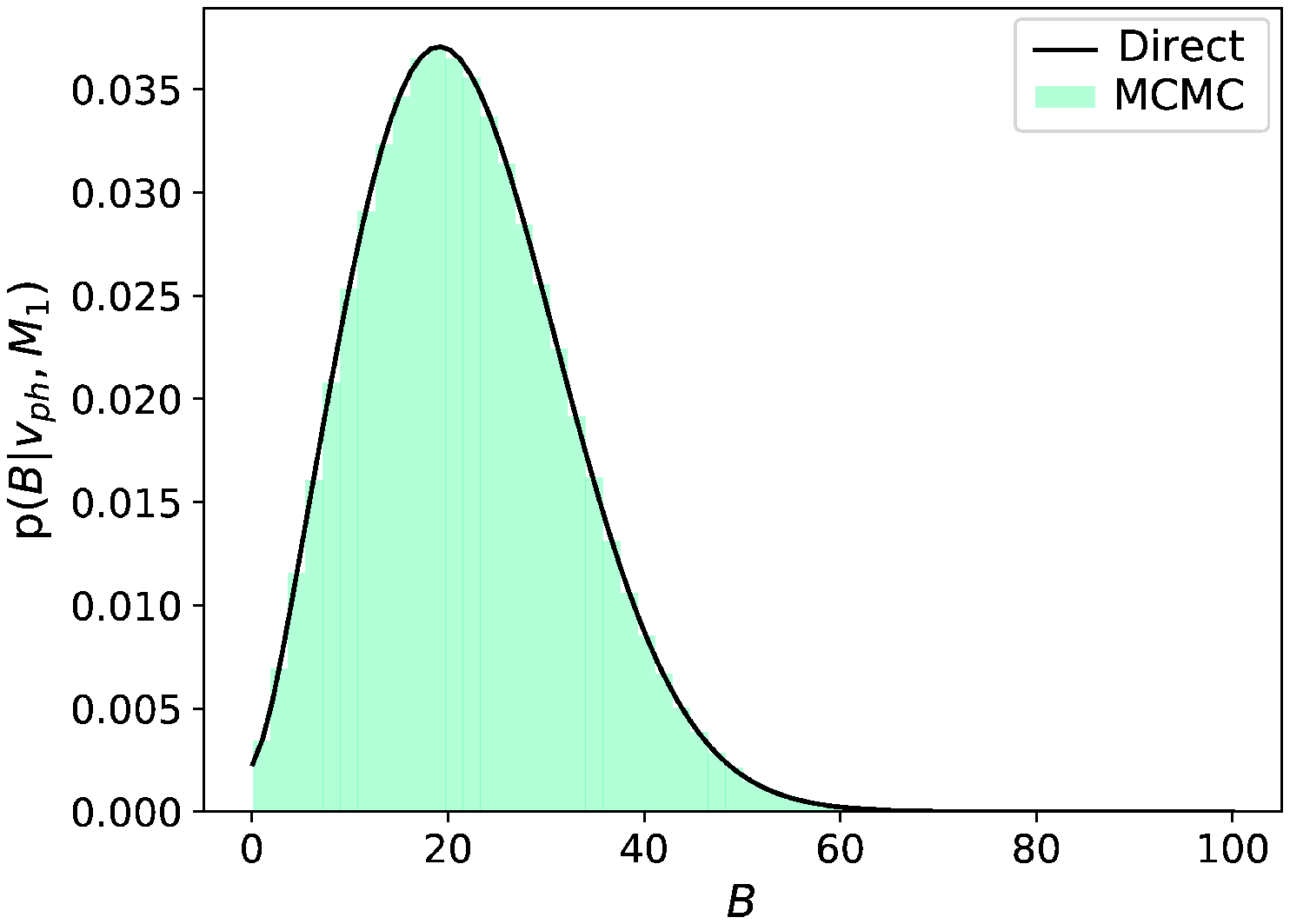} 
\caption{Solutions to the inversion of Eq.~(\ref{vphaseb0}) using different prior distributions for the density contrast. The different priors,  shown in the left column are the uniform prior, normal distribution, five different Cauchy functions and an exponential function, respectively. The  three rightmost columns show the corresponding posteriors for the density contrast, density of the waveguide, and magnetic field strength, respectively. Solid lines correspond to results obtained by direct integration of the posterior. Histograms are samples from the MCMC computations. \label{priorscontrast}}
\end{figure*}

The results of the prior analysis for different priors in density contrast are shown in Fig.~{\ref{priorscontrast}}. Each inversion with a different prior corresponds to a row of panels. The leftmost panels shows the used prior for density contrast, then we show the corresponding marginal posteriors for density contrast, density, and magnetic field strength. The results have been computed using both direct numerical integration and also Markov Chain Monte Carlo (MCMC) sampling of the posterior making use of {\em emcee} \citep{13emceepaper}, as explained in \cite{montesolis17}. As can be seen, the use of uniform, normal, Cauchy, or exponential priors for the density contrast influence the marginal posterior obtained for this parameter. Basically, what one gets as posterior is very similar to what we had as input in the prior. In all four cases, the marginal posterior for the density is unaffected. More importantly, the marginal posterior for the magnetic field strength is unaffected. The inference of magnetic field strength is therefore robust in front of changes in the employed prior distribution for density contrast.

\begin{figure*}
\center
\includegraphics[width = 0.24\textwidth]{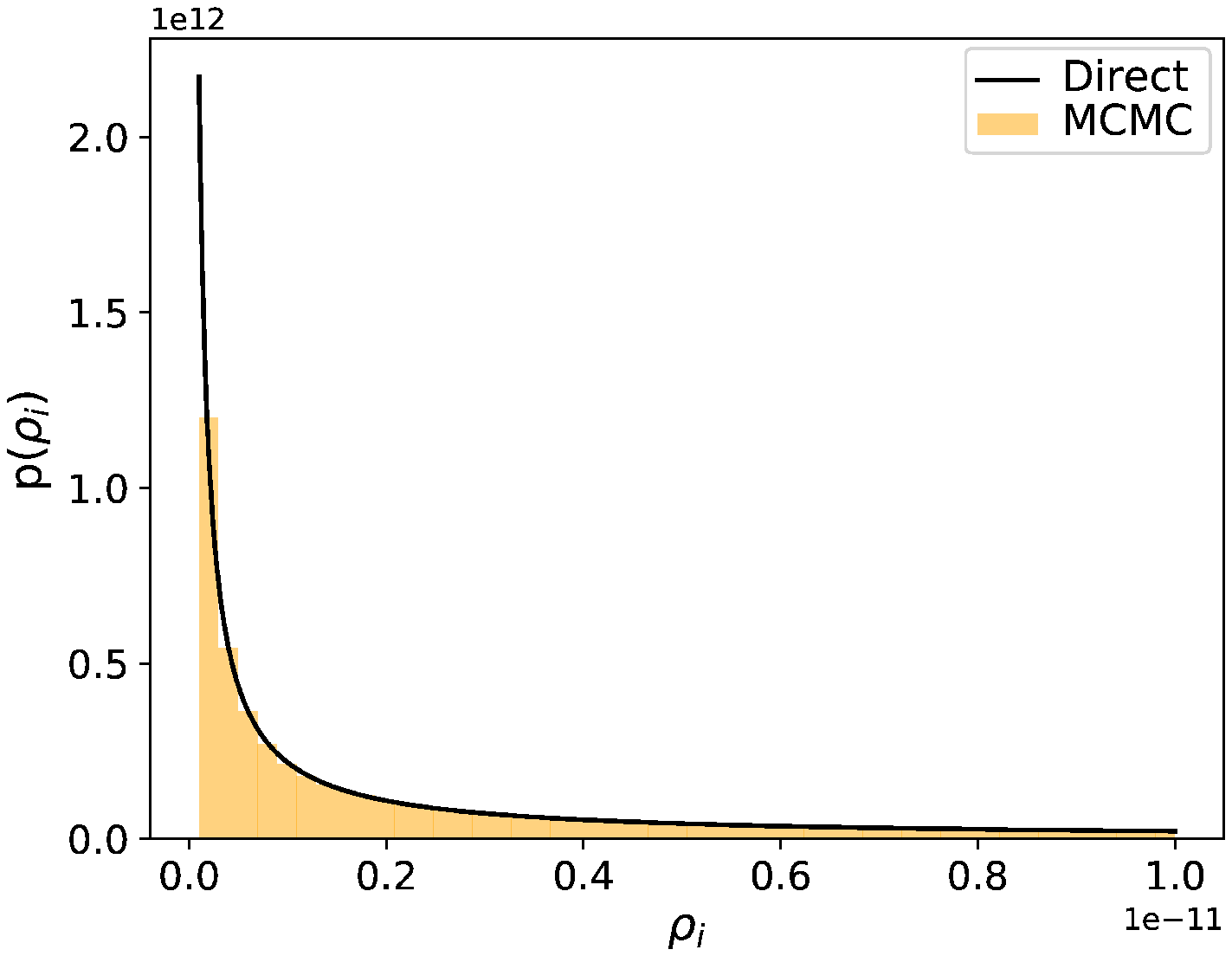} 
\includegraphics[width = 0.24\textwidth]{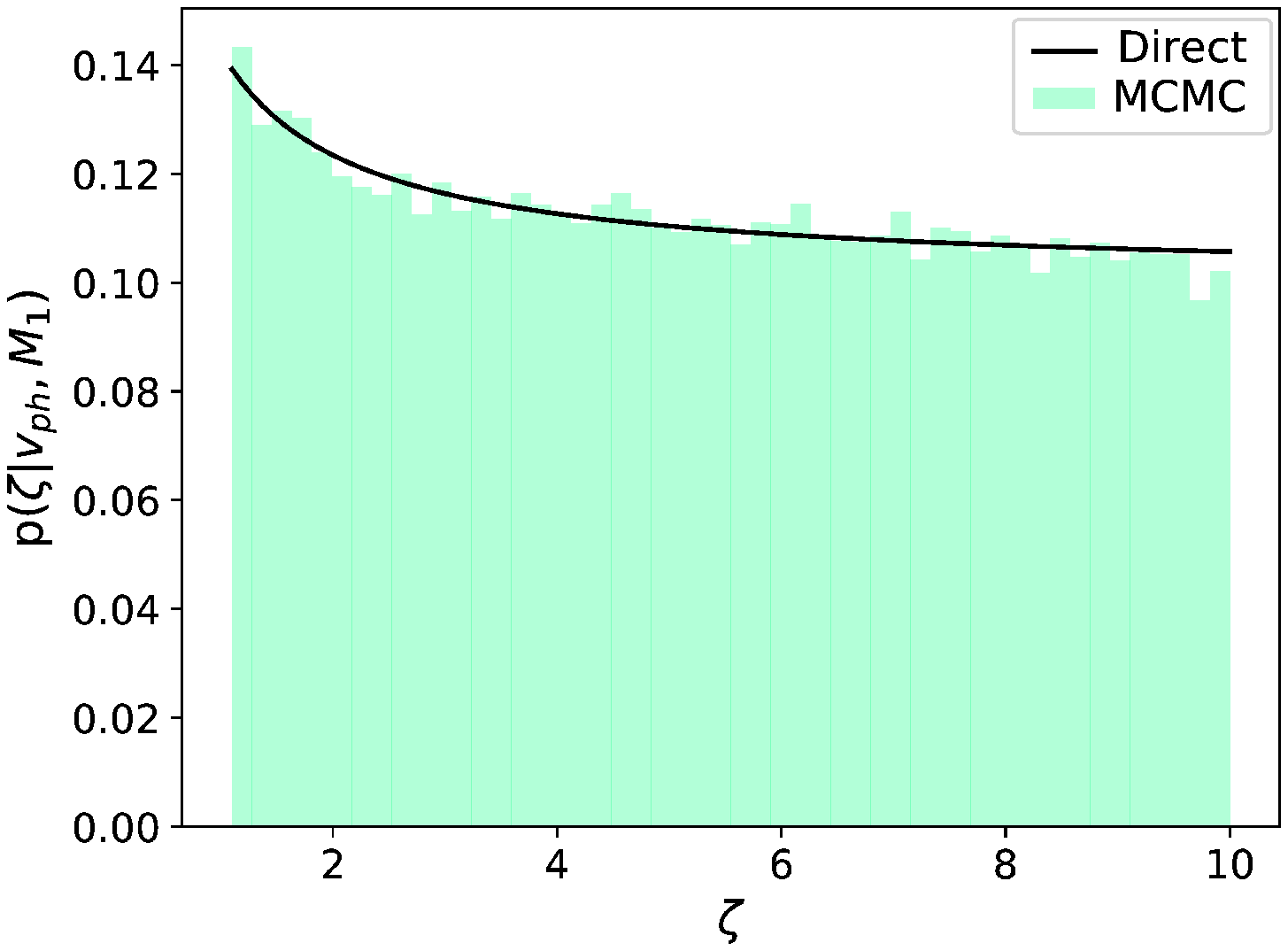} 
\includegraphics[width = 0.24\textwidth]{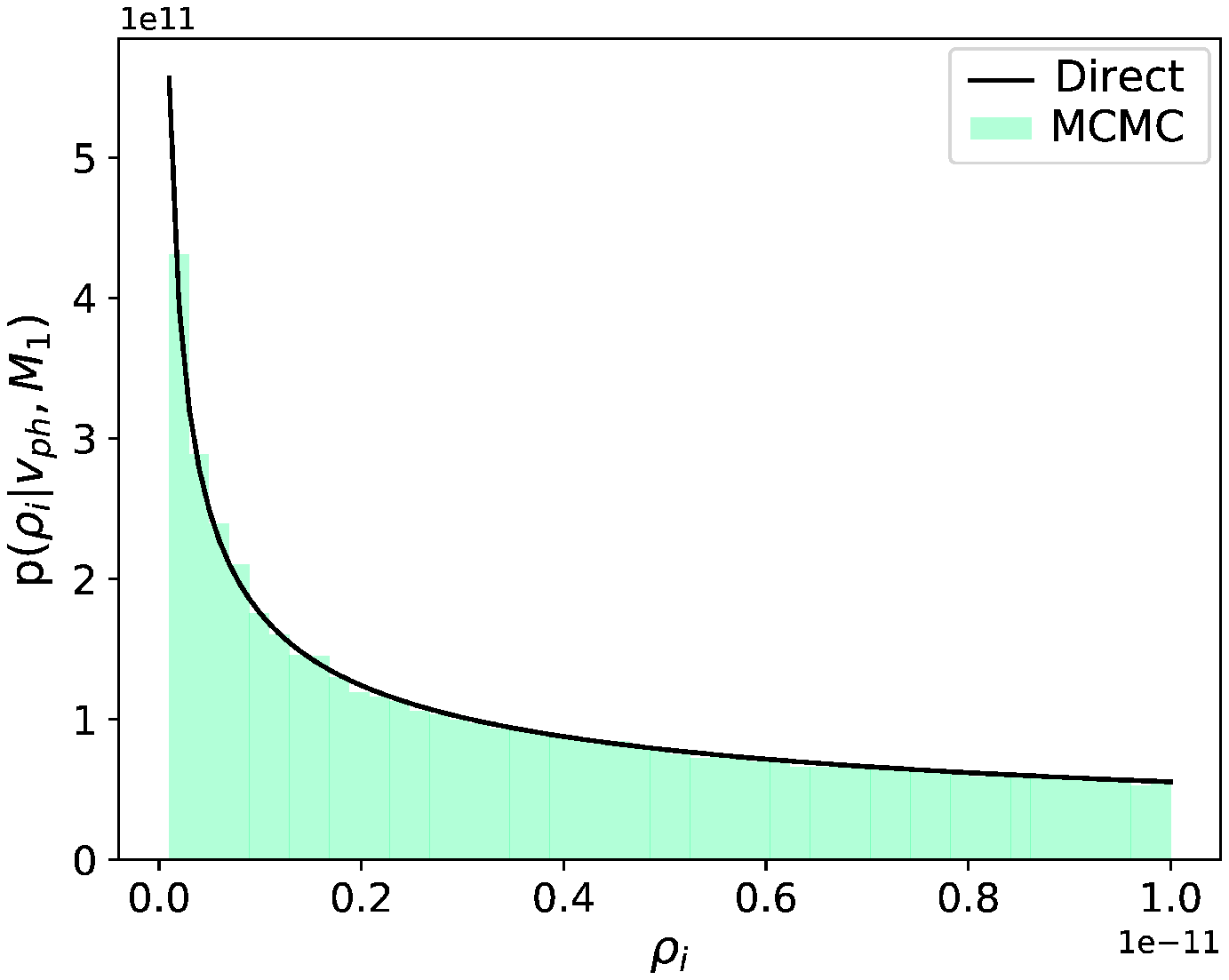} 
\includegraphics[width = 0.24\textwidth]{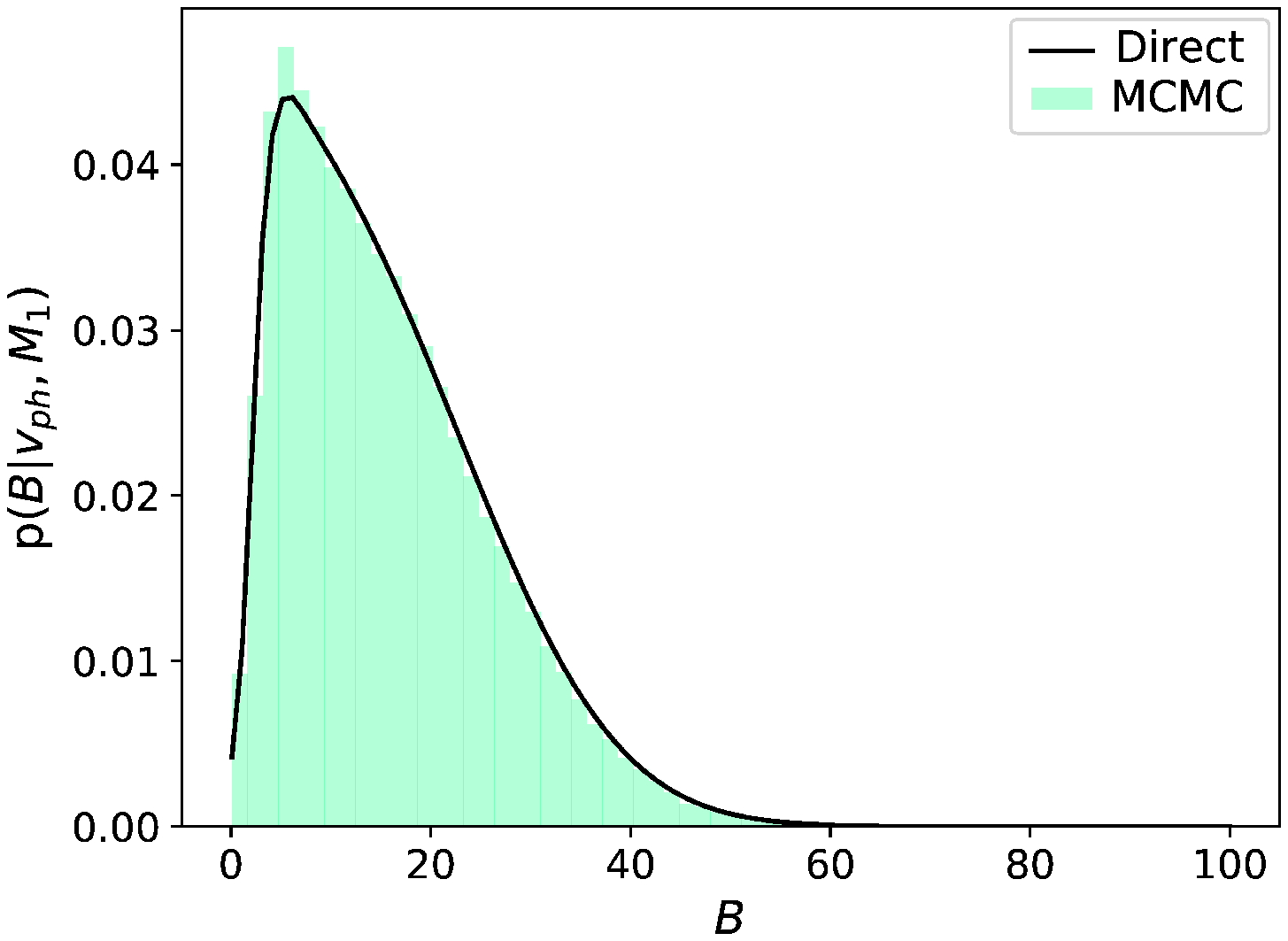} \\
\includegraphics[width = 0.24\textwidth]{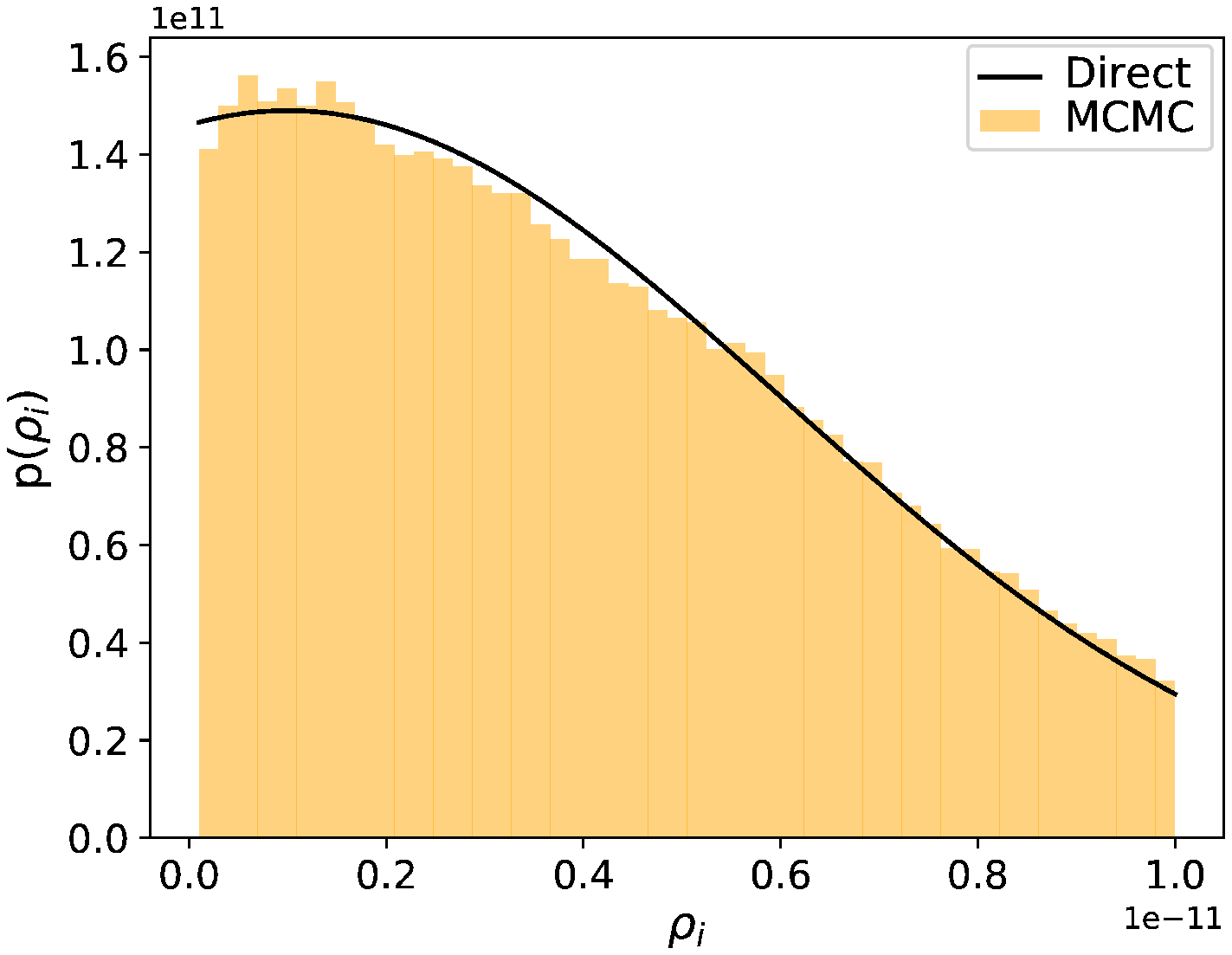} 
\includegraphics[width = 0.24\textwidth]{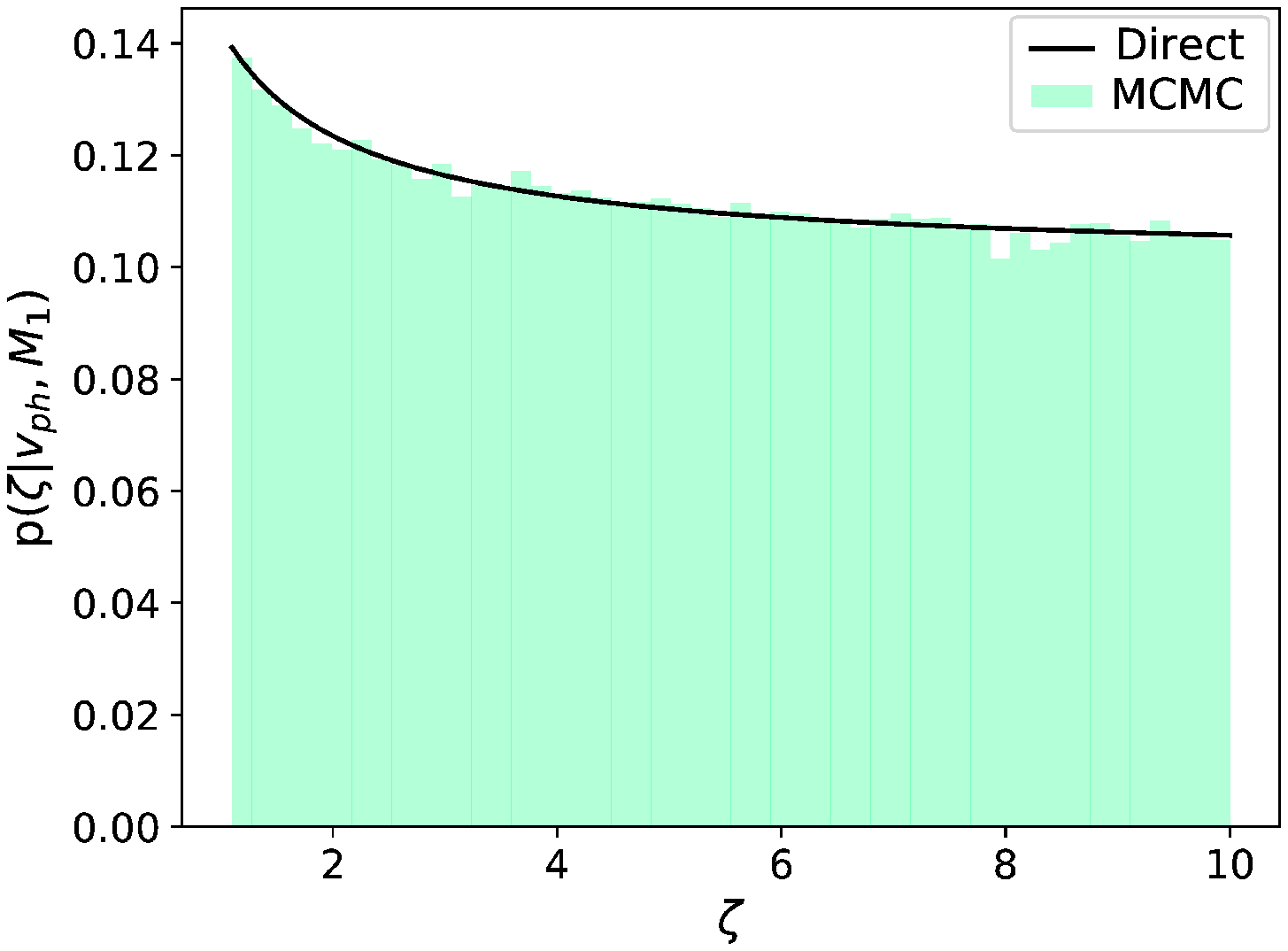} 
\includegraphics[width = 0.24\textwidth]{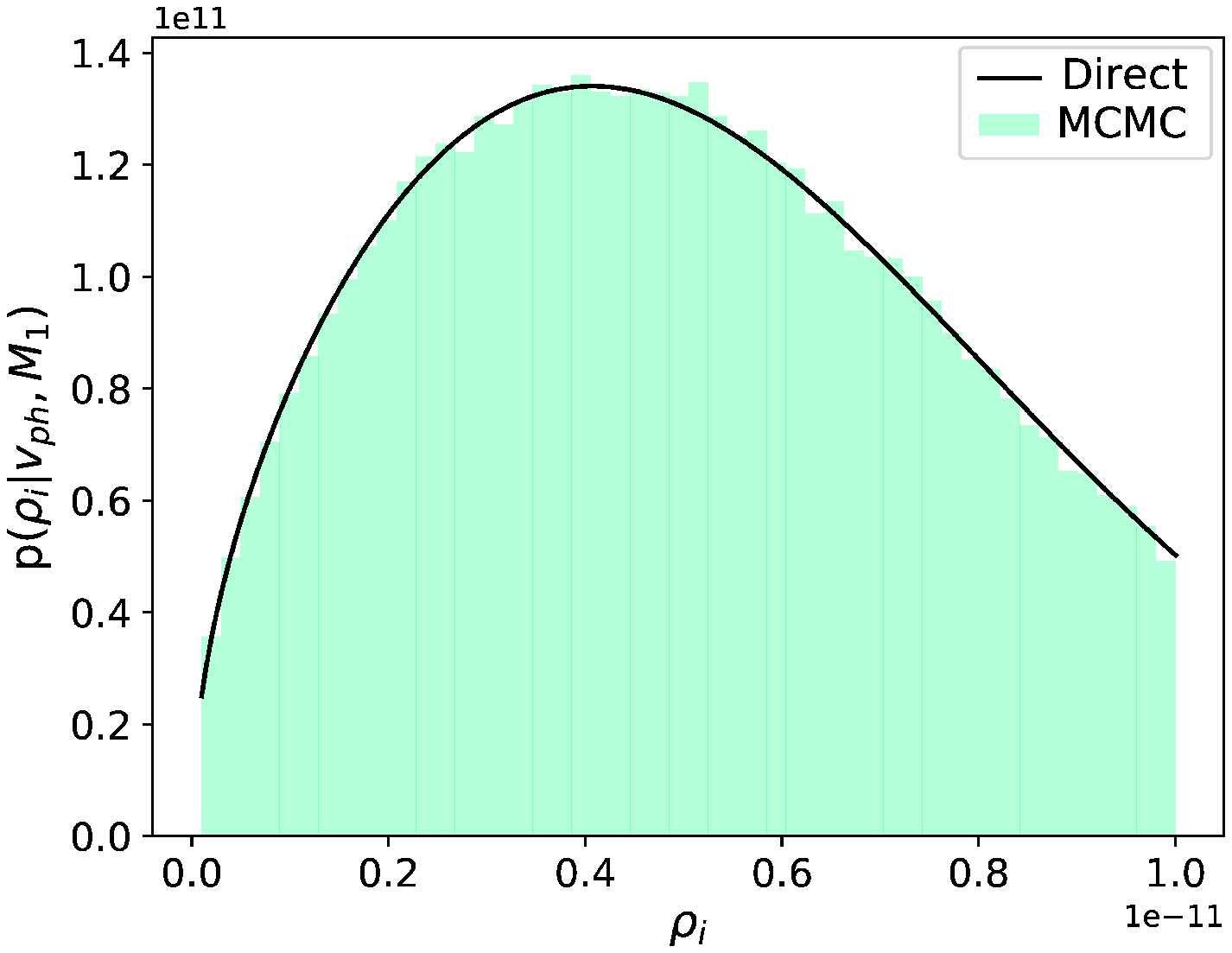} 
\includegraphics[width = 0.24\textwidth]{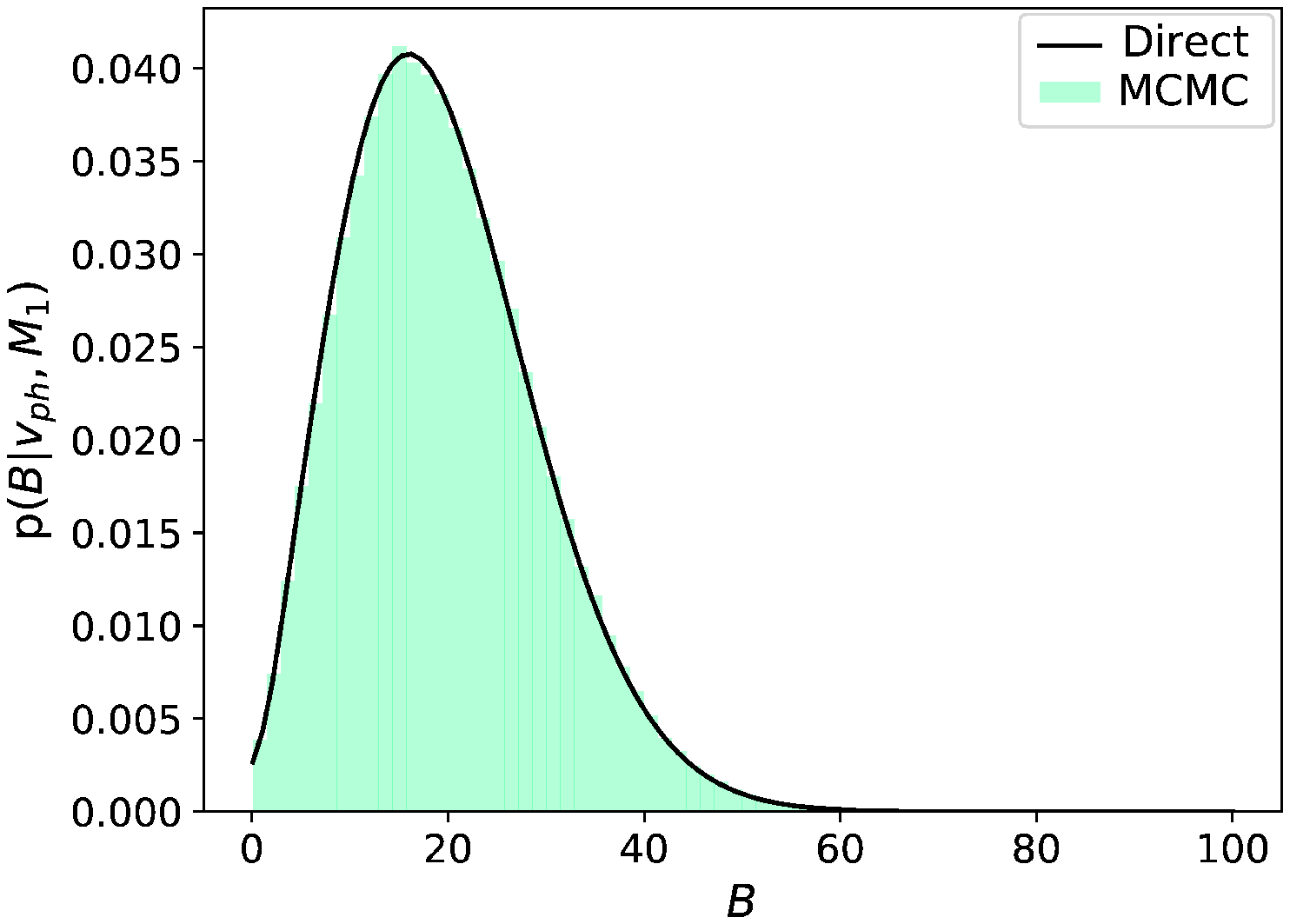}
\caption{Solutions to the inversion of Eq.~(\ref{vphaseb0}) using different prior distributions for the waveguide density. The different priors,  shown in the left column are the Jeffreys prior, and the normal distribution, respectively. The  three rightmost columns show the corresponding posteriors for the density contrast, density of the waveguide, and magnetic field strength, respectively. Solid lines correspond to results obtained by direct integration of the posterior. Histograms are samples from the MCMC computations. \label{priorsdensity}}
\end{figure*}

 The results of the prior analysis for different priors in density contrast are shown in Fig.~{\ref{priorsdensity}}. Again, each row shows results from the inversion using a different prior on density, indicated on the leftmost panel. Similarly to the previous case, changes in the prior information for the density do affect the marginal posterior obtained for this parameter, but do not influence significantly the marginal posteriors for the other two parameters of the problem, density contrast and magnetic field strength. The posterior for density contrast seems to be completely independent on the information on the density of the waveguide. Only in the case of a Jeffreys prior for the density, the posterior for the magnetic field strength slightly affected (see rightmost panel in Fig.~\ref{priorsdensity}).

In summary, what one is willing to accept a priori about the density and the density contrast affects their corresponding posteriors, but very little the inference of the magnetic field strength. The three parameters seem to be rather independent, since changes in the prior information do not seem to affect the regions of parameter space where the likelihood is high.

\section{Application to loop oscillation data}\label{appb}

We applied our Bayesian methods to existing data of transverse loop oscillations. Table~\ref{tab_B1} shows a comparison between magnetic field strength estimates from previous works and summaries of posteriors using Bayesian analysis. For each event, the length of the loop, oscillation period, and measured phase speed are shown. First, the summary of the posterior for the internal Alfv\'en speed is computed using the forward model (\ref{vphase}) and the full posterior (\ref{bayes1}), which is then marginalised. Then, the magnetic field strength posterior is computed using the forward model (\ref{vphaseb0}) and the Bayesian posterior (\ref{bayes2}) for a uniform prior on density over the range $\rho_{\rm i} \in [10^{-13}-10^{-11}]$ kg m$^{-3}$. The last column shows the more constrained results obtained using a Gaussian prior on density centered at the values estimated or assumed by the previous studies. \cite{nakariakov01} estimate a loop density of $10^{9.3\pm0.3}$ cm$^{-3}$, \cite{aschwanden02} list their estimated densities on their Table III, \cite{vandoorsselaere08} use an internal electron density of $9.8\pm0.3\times10^{14}$ m$^{-3}$, \cite{white12} and \cite{pascoe16} assume the loops have an electron number density of $10^{15}$ m$^{-3}$. When the posteriors computed with the Gaussian priors on density are summarised, the median and credible region at the 68\% credible interval agree well with the previous numerical estimates for the events presented by \cite{nakariakov01}, \cite{aschwanden02}, \cite{goossens02a}, and \cite{vandoorsselaere08}. 

We found a discrepancy between our posterior summaries and previous magnetic field strength estimates in the events analysed by \cite{white12} and \cite{pascoe16}. \cite{white12} obtained possible ranges of variation for the magnetic field strength according to the inequalities

\begin{equation}\label{whiteeq}
\frac{v_{\rm ph}}{\sqrt{2}}\sqrt{\mu_0\tilde{\mu} m_{\rm p} n_{\rm e}} \leq B \leq v_{\rm ph}  \sqrt{\mu_0\tilde{\mu} m_{\rm p} n_{\rm e}}.
\end{equation}
By inserting the values $\mu_0=4\pi \times10^{-7}$ H m$^{-1}$, $\tilde{\mu} = 1.27$, $m_{\rm p}= 1.6726 \times10^{-27}$ kg and $n_{\rm e} = 10^{15}$ m$^{-3}$ and using the estimated phase speeds in Table 1 by \cite{white12}, we obtain the magnetic field strength ranges shown in italics in Table~\ref{tab_B1}, which differ from the ranges in Table 3 by \cite{white12} but agree well with our Bayesian posterior summaries using a Gaussian prior on density. \cite{pascoe16}  find their magnetic field strength estimates and associated errors from the expression

\begin{equation}\label{pascoeeq}
B_0=C_{\rm A0} \sqrt{\mu_0\tilde{\mu} m_{\rm p} n_{\rm e}}.
\end{equation}
By inserting the values $\mu_0=4\pi \times10^{-7}$ H m$^{-1}$, $\tilde{\mu} = 1.27$, $m_{\rm p}= 1.6726 \times10^{-27}$ kg and $n_{\rm e} = 10^{15}$ m$^{-3}$ and using the estimated values for $C_{\rm A0}$ in their Table 3, we obtain the magnetic field strength values shown in italics in 
Table~\ref{tab_B1}, which differ from the values quoted in Table 3 by \cite{pascoe16} but agree well with our Bayesian posterior summaries using a Gaussian prior on density.
The source of the discrepancy between our results and those by \cite{white12} and \cite{pascoe16} could be a typo in one of the exponents in the square root term in Eqs.~(\ref{whiteeq}) and (\ref{pascoeeq}) or, equivalenty, the reported results in \cite{white12} and \cite{pascoe16} would apply for values of $n_e=10^{14}$ m$^{-3}$ rather than the 
stated $10^{15}$ m$^{-3}$.

A second application to loop oscillation data is shown in Table~\ref{tab_B2}. The list compiled by \cite{goddard16} from a catalogue by \cite{zimovets15} contains 120 events. Two of these events are not analysed, since no loop length measurement is available. In 52 cases,  estimates of damping times are given. For these cases, the Bayesian inference of the magnetic field strength is performed using uniform priors and for two cases. One without considering the damping, using Eqs.~(\ref{vphaseb0}) and (\ref{bayes2}), leading to the posterior summaries $B_{\rm no \hspace{0.05cm}damping}$. Another considering the damping, using Eqs.~(\ref{vphasedamp}), (\ref{damp2}) and (\ref{bayes3}), leading to the posterior summaries $B_{\rm damping}$. As can be appreciated, the differences in the posterior summaries are minimal, up to $\pm 1$ G at most in the median and/or the upper/lower error bars at the 68\% credible interval. Hence we confirm with our application to real data that considering the damping of the oscillations in the inference of the magnetic field strength has very little impact, at least when the thin tube and thin boundary approximations are used.  The analysis is completed by performing the inference without damping for 66 cases in which damping was not measured, using Eqs.~(\ref{vphaseb0}) and (\ref{bayes2}).

\begin{onecolumn}
\begin{longtable}{ccccccccc}	
\caption{Transverse loop oscillation data, previous inference results ($B_0$) and Bayesian posterior summaries using uniform ($B_{\rm u}$) and Gaussian ($B_{\rm G}$) priors on density. \label{tab_B1}}\\
\hline \hline
Event & Loop  & L & P& $v_{\rm ph}$ & $v_{\rm Ai}$ & $B_0$&$B_{\rm u}$ & $B_{\rm G}$ \\  
 & & (Mm) & (s) & (km s$^{-1}$) & (km s$^{-1}$) & (G)&(G) & (G)\\
\hline
\endfirsthead
\caption{continued.}\\
\hline\hline
Event & Loop  & L & P& $v_{ph}$ & $v_{Ai}$ & $B_0$&$B_u$ & $B_G$ \\  
 & & (Mm) & (s) & (km s$^{-1}$) & (km s$^{-1}$) & (G)&(G) & (G)\\
\hline
\endhead
\hline
\endfoot
\noalign{\smallskip}
\cite{nakariakov01}&&&&&&&&\\
\noalign{\smallskip}
\hline
\noalign{\smallskip}
1 & 1 & $130$ & $256$ & $1020\pm132$ & $800^{+122}_{-101}$ & $-$ &$22^{+6}_{-7}$ & $13^{+3}_{-3}$ \\ \noalign{\smallskip}
2 & 1 & $190$ & $360$ & $1030\pm410$ & $813^{+330}_{-317}$ & $13\pm 9$ &$21^{+12}_{-9}$ & $13^{+7}_{-6} $ \\\noalign{\smallskip}
\hline
\noalign{\smallskip}
\cite{aschwanden02} \\\cite{goossens02a}&&&&&&&&\\
\noalign{\smallskip}
\hline
\noalign{\smallskip}
a & 1 & $168$ & $261$ & $1287$ & $1011^{+127}_{-113}$ & 13& $28^{+7}_{-9}$ & $12^{+3}_{-3} $\\\noalign{\smallskip}
b & 1 & $72$ & $265$ & $543$ & $426^{+55}_{-47}$ & 6 &$12^{+3}_{-4}$ & $5^{+1}_{-1} $\\\noalign{\smallskip}
d & 1 & $174$ & $316$ & $1101$ & $863^{+110}_{-95}$ &11  &$24^{+5}_{-8}$ & $11^{+3}_{-3}$ \\\noalign{\smallskip}
f & 1 & $204$ & $277$ & $1473$ & $1156^{+145}_{-125}$ &16 & $32^{+7}_{-10}$ & $15^{+4}_{-4} $\\\noalign{\smallskip}
g & 1 & $162$ & $272$ & $1191$ & $936^{+118}_{-103}$ & 10 &$26^{+6}_{-9}$ & $10^{+2}_{-2} $\\\noalign{\smallskip}
a & 3 & $390$ & $522$ & $1494$ & $1173^{+148}_{-130}$ & 11 &$32^{+8}_{-10}$ & $11^{+2}_{-3} $\\\noalign{\smallskip}
a & 4 & $258$ & $435$ & $1186$ & $931^{+118}_{-103}$ & 13 &$25^{+7}_{-8}$ & $12^{+3}_{-3} $\\\noalign{\smallskip}
c & 5 & $166$ & $143$ & $2322$ & $1823^{+230}_{-200}$ & 27 &$50^{+12}_{-16}$ & $25^{+6}_{-6}$ \\\noalign{\smallskip}
a & 10 & $406$ & $423$ & $1920$ & $1508^{+188}_{-165}$ &20 & $42^{+10}_{-13}$ & $19^{+5}_{-5} $\\\noalign{\smallskip}
a & 16 & $192$ & $185$ & $2076$ & $1631^{+205}_{-180}$ & 15 &$45^{+11}_{-14}$ & $14^{+3}_{-3} $\\\noalign{\smallskip}
a & 17 & $146$ & $396$ & $737$ & $579^{+73}_{-63}$ & 6 &$16^{+4}_{-5}$ & $5^{+1}_{-1} $\\\noalign{\smallskip}
\hline
\noalign{\smallskip}
\cite{vandoorsselaere08}&&&&&&&&\\
\noalign{\smallskip}
\hline
\noalign{\smallskip}
1 & 1 & $390$ & $296\pm24$ & $2600\pm500$ & $2045^{+425}_{-402}$ & $39\pm8$ &$55^{+18}_{-19}$ & $32^{+6}_{-6} $\\\noalign{\smallskip}
\hline
\noalign{\smallskip}
\cite{white12}&&&&&&&&\\
\noalign{\smallskip}
\hline
\noalign{\smallskip}
1 & 1 & $121\pm2$ & $225\pm40$ & $1080\pm220$ & $851^{+185}_{-178}$  & $3.9-5.6$&$23^{+8}_{-8}$ & $14^{+4}_{-4}$ \\\noalign{\smallskip}
    &   &                      &                    &                        &                                    & $\mathit{(12.5-17.6)}$            &                           &   \\\noalign{\smallskip}
1 & 2  & $111\pm6$ & $215\pm5$ & $1030\pm110$ & $808^{+108}_{-93}$  &$3.8-5.3$ &$22^{+6}_{-7}$ & $13^{+3}_{-3}$ \\\noalign{\smallskip}
	     &   &                      &                    &                        &                                    & $\mathit{(11.9-16.8)}$            &                           &   \\\noalign{\smallskip}
1 & 3  & $132$ & $213\pm9$ & $1240\pm140$ & $973^{+133}_{-118}$  &$4.5-6.4$ &$27^{+7}_{-9}$ & $16^{+4}_{-4}$ \\\noalign{\smallskip}
	     &   &                      &                    &                        &                                    & $\mathit{(14.3-20.3)}$            &                           &   \\\noalign{\smallskip}
1 & 4  & $113\pm4$ & $216\pm30$ & $1050\pm170$ & $826^{+148}_{-138}$  &$3.8-5.4$ &$22^{+7}_{-7}$ & $13^{+4}_{-4}$\\ \noalign{\smallskip}
	     &   &                      &                    &                        &                                    & $\mathit{(12.1-17.2)}$            &                           &   \\\noalign{\smallskip}
2 & 1  & $396$ & $520\pm5$ & $1520\pm150$ & $1193^{+148}_{-130}$  &$5.6-7.9$ &$33^{+8}_{-10}$ & $20^{+5}_{-5}$ \\\noalign{\smallskip}
	     &   &                      &                    &                        &                                    & $\mathit{(17.6-24.8)}$            &                           &   \\\noalign{\smallskip}
2 & 2  & $374$ & $596\pm50$ & $1260\pm160$ & $991^{+145}_{-135}$  &$4.6-6.5$ &$27^{+7}_{-8}$ & $16^{+4}_{-4}$ \\\noalign{\smallskip}
	     &   &                      &                    &                        &                                    & $\mathit{(14.6-20.6)}$            &                           &   \\\noalign{\smallskip}
3 & 1  & $279\pm3$ & $212\pm20$ & $2630\pm360$ & $2068^{+322}_{-297}$  & $9.6-13.6$&$56^{+15}_{-18}$ & $34^{+9}_{-8}$\\ \noalign{\smallskip}
	     &   &                      &                    &                        &                                    & $\mathit{(30.4-42.9)}$            &                           &   \\\noalign{\smallskip}
3 & 2  & $240\pm4$ & $256\pm20$ & $1880\pm250$ & $1478^{+225}_{-208}$  & $6.8-9.7$&$40^{+11}_{-13}$ & $24^{+6}_{-6}$ \\\noalign{\smallskip}
	     &   &                      &                    &                        &                                    & $\mathit{(21.7-30.7)}$            &                           &   \\\noalign{\smallskip}
3 & 3  & $241$ & $135\pm9$ & $3570\pm430$ & $2806^{+400}_{-360}$  & $13.0-18.4$&$77^{+20}_{-25}$ & $46^{+12}_{-11}$ \\\noalign{\smallskip}
	     &   &                      &                    &                        &                                    & $\mathit{(41.2-58.3)}$            &                           &   \\\noalign{\smallskip}
3 & 4  & $159\pm6$ & $115\pm2$ & $2770\pm280$ & $2176^{+275}_{-243}$  & $10.1-14.3$&$60^{+14}_{-19}$ & $36^{+9}_{-9}$ \\\noalign{\smallskip}
	     &   &                      &                    &                        &                                    & $\mathit{(32.0-45.3)}$            &                           &   \\\noalign{\smallskip}
3 & 5  & $132$ & $103\pm8$ & $2560\pm330$ & $2013^{+300}_{-275}$  &$9.4-13.2$ &$55^{+15}_{-17}$ & $33^{+8}_{-8}$ \\\noalign{\smallskip}
	     &   &                      &                    &                        &                                    & $\mathit{(29.6-41.8)}$            &                           &   \\\noalign{\smallskip}
\hline
\newpage
\noalign{\smallskip}
\cite{pascoe16}\\
\noalign{\smallskip}
\hline
\noalign{\smallskip}

43 & 4 (Loop \#1)& $222\pm31$ & $284\pm1$ & $1564\pm222$ & $1231^{+198}_{-183}$  &$9.38\pm2.56$ & $34^{+9}_{-11}$& $27^{+7}_{-7}$ \\\noalign{\smallskip}
     &   &                      &                    &                        &                                    & $\mathit{(30\pm8)}$            &                           &   \\\noalign{\smallskip}
31 & 1 (Loop \#2)& $162\pm31$ & $459\pm1$ & $706\pm136$ & $556^{+115}_{-110}$  & $4.37\pm0.88$ &$15^{+5}_{-5}$ & $13^{+4}_{-4}$ \\\noalign{\smallskip}
     &   &                      &                    &                        &                                    & $\mathit{(14\pm3)}$            &                           &   \\\noalign{\smallskip}
32 & 1 (Loop \#3)& $234\pm31$ & $250\pm1$ & $1871\pm252$ & $1456^{+228}_{-208}$ & $17.06\pm2.63$& $40^{+11}_{-13}$ & $58^{+15}_{-14}$ \\\noalign{\smallskip}
     &   &                      &                    &                        &                                    & $\mathit{(54\pm8)}$            &                           &   \\\noalign{\smallskip}

\end{longtable}

\begin{longtable}{ccccccccc}
\caption{Transverse loop oscillation data from \cite{goddard16} and Bayesian posterior summaries with and without damping. \label{tab_B2}}\\
\hline\hline
Event & Loop  & L & P& $\tau_{\rm d}$ & $v_{\rm ph}$ & $v_{\rm Ai}$ & $B_{\rm no \mbox{\hspace{0.05cm}}damping}$& $B_{\rm damping}$ \\  
 & & (Mm) & (s) & (s) & (km s$^{-1}$) & (km s$^{-1}$) & (G) & (G)\\
\hline
\endfirsthead
\caption{continued.}\\
\hline\hline
Event & Loop  & L & P& $\tau_{\rm d}$ & $v_{\rm ph}$ & $v_{\rm Ai}$ & $B_{\rm no \mbox{\hspace{0.05cm}}damping}$& $B_{\rm damping}$ \\  
 & & (Mm) & (s) & (s) & (km s$^{-1}$) & (km s$^{-1}$) & (G) & (G)\\
\hline
\endhead
\hline
\endfoot
\noalign{\smallskip}
1& 1 & $232$ & $205\pm4$ & $320\pm67$ &$2261\pm40$ & $1748^{+165}_{-63}$&   $50^{+9}_{-16}$  & $50^{+10}_{-16}$\\\noalign{\smallskip}
1& 2 & $78$ & $247\pm3$ &$646\pm167$& $633\pm8$ & $489^{+45}_{-15}$  & $14^{+3}_{-5}$ &$14^{+3}_{-4}$ \\\noalign{\smallskip}
2& 1 & $156$ & $398\pm4$ &-& $783\pm7$ & $605^{+58}_{-18}$  & $17^{+3}_{-5}$ & -\\\noalign{\smallskip}
3& 1 & $213$ & $148\pm2$ &$528\pm108$& $2886\pm35$ & $2228^{+210}_{-73}$  & $64^{+12}_{-20}$&  $65^{+13}_{-20}$ \\\noalign{\smallskip} 
3& 2 & $262$ & $217\pm5$ &$247\pm28$& $2413\pm53$ & $1868^{+173}_{-73}$  & $54^{+10}_{-17}$  & $53^{+10}_{-17}$\\\noalign{\smallskip}
3&3& $311$ & $242\pm6$ &-& $2566\pm64$ & $1988^{+185}_{-83}$  & $57^{+11}_{-18}$ &-\\\noalign{\smallskip}
4& 1 & $183$ & $137\pm2$ &$431\pm90$& $2664\pm35$ & $2056^{+195}_{-65}$  & $59^{+11}_{-19}$  & $60^{+11}_{-19}$ \\\noalign{\smallskip}
4& 2 & $181$ & $208\pm2$ &$446\pm60$& $1739\pm15$ & $1341^{+128}_{-40}$  & $38^{+7}_{-12}$ &  $39^{+7}_{-12}$ \\\noalign{\smallskip}
5&1&$438$ & $422\pm4$ &-& $2077\pm18$ & $1601^{+153}_{-48}$  & $57^{+11}_{-18}$ &- \\\noalign{\smallskip}
6&1&$430$ & $483\pm16$ &-& $1781\pm58$ & $1383^{+128}_{-65}$  & $40^{+7}_{-13}$ &- \\\noalign{\smallskip}
7& 1 & $162$ & $101\pm1$ &$434\pm78$& $3195\pm38$ & $2466^{+233}_{-78}$  & $71^{+13}_{-23}$ &  $72^{+14}_{-23}$ \\\noalign{\smallskip}
8& 1 & $207$ & $224\pm4$ &$600\pm60$& $1845\pm35$ & $1426^{+135}_{-50}$  & $41^{+8}_{-13}$  & $41^{+8}_{-13}$ \\\noalign{\smallskip}
9& 1 & $264$ & $308\pm10$ &$305\pm59$& $1712\pm57$ & $1331^{+123}_{-65}$ & $38^{+7}_{-11}$ &  $37^{+7}_{-11}$\\\noalign{\smallskip}
9& 2 & $326$ & $537\pm8$ &$710\pm286$ &$1214\pm19$ & $938^{+88}_{-33}$  & $27^{+5}_{-9}$  & $26^{+5}_{-8}$ \\\noalign{\smallskip}
10& 1 & $397$ & $688\pm10$ & $481\pm65$&$1155\pm17$ & $891^{+85}_{-30}$  & $25^{+5}_{-8}$ & $25^{+5}_{-8}$ \\\noalign{\smallskip}
10&2& $279$ & $509\pm10$ &-& $1097\pm21$ & $848^{+80}_{-30}$  & $24^{+5}_{-7}$ &- \\\noalign{\smallskip}
11&1& $78$ & $238\pm4$ &-& $657\pm12$ & $509^{+48}_{-20}$  & $15^{+3}_{-5}$ & -\\\noalign{\smallskip}
11&2& $95$ & $231\pm7$ &-& $823\pm24$ & $639^{+60}_{-28}$  & $18^{+3}_{-5}$ & -\\\noalign{\smallskip}
11& 3 & $118$ & $156\pm3$ &$530\pm90$& $1513\pm29$ & $1171^{+108}_{-43}$  & $34^{+6}_{-11}$ &$34^{+7}_{-11}$ \\ \noalign{\smallskip}
11&4& $125$ & $229\pm2$ &-& $1094\pm11$ & $843^{+80}_{-25}$ & $24^{+5}_{-7}$ &-   \\\noalign{\smallskip}
11&5& $135$ & $305\pm4$ &-& $884\pm10$ & $681^{+65}_{-20}$&  $19^{+4}_{-6}$ & -  \\\noalign{\smallskip}
11&6& $160$ & $368\pm13$ &-& $870\pm30$ & $676^{+63}_{-33}$ & $19^{+3}_{-6}$ & -   \\\noalign{\smallskip}
12&1& $148$ & $334\pm4$ &-& $887\pm11$ & $684^{+65}_{-20}$& $19^{+4}_{-6}$ & -   \\\noalign{\smallskip}
15&1& $174$ & $458\pm22$ &-& $759\pm37$ & $591^{+58}_{-35}$& $17^{+3}_{-5}$ & -  \\\noalign{\smallskip}
16&1& $242$ & $157\pm2$ &-& $3079\pm47$ & $2378^{+223}_{-80}$& $68^{+13}_{-21}$ & - \\ \noalign{\smallskip}
16& 2 & $146$ & $141\pm4$ &$161\pm38$& $2071\pm62$ & $1608^{+148}_{-75}$  & $46^{+9}_{-14}$  & $45^{+9}_{-14}$ \\\noalign{\smallskip}
16&3& $318$ & $314\pm11$ &-& $2027\pm74$ & $1576^{+148}_{-80}$&  $45^{+9}_{-14}$ & -  \\\noalign{\smallskip}
17& 1 & $153$ & $124\pm2$ &$599\pm275$& $2464\pm48$ & $1906^{+178}_{-70}$  & $54^{+11}_{-17}$ & $55^{+11}_{-17}$  \\\noalign{\smallskip}
18&1& $289$ & $431\pm19$ &-& $1342\pm60$ & $1046^{+98}_{-63}$& $29^{+6}_{-9}$ & - \\\noalign{\smallskip}
18& 2 & $284$ & $571\pm7 $&$732\pm208$ &$994\pm11$ & $766^{+73}_{-23}$  & $22^{+4}_{-7}$ & $22^{+4}_{-7}$  \\\noalign{\smallskip}
18&3& $393$ & $781\pm10$ &-& $1006\pm13$ & $776^{+75}_{-25}$& $22^{+5}_{-7}$ & - \\\noalign{\smallskip}
19&1& $123$ & $584\pm12$ &-& $421\pm9$ & $326^{+30}_{-13}$ & $9^{+2}_{-3}$ & -  \\\noalign{\smallskip}
19& 2 & $348$ & $676\pm7$ &$993\pm86$& $1029\pm11$ & $793^{+75}_{-25}$  & $23^{+4}_{-7}$ &   $23^{+4}_{-7}$ \\\noalign{\smallskip}
20& 1 & $253$ & $322\pm14$&$971\pm460$& $1573\pm68$ & $1226^{+115}_{-70}$ & $35^{+7}_{-11}$ &  $35^{+7}_{-11}$ \\\noalign{\smallskip}
21&1& $499$ & $429\pm121$ &-& $2326\pm654$ & $1833^{+532}_{-517}$ & $48^{+20}_{-19}$ & - \\ \noalign{\smallskip}
22&1& $288$ & $162\pm7$ &-& $3556\pm145$ & $2768^{+257}_{-153}$ & $78^{+15}_{-25}$ & -  \\\noalign{\smallskip}
23& 1 & $365$ & $922\pm24$ &$1151\pm93$& $792\pm21$ & $614^{+58}_{-25}$  & $17^{+3}_{-5}$  & $17^{+3}_{-6}$ \\\noalign{\smallskip}
24& 1 & $432$ & $1072\pm18$ &$1646\pm256$& $806\pm14$ & $625^{+58}_{-23}$  & $18^{+3}_{-6}$ &  $18^{+3}_{-6}$ \\ \noalign{\smallskip}
24&2& $427$ & $987\pm17$ &-& $865\pm15$ & $669^{+63}_{-23}$ &  $19^{+3}_{-6}$ &- \\\noalign{\smallskip}
24& 3 & $538$ & $1228\pm35$ &$2101\pm386$& $877\pm25$ & $681^{+63}_{-30}$  & $19^{+4}_{-6}$ & $19^{+4}_{-6}$ \\\noalign{\smallskip}
25& 1 & $156$ & $308\pm7$ &$480\pm300$& $1014\pm22$ & $786^{+73}_{-33}$  & $22^{+5}_{-7}$  & $22^{+4}_{-7}$  \\\noalign{\smallskip}
25&2& $264$ & $438\pm10$ &-& $1205\pm26$ & $933^{+85}_{-38}$ &  $27^{+5}_{-9}$ & -  \\\noalign{\smallskip}
26& 1 & $473$ & $717\pm8$ &$1123\pm270$& $1319\pm14$ & $1018^{+95}_{-33}$ & $29^{+5}_{-9}$ & $29^{+6}_{-9}$  \\\noalign{\smallskip}
26&2& $185$ & $751\pm11$ &-& $493\pm7$ & $381^{+35}_{-13}$& $11^{+2}_{-3}$ &-  \\\noalign{\smallskip}
27&1& $244$ & $917\pm24$ &-& $532\pm14$ & $414^{+38}_{-18}$&  $12^{+2}_{-4}$ & -  \\\noalign{\smallskip}
29& 1 & $154$ & $223\pm3$ &$470\pm37$& $1384\pm19$ & $1068^{+100}_{-35}$  & $31^{+5}_{-10}$  & $31^{+6}_{-9}$ \\\noalign{\smallskip}
31& 1 & $162$ & $460\pm2$ &$1453\pm121$& $704\pm4$ & $544^{+50}_{-18}$  & $15^{+3}_{-5}$  & $16^{+3}_{-5}$ \\\noalign{\smallskip}
31& 2 & $138$ & $575\pm5$ &$1054\pm141$& $480\pm5$ & $371^{+35}_{-13}$  & $11^{+2}_{-3}$ &  $11^{+2}_{-3}$ \\\noalign{\smallskip}
31&3& $532$ & $694\pm7$ &-& $1534\pm16$ & $1183^{+113}_{-38}$&  $34^{+6}_{-10}$ & - \\ \noalign{\smallskip}
32& 1 & $234$ & $257\pm1$ & $933\pm73$&$1822\pm9$ & $1403^{+135}_{-40}$  & $40^{+8}_{-13}$ &  $41^{+8}_{-13}$ \\\noalign{\smallskip}
32& 2 & $233$ & $203\pm1$ &$1147\pm291$& $2298\pm14$ & $1771^{+170}_{-53}$  & $51^{+9}_{-16}$ & $52^{+10}_{-17}$  \\\noalign{\smallskip}
33&1& $314$ & $281\pm5$ &-& $2232\pm38$ & $1726^{+160}_{-60}$&  $50^{+9}_{-16}$ & -  \\\noalign{\smallskip}
33&2& $407$ & $391\pm6$ &-& $2081\pm32$ & $1608^{+150}_{-55}$ & $46^{+9}_{-15}$ & - \\\noalign{\smallskip}
34& 1 & $333$ & $597\pm16$ &$1002\pm62$& $1116\pm30$ & $866^{+80}_{-38}$  & $25^{+5}_{-8}$  & $25^{+5}_{-8}$ \\\noalign{\smallskip}
35&1& $327$ & $527\pm8$ &-& $1241\pm18$ & $958^{+90}_{-33}$& $27^{+5}_{-9}$ &-  \\\noalign{\smallskip}
35&2& $312$ & $346\pm6$ &-& $1802\pm31$ & $1393^{+130}_{-50}$& $40^{+8}_{-12}$ & -  \\\noalign{\smallskip}
36&1& $282$ & $401\pm6$ &-& $1407\pm21$ & $1086^{+103}_{-35}$&$31^{+6}_{-9}$ & -  \\\noalign{\smallskip}
37&1& $358$ & $496\pm13$ &-& $1443\pm38$ & $1118^{+103}_{-48}$ &  $32^{+6}_{-10}$ & - \\\noalign{\smallskip}
38&1& $224$ & $182\pm2$ &-& $2456\pm24$ & $1893^{+180}_{-58}$&  $54^{+10}_{-17}$ & -  \\\noalign{\smallskip}
38& 2 & $270$ & $312\pm5$ &$914\pm330$ &$1731\pm27$ & $1338^{+125}_{-48}$  & $38^{+7}_{-12}$ &  $39^{+7}_{-12}$  \\\noalign{\smallskip}
38&3& $424$ & $785\pm13$ &-& $1081\pm17$ & $836^{+78}_{-30}$&  $24^{+4}_{-8}$ &-   \\\noalign{\smallskip}
38&4& $478$ & $644\pm11$ &-& $1484\pm25$ & $1146^{+108}_{-40}$&  $33^{+6}_{-10}$ & - \\ \noalign{\smallskip}
39&1& $402$ & $647\pm6$ &-& $1242\pm12$ & $958^{+90}_{-30}$ &  $27^{+5}_{-9}$ & -  \\\noalign{\smallskip}
39&2& $334$ & $641\pm7$ &-& $1042\pm12$ & $803^{+78}_{-25}$ &  $23^{+5}_{-7}$ & - \\\noalign{\smallskip}
39&3& $376$ & $754\pm22$ &-& $997\pm29$ & $773^{+73}_{-35}$ &$22^{+4}_{-7}$ &- \\\noalign{\smallskip}
39&4& $454$ & $858\pm10$ &-& $1058\pm13$ & $816^{+78}_{-25}$&  $23^{+5}_{-7}$ & -  \\\noalign{\smallskip}
40&1& $171$ & $341\pm4$ &-& $1004\pm11$ & $773^{+75}_{-23}$& $22^{+4}_{-7}$ & - \\\noalign{\smallskip}
40& 2 & $347$ & $337\pm2$ &$1490\pm205$& $2062\pm11$ & $1588^{+153}_{-48}$  & $46^{+9}_{-14}$ &  $47^{+9}_{-14}$ \\\noalign{\smallskip}
40&3& $325$ & $355\pm42$ &-& $1830\pm216$ & $1438^{+203}_{-183}$& $40^{+10}_{-13}$ & - \\ \noalign{\smallskip}
40& 4 & $258$ & $332\pm2$ & $439\pm65$&$1555\pm11$ & $1198^{+115}_{-35}$  & $34^{+7}_{-11}$ & $34^{+6}_{-11}$ \\\noalign{\smallskip}
40&5& $297$ & $325\pm1$ &-& $1827\pm7$ & $1408^{+135}_{-43}$& $40^{+8}_{-13}$ & -   \\\noalign{\smallskip}
40&6& $425$ & $416\pm2$ &-& $2044\pm12$ & $1576^{+150}_{-48}$& $46^{+8}_{-15}$ & -  \\\noalign{\smallskip}
40& 7 & $353$ & $343\pm4$ &$850\pm164$& $2057\pm22$ & $1586^{+153}_{-48}$  & $46^{+9}_{-14}$ &  $46^{+9}_{-14}$ \\\noalign{\smallskip}
40& 8 & $238$ & $260\pm5$ &$541\pm130$ &$1832\pm34$ & $1416^{+133}_{-50}$  & $40^{+8}_{-13}$  & $41^{+8}_{-13}$ \\\noalign{\smallskip}
40& 9 & $473$ & $371\pm3$ &$789\pm160$& $2551\pm21$ & $1966^{+188}_{-58}$  & $56^{+11}_{-17}$  & $57^{+11}_{-17}$\\\noalign{\smallskip}
40&10& $238$ & $376\pm2$ &-& $1265\pm6$ & $976^{+93}_{-30}$& $28^{+5}_{-9}$ & -  \\\noalign{\smallskip}
40&11& $220$ & $286\pm2$ &-& $1541\pm13$ & $1188^{+113}_{-35}$& $34^{+7}_{-11}$ & - \\\noalign{\smallskip}
43& 1 & $363$ & $428\pm4$ & $452\pm87$&$1695\pm17$ & $1308^{+123}_{-43}$  & $38^{+7}_{-12}$ &  $37^{+7}_{-12}$ \\\noalign{\smallskip}
43& 2 & $241$ & $216\pm2$ &$566\pm55$& $2231\pm19$ & $1721^{+163}_{-53}$  & $50^{+9}_{-16}$ & $50^{+10}_{-15}$ \\\noalign{\smallskip}
43& 3 & $368$ & $501\pm5$ &$902\pm109$& $1469\pm14$ & $1133^{+108}_{-35}$  & $33^{+6}_{-11}$ & $33^{+6}_{-10}$  \\\noalign{\smallskip}
43&4& $222$ & $310\pm2$ &-& $1434\pm8$ & $1106^{+105}_{-33}$&  $32^{+6}_{-10}$ & -  \\\noalign{\smallskip}
43& 5 & $260$ & $270\pm1$ &$840\pm120$ &$1926\pm9$ & $1483^{+143}_{-43}$  & $43^{+8}_{-14}$  & $44^{+8}_{-13}$  \\\noalign{\smallskip}
44& 1 & $295$ & $434\pm4$ & $945\pm185$&$1360\pm11$ & $1048^{+100}_{-33}$  & $30^{+5}_{-9}$  & $30^{+6}_{-9}$ \\\noalign{\smallskip}
44& 2 & $512$ & $587\pm11$ &$877\pm298$& $1745\pm34$ & $1351^{+125}_{-50}$ & $39^{+7}_{-13}$ & $38^{+7}_{-12}$ \\\noalign{\smallskip}
44& 3 & $352$ & $417\pm8$ &$540\pm180$& $1688\pm34$ & $1306^{+123}_{-50}$  & $38^{+7}_{-12}$ & $37^{+7}_{-12}$ \\\noalign{\smallskip}
44&4& $202$ & $145\pm3$ &-& $2794\pm58$ & $2163^{+200}_{-83}$&  $62^{+12}_{-19}$ & - \\\noalign{\smallskip}
45& 1 & $92$ & $149\pm2$ & $469\pm100$&$1237\pm20$ & $956^{+90}_{-33}$  & $27^{+5}_{-9}$ & $28^{+5}_{-9}$  \\\noalign{\smallskip}
46&1& $430$ & $724\pm14$ &-& $1188\pm23$ & $918^{+85}_{-33}$& $26^{+5}_{-8}$ & -   \\\noalign{\smallskip}
46&2& $498$ & $659\pm7$ &-& $1510\pm15$ & $1166^{+110}_{-38}$& $34^{+6}_{-11}$ & -  \\\noalign{\smallskip}
46&3& $384$ & $594\pm6$ &-& $1293\pm13$ & $998^{+93}_{-33}$&$29^{+5}_{-9}$ &-  \\\noalign{\smallskip}
47&1& $225$ & $316\pm8$ &-& $1423\pm38$ & $1103^{+103}_{-48}$& $32^{+6}_{-10}$ -  \\\noalign{\smallskip}
47&2& $222$ & $301\pm7$ &-& $1474\pm35$ & $1141^{+108}_{-45}$& $33^{+6}_{-11}$ & -  \\\noalign{\smallskip}
48& 1 & $540$ & $917\pm10$ &$1319\pm936$& $1178\pm12$ & $908^{+88}_{-28}$  & $26^{+5}_{-9}$ &  $26^{+5}_{-8}$ \\\noalign{\smallskip}
48& 2 & $588$ & $946\pm7$ &$1598\pm130$& $1244\pm9$ & $958^{+93}_{-28}$  & $27^{+5}_{-9}$  &  $27^{+5}_{-8}$ \\\noalign{\smallskip}
48& 3 & $597$ & $965\pm13$ &$946\pm185$ &$1238\pm16$ & $956^{+90}_{-33}$  & $27^{+5}_{-9}$ & $27^{+5}_{-8}$ \\\noalign{\smallskip}
48&4& $426$ & $554\pm14$ &-& $1538\pm38$ & $1191^{+110}_{-48}$ & $34^{+7}_{-11}$ & - \\\noalign{\smallskip}
48&5& $471$ & $950\pm13$ &-& $992\pm13$ & $766^{+73}_{-25}$& $22^{+4}_{-7}$ &- \\\noalign{\smallskip}
49&1& $484$ & $907\pm28$ &-& $1067\pm33$ & $828^{+78}_{-38}$&  $23^{+5}_{-7}$ & -  \\\noalign{\smallskip}
49&2& $197$ & $464\pm8$ &-& $850\pm15$ & $656^{+63}_{-23}$&  $19^{+3}_{-6}$ &-  \\\noalign{\smallskip}
49& 4 & $386$ & $627\pm10$ &$923\pm155$ &$1231\pm20$ & $951^{+90}_{-33}$  & $27^{+5}_{-9}$  & $27^{+5}_{-8}$ \\\noalign{\smallskip}
49& 5 & $191$ & $482\pm11$ &$562\pm73$& $793\pm18$ & $614^{+58}_{-23}$  & $17^{+3}_{-5}$  &  $17^{+3}_{-6}$ \\\noalign{\smallskip}
52&1& $183$ & $356\pm7$ &-& $1029\pm21$ & $796^{+75}_{-30}$&  $23^{+4}_{-7}$ & - \\\noalign{\smallskip}
53&1& $420$ & $569\pm13$ &-& $1477\pm34$ & $1143^{+108}_{-45}$&  $33^{+6}_{-11}$ & - \\ \noalign{\smallskip}
54&1& $408$ & $500\pm4$ &-& $1633\pm14$ & $1258^{+120}_{-38}$&  $36^{+7}_{-11}$ & - \\\noalign{\smallskip}
54&2& $400$ & $448\pm6$ &-& $1787\pm24$ & $1378^{+133}_{-45}$&  $40^{+7}_{-13}$ & - \\\noalign{\smallskip}
54&3& $238$ & $139\pm3$ &-& $3420\pm74$ & $2648^{+245}_{-103}$&  $76^{+15}_{-24}$ & - \\\noalign{\smallskip}
54&4& $355$ & $226\pm8$ &-& $3139\pm108$ & $2441^{+225}_{-123}$& $69^{+14}_{-21}$ & - \\\noalign{\smallskip}
54& 5 & $257$ & $288\pm6$ &$1183\pm194$ &$1785\pm37$ & $1381^{+130}_{-53}$& $40^{+7}_{-13}$ & $41^{+8}_{-13}$ \\\noalign{\smallskip}
55&1& $405$ & $518\pm14$ &-& $1564\pm44$ & $1213^{+113}_{-53}$& $35^{+7}_{-11}$ & -  \\\noalign{\smallskip}
55&2& $477$ & $392\pm10$ &-& $2431\pm63$ & $1883^{+175}_{-78}$ & $54^{+11}_{-17}$ & -  \\\noalign{\smallskip}
56& 1 & $403$ & $544\pm8$ &$1243\pm283$& $1481\pm23$ & $1143^{+108}_{-38}$  & $33^{+6}_{-10}$ &  $33^{+6}_{-10}$ \\\noalign{\smallskip}
56& 2 & $314$ & $713\pm8$ &$1177\pm178$& $881\pm10$ & $681^{+63}_{-23}$  & $19^{+4}_{-6}$ & $19^{+4}_{-6}$  \\\noalign{\smallskip}
56&3& $205$ & $193\pm10$ &-& $2122\pm105$ & $1656^{+158}_{-105}$&  $47^{+9}_{-15}$ & -  \\ \noalign{\smallskip}
56&4& $501$ & $863\pm20$ &-& $1161\pm27$ & $898^{+85}_{-35}$&  $25^{+5}_{-8}$ & -  \\\noalign{\smallskip}
56& 5 & $431$ & $810\pm10$ &$1450\pm308$& $1064\pm13$ & $821^{+78}_{-25}$ & $23^{+5}_{-7}$ & $23^{+4}_{-8}$ \\\noalign{\smallskip}
56&6& $392$ & $455\pm12$ &-& $1722\pm45$ & $1336^{+123}_{-58}$&  $38^{+7}_{-12}$ & - \\ \noalign{\smallskip}
56& 7 & $457$ & $850\pm33$ &$818\pm236$& $1117\pm45$ & $868^{+83}_{-48}$ & $25^{+5}_{-8}$ & $24^{+5}_{-8}$  \\\noalign{\smallskip}
56&8& $379$ & $638\pm9$ &-& $1187\pm17$ & $916^{+88}_{-30}$ & $26^{+5}_{-8}$ & - \\\noalign{\smallskip}
\end{longtable}
\end{onecolumn}

\end{appendix}

\end{document}